\documentclass[final,1p,times,twocolumn]{elsarticle}

\usepackage{graphicx}

\usepackage{amssymb}
\usepackage{color}
\usepackage{amssymb}
\usepackage{amsmath}
\usepackage{epsfig}
\usepackage{tabularx}
\usepackage{multirow}
\usepackage{subfigure}
\usepackage{algorithm}
\usepackage{algorithmic}
\usepackage{psfrag}
\usepackage{booktabs}
\usepackage{xspace}
\usepackage[bottom]{footmisc}
\usepackage{rotating}
\usepackage{rotating}
\usepackage{nomencl}
\usepackage{ifthen}
\usepackage{subfigmat}
\usepackage{captcont}
\usepackage{wrapfig}
\usepackage{clrscode}
\usepackage{paralist}

\renewcommand{\nomgroup}[1]{
\ifthenelse{\equal{#1}{S}}{\item[\textbf{Subscripts}]}{
\ifthenelse{\equal{#1}{G}}{\item[\textbf{Greek symbols}]}{}}}
\usepackage{rotating}

\journal{Elsevier}

\begin{document}
\newcommand{\etal}{\emph{et al.}}
\newcommand{\uVech}{\boldsymbol{u}_h}
\newcommand{\vVech}{\boldsymbol{v}_h}
\newcommand{\uvec}{{\bf u}}
\newcommand{\wvec}{{\boldsymbol{w}}}
\newcommand{\kvec}{{\boldsymbol{z}}}
\newcommand{\dvec}{{\boldsymbol{d}}}
\newcommand{\evec}{{\boldsymbol{e}}}
\newcommand{\pvec}{{\boldsymbol{p}}}
\newcommand{\qvec}{{\boldsymbol{q}}}
\newcommand{\zvec}{{\bf z}}
\newcommand{\Wvec}{{\bf W}}
\newcommand{\Uvec}{{\bf U}}
\newcommand{\vvec}{{\bf v}}
\newcommand{\Fvec}{{\bf F}}
\newcommand{\Fvecc}{{\Fvec^\mathrm{c}}}
\newcommand{\Fvecv}{{\Fvec^\mathrm{v}}}
\newcommand{\Fcij}[1]{{F_{#1}^\mathrm{c}}}
\newcommand{\Fvij}[1]{{F_{#1}^\mathrm{v}}}
\newcommand{\numFcij}[1]{{\widehat{F}_{#1}^\mathrm{c}}}
\newcommand{\numFvij}[1]{{\widehat{F}_{#1}^\mathrm{v}}}
\newcommand{\fvec}{{\bf f}}
\newcommand{\Ivec}{{\bf I}}
\newcommand{\intV}[2]{\int_{#1} #2 \ud \Omega} 
\newcommand{\alert}[1]{{\color{red}#1}}
\newcommand{\Pn}[1]{$\mathbb{P}^{#1}$}
\newcommand{\Pol}{\mathbb{P}}

\newcommand{\Tr}{\mathrm{Tr}}
\newcommand{\T}{\kappa}
\newcommand{\f}{\sigma}
\newcommand{\Kl}{\kappa_{\ell}}
\newcommand{\Klpo}{\kappa_{\ell+1}}
\newcommand{\Tpr}{\kappa'}
\newcommand{\lnt}{\widetilde{T}}
\newcommand{\lnp}{\widetilde{p}}

\newcommand{\jump}[1]{\ldbracket #1 \rdbracket}
\newcommand{\averg}[1]{\lgbracket #1 \rgbracket}

\newcommand{\ko}{$k$-$\omega$}
\newcommand{\boldko}{$\boldsymbol{k}$-$\boldsymbol{\omega}$}
\newcommand{\lomega}{\widetilde{\omega}}
\newcommand{\romega}{\widetilde{\omega}_{r}}
\newcommand{\romegao}{\widetilde{\omega}_{r0}}
\newcommand{\lk}{\overline{k}}
\newcommand{\mut}{\nu_{t}}
\newcommand{\lmut}{\overline{\mu}_{t}}
\newcommand{\turb}{{\rm Tu}}
\newcommand{\tsca}{\ell}

\newcommand{\Poly}[2]{\mathbb{P}_{#1}^{#2}}
\newcommand{\ST}{\,:\,}

\newcommand{\JVER}{\hat{j}}
\newcommand{\BETA}{\boldsymbol{\beta}}
\newcommand{\NABLA}{\boldsymbol{\nabla}}
\newcommand{\NABLAH}{\boldsymbol{\nabla}_h}
\newcommand{\LAMBDA}{\boldsymbol{\lambda}}
\newcommand{\ZERO}{{\bf 0}}
\newcommand{\SIGMA}{\boldsymbol{\sigma}}
\newcommand{\TAU}{\boldsymbol{\tau}}
\newcommand{\EPSILON}{\boldsymbol{\epsilon}}
\newcommand{\ZETA}{\boldsymbol{\zeta}}
\newcommand{\PHI}{\boldsymbol{\phi}}
\newcommand{\PSI}{\boldsymbol{\psi}}
\newcommand{\VARPHI}{\boldsymbol{\varphi}}
\newcommand{\U}{{\bf u}}
\newcommand{\Ui}[1]{u_{#1}}
\newcommand{\Q}{{\bf q}}
\newcommand{\Qi}[1]{q_{#1}}
\newcommand{\xvec}{{\bf x}}
\newcommand{\rvec}{{\bf r}}
\newcommand{\V}{{\bf v}}
\newcommand{\N}{{\bf n}}
\newcommand{\G}{{\bf g}}
\newcommand{\W}{{\bf w}}
\newcommand{\Wi}[1]{w_{#1}}
\newcommand{\Wdofi}[1]{W_{#1}}
\newcommand{\E}{{\bf e}}
\newcommand{\bS}{{\bf s}}
\newcommand{\Si}[1]{s_{#1}}
\newcommand{\X}{{\bf x}}
\newcommand{\rf}[1]{{\bf r}_f(#1)}
\newcommand{\F}{{\bf f}}
\newcommand{\Z}{{\bf z}}
\newcommand{\bF}{{\bf F}}
\newcommand{\bM}{{\bf M}}
\newcommand{\bR}{{\bf R}}
\newcommand{\bU}{{\bf U}}
\newcommand{\bW}{{\bf W}}
\newcommand{\bv}{{\bf v}}
\newcommand{\bA}{{\bf A}}
\newcommand{\bK}{{\bf K}}
\newcommand{\Jm}{{\boldsymbol{J}}}
\newcommand{\Gm}{{\boldsymbol{G}}}
\newcommand{\Mm}{{\boldsymbol{M}}}
\newcommand{\fv}{{\boldsymbol{f}}}
\newcommand{\gv}{{\boldsymbol{g}}}
\newcommand{\SysM}[1]{\mathbf{J}_{#1}}
\newcommand{\B}{{\bf b}}
\newcommand{\bI}{{\bf I}}

\newcommand{\WX}{\widehat{\X}}
\newcommand{\Wx}{\widehat{x}}
\newcommand{\Wf}{\widehat{f}}
\newcommand{\WK}{\widehat{K}}
\newcommand{\WKs}{{\widehat{S}}}
\newcommand{\WKt}{{\widehat{T}}}
\newcommand{\WKq}{{\widehat{Q}}}
\newcommand{\WKh}{{\widehat{H}}}



\newcommand{\cons}{\mathbf{u}}
\newcommand{\prim}{\mathbf{v}}
\newcommand{\vel}{\boldsymbol{v}}
\newcommand{\avel}{\boldsymbol{a}}
\newcommand{\vmu}{\mu}
\newcommand{\vnu}{\nu}
\newcommand{\vk}{k}
\newcommand{\vlam}{\lambda}
\newcommand{\temp}{T}
\newcommand{\pres}{P}
\newcommand{\dens}{\rho}
\newcommand{\ener}{\mathrm{e}}
\newcommand{\tote}{E}
\newcommand{\enth}{h}
\newcommand{\toth}{H}
\newcommand{\totr}{H_r}
\newcommand{\hflx}{\boldsymbol{q}}
\newcommand{\stress}{\boldsymbol{T}}
\newcommand{\cflx}{\mathcal{F}_{c}}
\newcommand{\Gflx}{\boldsymbol{\mathcal{F}}}
\newcommand{\Cflx}{\boldsymbol{\mathcal{F}}_{c}}
\newcommand{\cflxn}{\boldsymbol{\mathbf{f}}_{c}}
\newcommand{\Vflx}{\boldsymbol{\mathcal{F}}_{v}}
\newcommand{\nvec}{\boldsymbol{n}}
\newcommand{\avec}{\boldsymbol{a}}
\newcommand{\data}{{\boldsymbol{\alpha}}}
\newcommand{\numf}[1]{\widehat{#1}}
\newcommand{\cjac}{{\bf A}_c}
\newcommand{\vjac}{{\bf D}_v}
\newcommand{\bcd}{\phi}
\newcommand{\rin}{g}
\newcommand{\Rin}{\boldsymbol{g}}

\newcommand{\nele}{{N_e}}
\newcommand{\nbas}{{N_b}}
\newcommand{\nqua}{{N_q}}

\newcommand{\bfcg}{N}               
\newcommand{\bfdg}[2]{N_{#1#2}}     
\newcommand{\bfun}{N}               
\newcommand{\wbfun}{\widehat{N}}    

\newcommand{\mpuf}{u}               
\newcommand{\mptf}{v}               
\newcommand{\mpud}{U}               
\newcommand{\mptd}{V}               
\newcommand{\mpug}{\boldsymbol{z}}  
\newcommand{\mptg}{\boldsymbol{w}}  
\newcommand{\mpgl}[1]{\boldsymbol{r}\left(#1\right)}   
\newcommand{\mpll}[1]{\boldsymbol{r}_e\left(#1\right)} 
\newcommand{\mpfs}{{V_h}}                
\newcommand{\mpgs}{{\boldsymbol{V}_h}}   
\newcommand{\mpugb}{{(\mpug\cdot\nvec)_b}}
\newcommand{\sflx}{\boldsymbol{f}}  
\newcommand{\sflxn}{{f_{n}}}        
\newcommand{\sflxnb}{{f_{nb}}}      
\newcommand{\sdm}{A}                
\newcommand{\normal}{\mathbf{n}}

\newcommand{\nsuf}{\mathbf{u}}                   
\newcommand{\nstf}{\mathbf{v}}                   
\newcommand{\nsug}{\boldsymbol{\mathcal{Z}}}     
\newcommand{\nstg}{\boldsymbol{\mathcal{W}}}     
\newcommand{\nsgl}[1]{\boldsymbol{\mathcal{R}}\left(#1\right)}   
\newcommand{\nsll}[1]{\boldsymbol{\mathcal{R}}_e\left(#1\right)} 
\newcommand{\nsfs}{{\mathbf{V}_h}}               
\newcommand{\nsgs}{{\boldsymbol{\mathcal{V}}_h}} 
\newcommand{\nsugb}{(\partial_n\nsuf)_b}

\newcommand{\rnum}{{\mathbb{R}}}
\newcommand{\nnum}{{\mathbb{N}}}
\newcommand{\Px}[2]{{\mathbb{P}_{#1}(#2)}}
\newcommand{\Pk}[1]{{\mathbb{P}_{k}\left(#1\right)}}

\newcommand{\JWGRAD}[1]
           {(J^{-1}\widehat{\boldsymbol{\nabla}})#1}
\newcommand{\WGRAD}[1]{\widehat{\boldsymbol{\nabla}}#1}
\newcommand{\GRAD}[1]{\boldsymbol{\nabla}#1}
\newcommand{\GRADU}[1]{\boldsymbol{\nabla_u}#1}
\newcommand{\DIV}[1]{\boldsymbol{\nabla}\cdot#1}
\newcommand{\LAP}{\nabla^2}
\newcommand{\eqbydef}{\stackrel{\mathrm{def}}{=}}
\newcommand{\diad}{:}
\newcommand{\tria}{\odot}

\newcommand{\ldbracket}{\left[\hspace{-1.5pt}\left[}
\newcommand{\rdbracket}{\right]\hspace{-1.5pt}\right]}
\newcommand{\lgbracket}{\left\{\hspace{-2.0pt}\left\{}
\newcommand{\rgbracket}{\right\}\hspace{-2.0pt}\right\}}

\newcommand{\jumpv}[1]{\ldbracket #1 \rdbracket_{\star}}
\newcommand{\jumps}[1]{\ldbracket #1 \rdbracket}
\newcommand{\jumpsb}[1]{\ldbracket #1 \rdbracket_{0}}
\newcommand{\avg}[1]{\left\lbrace #1 \right\rbrace}
\newcommand{\ovel}{{\boldsymbol \omega}}

\newcommand{\deriv}[2]{\frac{\mathrm{d} #1}{\mathrm{d} #2}}
\newcommand{\pderiv}[2]{\frac{\partial #1}{\partial #2}}
\newcommand{\partialt}[1]{\frac{\partial #1}{\partial t}}
\newcommand{\partialn}[1]{\frac{\partial #1}{\partial\N}}
\newcommand{\partialx}[1]{\frac{\partial #1}{\partial x}}

\newcommand{\ud}{~\mathrm{d}}
\newcommand{\uD}{~\mathrm{D}}
\newcommand{\nablah}{\nabla_{h}}

\newcommand{\grid}{\mathcal{T}_h}           
\newcommand{\intg}{\mathcal{E}_h^0}         
\newcommand{\boug}{\mathcal{E}_h^\partial}  
\newcommand{\allg}{\mathcal{E}_h}           

\newcommand{\dom}{\Omega}                   
\newcommand{\bou}{{\partial \Omega}}        
\newcommand{\infl}{{\partial \Omega_{in}}}  
\newcommand{\outf}{{\partial \Omega_{out}}} 
\newcommand{\Dbc}{{\partial \Omega_{0}}}    
\newcommand{\Nbc}{{\partial \Omega_{n}}}    
\newcommand{\calc}{\mathcal{C}}             

\newcommand{\domh}{{\Omega_h}}            
\newcommand{\inth}{{\Gamma_h^0}}          
\newcommand{\bouh}{{\partial \Omega_h}}   
\newcommand{\allh}{{\Gamma_h}}            
\newcommand{\dome}{{K}}                   
\newcommand{\inte}{{\partial K^0}}        
\newcommand{\boue}{{\partial K^\partial}} 

\newcommand{\infh}{{\partial \Omega_{h,in}}}  
\newcommand{\oufh}{{\partial \Omega_{h,out}}} 
\newcommand{\Dbch}{{\partial \Omega_{h,0}}}   
\newcommand{\Nbch}{{\partial \Omega_{h,n}}}   

\newcommand{\intD}[2]{\int_{#1} #2 \ud{\bf x}}              
\newcommand{\intWD}[2]{\int_{#1} #2 \ud{\widehat{\bf x}}}   
\newcommand{\intSEG}[1]{\int_{-1}^{1} #1 \ud\xi}            

\newcommand{\Jaco}[3]{P_{#3}^{#1,#2}}          
\newcommand{\Lege}[1]{L_{#1}}                  
\newcommand{\Moda}[1]{B_{#1}}                  
\newcommand{\Lagr}[2]{H_{#1}^{#2}}             

\newcommand{\jmap}{{\bf J}}                      
\newcommand{\matr}{{\bf A}}                      
\newcommand{\mpUd}{{\bf U}}                      
\newcommand{\mass}{{\bf M}}                      
\newcommand{\resid}{{\bf R}}                     
\newcommand{\residU}{{\bf R}\left(\mpUd\right)}  
\newcommand{\pmat}{{\bf P}}                      
\newcommand{\Pij}[1]{{P}_{#1}}            
\newcommand{\Mp}{{\bM_{\pmat}}}           
\newcommand{\invMp}{\bM_\pmat^{-1}}       
\newcommand{\M}{{\bM}}                    
\newcommand{\invM}{\bM^{-1}}              

\newcommand{\FRAC}[2]{\frac{\partial}{\partial #1}\left( #2 \right)}
\newcommand{\FRACs}[2]{\frac{\partial #1}{\partial #2}}

\newcommand{\intB}[2]{\int_{#1} #2 } 
\newcommand{\intO}[1]{\displaystyle \int_{\Omega} #1 }
\newcommand{\sumT}{\displaystyle\sum_{\T\in\Th}}
\newcommand{\sumF}{\displaystyle\sum_{{\sigma}\in\Ff}}
\newcommand{\Ffb}{\Ff^{\rm b}}
\newcommand{\Ffi}{\Ff^{\rm i}}
\newcommand{\intT}[1]{\displaystyle \int_{\T} #1}
\newcommand{\intF}[1]{\displaystyle \int_{\f} #1}
\newcommand{\Imat}{{\bf Id}}
\newcommand{\Th}{\mathcal{T}_h}
\newcommand{\Tf}{\mathcal{T}_h}
\newcommand{\Ff}{\mathcal{F}_{h}}
\newcommand{\Ft}{\mathcal{F}_{\partial \T}}
\newcommand{\sumf}{\sum_{{f}\in\mathcal{F}}}
\newcommand{\ie}{\emph{i.e.}}
\newcommand{\eg}{\emph{e.g.}}

\newcommand{\diss}{\varepsilon}
\newcommand{\disR}{\varepsilon^{\mathrm{R}}}
\newcommand{\disS}{\varepsilon^{\mathrm{S}}}
\newcommand{\disRS}{\varepsilon^{\mathrm{R|S}}}

\newcommand{\lR}{l^{\mathrm{R}}}
\newcommand{\lS}{l^{\mathrm{S}}}
\newcommand{\lRS}{l^{\mathrm{R|S}}}
\newcommand{\lX}{\tilde{l}}

\newcommand{\tR}{t^{\mathrm{R}}}
\newcommand{\tS}{t^{\mathrm{S}}}
\newcommand{\tRS}{t^{\mathrm{R|S}}}
\newcommand{\tX}{\tilde{t}}

\newcommand{\lmutR}{\lmut^{\mathrm{R}}}
\newcommand{\lmutS}{\lmut^{\mathrm{S}}}
\newcommand{\lmutRS}{\lmut^{\mathrm{R|S}}}

\newcommand{\flt}{\Delta}
\newcommand{\cus}{C_1}
\newcommand{\cds}{C_2}

\newcommand{\xomega}{\hat{\omega}}

\newcommand{\Cp}{$C_p$}
\newcommand{\Cf}{$C_f$}
\newcommand{\Cd}{$C_d$}
\newcommand{\Cl}{$C_l$}
\newcommand{\Cm}{$C_m$}
\newcommand{\Mach}{\mathrm{M}}
\newcommand{\Reynolds}{\mathrm{Re}}
\newcommand{\Pmat}{{\bf P}}
\newcommand{\Jvec}{{\bf J}}

\newcommand{\IntOp}{\mathcal{I}}
\newcommand{\IntOpB}{{\bf{\mathcal{I}}}}
\newcommand{\card}{\mathrm{card}}

\maketitle    
\makenomenclature



\begin{frontmatter}

\title{\emph{p}-Multigrid matrix-free discontinuous Galerkin \\solution strategies for the under-resolved simulation of \\incompressible turbulent flows}
%
%


\author[label1,label3]{M.~Franciolini\corref{cor1}}
\author[label2]{L.~Botti}
\author[label2]{A.~Colombo}
\author[label3]{A.~Crivellini}

\address[label1]{\ USRA Fellow, NASA Ames Research Center, \\Mountain View, CA, 94035, United States}
\address[label2]{\ Dipartimento di Ingegneria e Scienze Applicate, \\ Universit\`a degli Studi di Bergamo, 24044 Dalmine (BG), Italy}
\address[label3]{\ Dipartimento di Ingegneria Industriale e Scienze Matematiche, \\ Universit\`a Politecnica delle Marche, Ancona 60131, Italy}

\cortext[cor1]{Corresponding author: matteo.franciolini@nasa.gov}
   
\begin{abstract}
In recent years several research efforts focused on the development of high-order discontinuous Galerkin (dG) methods for scale resolving simulations of turbulent flows. Nevertheless, in the context of incompressible flow computations, the computational expense of solving large scale equation systems characterized by indefinite Jacobian matrices has often prevented from dealing with industrially-relevant computations. In this work we seek to improve the efficiency of Rosenbrock-type linearly-implicit Runge-Kutta methods by devising robust, scalable and memory-lean solution strategies. In particular, we introduce memory saving \emph{p}-multigrid preconditioners coupling matrix-free and matrix-based Krylov iterative smoothers. The \emph{p}-multigrid preconditioner relies on cheap block-diagonal smoother's preconditioners on the fine space to reduce assembly costs and memory allocation, and ensures an adequate resolution of the coarsest space of the multigrid iteration using Additive Schwarz precondioned smoothers to obtain satisfactory convergence rates and optimal parallel efficiency of the method. In addition, the use of specifically crafted rescaled-inherited coarse operators to overcome the excess of stabilization provided by the standard inheritance of the fine space operators is explored.
Extensive numerical validation is performed. 
The Rosenbrock formulation is applied to test cases of growing complexity: the laminar unsteady flow around a two-dimensional cylinder at $Re=200$ and around a sphere at $Re=300$, the transitional flow problem  of the ERCOFTAC T3L test case suite with different levels of free-stream turbulence. As proof of concept, the numerical solution of the Boeing Rudimentary Landing Gear test case at $Re=10^6$ is reported. A good agreement of the solutions with experimental data is documented, as well as strong memory savings and execution time gains with respect to state-of-the art solution strategies.
\end{abstract}


\end{frontmatter}




\section{Introduction}
In recent years the increasing availability of High Performance Computing (HPC) resources strongly promoted the widespread of Large Eddy Simulation (LES) turbulence modelling approaches. In particular, Implicit LES (ILES) based on discontinuous Galerkin (dG) spatial discretizations showed very promising results due to the favourable dispersion and dissipation properties of the method~\cite{Bassi.Botti.ea_LES_DNS:2015}. The high potential of dG approximations for the under-resolved simulation of turbulent flows has been demonstrated in the literature for those moderate Reynolds numbers conditions where Reynolds-averaged Navier--Stokes (RANS) approaches are known to fall short, \emph{e.g.}, massively separated flows~\cite{chapelier2014evaluation, wiart2015implicit}.

Research on this topic is growing fast and several efforts focused on devising efficient time integration strategies suited for massively parallel architectures. Indeed, the inherently unsteady nature of LES/ILES and the need to reduce time-to-results pose serious challenges to the achievement of cost effective scale-resolving computations and the ability to fruitfully exploit large computational facilities. In this context high-order implicit time integration schemes are attractive to overcome the strict stability limits of explicit methods when dealing with high-degree polynomial approximations,~\cite{Bassi20071529, uranga2011implicit, bassi2015linearly}. Nevertheless, implicit schemes require to solve large non-linear/linear systems of equations. The sparsity pattern of the global system matrix imposes the use of iterative methods, indeed the number of non-zeros scales as $k^{2d}$, where $k$ is the degree of dG polynomial spaces and $d$ is the number of dimensions of the problem. As a result high-order accurate computations for industrially relevant application turns out to be highly memory intensive and expensive from the CPU time point of view, even when employing state-of-the-art iterative solvers and modern HPC facilities.

Previous studies considered the possibility of using memory-saving implementations of the iterative solver in the contex of discontinuous Galerkin discretizations. In~\cite{Crivellini201181}, a matrix-free GMRES solver was used to solve steady compressible flows. The algorithm showed to be competitive on the overall computational efficiency for sufficiently high-order polynomial approximations when using ILU(0) and Additive Schwarz preconditioners. However, the use of full-matrix operators was still required for preconditioning purposes, and thus only a limited reduction of the overall memory footprint has been achieved. In~\cite{sarshar2017numerical} a matrix-free approach is employed in the context of several time integration strategies with applications to unsteady, laminar two-dimensional problems. Recently the use of a matrix-free implementation was explored and compared to a matrix-based approach in the context of incompressible unsteady turbulent flows, see ~\cite{Franciolini2017276,crivellini2017matrix}. In particular the solution of the Rayleigh--Benard convection problem and turbulent channel flows at moderately high Reynolds numbers was considered coupling matrix-free with cheap element-wise Block-Jacobi (EWBJ) preconditioners. The algorithm showed considerable memory savings: being the use of off-diagonal jacobian blocks not required for time stepping purposes and preconditioning, the overall memory footprint could be significantly reduced. Unfortunately, when dealing with stiff problems, \emph{e.g.}, stretched elements, low Mach flows or large time step, a severe convergence degradation is observed when using EWBJ preconditioners: the solution is achieved only after a considerable number of iterations. Bearing that in mind, it is trivial to highlight that one of the main challenges to obtain a memory efficient implicit solution strategy for complex unsteady flow problems is the implementation of an efficient and memory saving preconditioning method to be coupled to matrix-free iterative solvers. For example, in~\cite{diosady2017tensor} the use of a matrix-free implementation is coupled with preconditioner operators expressed in separable tensor product form, whose arithmetic complexity scales more favourably with the order of accuracy of the scheme than a standard block operator. Other implementations in the same line exist in literature, see for example~\cite{pazner2018approximate}, where the best Kroneker product approximation of the block diagonal portion of the Jacobian is solved through the use of a matrix-free Singular Value Decomposition. In~\cite{diosady2019scalable} the same objective is obtained through the solution of an optimization problem. Despite being memory saving and capable of reducing the computational complexity of the algorithm due to the use of tensor product matrices, the main drawback of those strategies is the fact that they are based on approximations of the EWBJ preconditioner, and they fail to solve efficiently complex flow problems involving stiff computational meshes.

On the other hand, multilevel methods have been considered in the past as an efficient way to solve both linear and non-linear problems arising from high-order discontinuous Galerkin discretizations. Such methods were first proposed in a dG context by Helenbrook \etal~\cite{helenbrook2003analysis}, Bassi and Rebay~\cite{bassi2003numerical}, Fidkowski \etal~\cite{fidkowski2005p}. Those authors focused on the analysis of a \emph{p}-multigrid (\emph{p}-MG) non-linear solver, proving convergence properties and performance in the context of compressible flows using element- or line-Jacobi smoothing. Several authors also considered multigrid operators built on agglomerated coarse grids, such as \emph{h}-multigrid, see for example~\cite{prill2009smoothed, wallraff2015multigrid, antonietti2015multigrid}. The possibility of using multigrid operators as a preconditioner of an iterative solver was also explored in the context of steady compressible flows, see for example~\cite{shahbazi2009multigrid, diosady2009preconditioning}. In these works, the algorithm is reported as the most efficient and scalable if compared to single-grid preconditioners, and a large reduction in the number of iteration to reach convergence was achieved. More recently, an \emph{h}-multigrid preconditioner was proposed in~\cite{botti2017h} in the context of steady and unsteady incompressible flows. In this latter work a specific treatment for inherited dG discrezations of viscous terms on coarse levels was introduced, significantly improving the performance of the multigrid iteration.

The present work aims to devise a memory saving preconditioning strategy to be coupled with a matrix-free iterative solver for the solution of complex flow problems, extending and generalizing the techniques proposed in~\cite{Franciolini2017276, Crivellini201181, crivellini2019implicit} to deal with stiff unsteady turbulent flow problems. The time-accurate numerical framework relies on linearly-implicit Runge--Kutta schemes of the Rosenbrock type. This class of schemes requires the solution of a linear systems at each stage while the matrix is assembled only once per time step. For the linear systems solution we rely on a matrix-free implementation of the Flexible GMRES (FGMRES) method, coupled with a memory saving \emph{p}-MG preconditioner. In particular, the \emph{p}-multigrid iteration is built using a memory saving smoother on finest level, and standard matrix-based GMRES smothers on coarse levels. The numerical experiments show that this technique allows to retain the memory footprint reduction presented in~\cite{Franciolini2017276}, while improving the computational efficiency on stiff problems. Finally, the use of a rescaled-inherited approach for the coarse space operators proposed in~\cite{botti2017h} in the context of $h$-multigrid is here assessed for the $p$ version of the multigrid solver on the iterative performance. While the use of rescaled coarse spaces is recommended to maintain acceptable convergence rates of $h$-multigrid solvers, smaller advantages have been observed in our experiments, which vanish for convection dominated cases such as the under-resolved simulations of turbulent flows.

The paper presents an extensive validation of the numerical strategy on test cases involving high-order accurate ILES of complex flow configurations using unstructured meshes made of severely stretched and curved elements. The effectiveness of the proposed coupling between a matrix-free linear solver and a matrix-free \emph{p}-multigrid preconditioner is proved by comparing computational time, memory footprint of the solver as well as the algorithmic scalability of the preconditioning strategy on HPC facilities using a domain decomposition parallelization method.

The paper is structured as follows. Section~\ref{sec:NumFrwrk} describes the space and time discretization and presents the multigrid framework here employed, with particular attention to the coarse spaces assembly and the intergrid transfer operators. Section~\ref{sec:INSResults} reports a thorough assessment of the stabilization scaling on test cases of growing complexity involving unsteady flows: the unsteady flow over a two dimensional cylinder at $Re=200$, and the unsteady flow over a sphere at $Re=300$. Section~\ref{crivellini_sec:results} demonstrates the advantages of using the proposed solver for the solution of the T3L1 flow problem of the ERCOFTAC test case suites, \emph{i.e.}, the incompressible turbulent flow over a rounded leading-edge flat plate at $Re=3450$ with different levels of free-stream turbulence. After a brief physical discussion of the solution accuracy, significant memory savings as well as improvements in computational efficiency with respect to matrix-based methods are documented. As proof of concept, the solution of the Boeing Rudimentary Landing Gear test case at $Re=10^6$ is reported, including a favourable agreement with experimental data. \ref{app:stab} and \ref{sec:PoissResults} report details of the rescaled-inherited approach and an assessment on the Poisson problem.

\section{The numerical framework}\label{sec:NumFrwrk}
In this section the space and time discretizations of the incompressible Navier--Stokes (INS) equations are briefly introduced together with a detailed description of the main building blocks of the \emph{p}-multigrid preconditioner. 

\subsection{The dG discretization}
We consider the unsteady INS equations in conservation form with Dirichlet and Neumann 
boundary conditions in a fixed Cartesian reference frame,
\begin{subequations}
  \label{eq:ns}
  \begin{alignat}{2}
    \label{eq:ns.momentum}
    \partial_t \mathbf{u}
    + \DIV \left( \uvec \otimes \uvec + p \Ivec \right)
    -\nu \DIV \left[ \left(\nabla \otimes \uvec + (\nabla \otimes \uvec)^t \right) - \frac23 \left( \nabla \cdot \uvec \right) \Ivec \right] = 0
    &\qquad& \text{in $\Omega\times (0,t_F)$}, \\
    \label{eq:ns.mass}
    \DIV\mathbf{u} = 0
    &\qquad& \text{in $\Omega\times (0,t_F)$}, \\
    \label{eq:ns.bcd}
    \mathbf{u}= \mathbf{f} 
    &\qquad& \text{on $\partial\Omega_D \times (0,t_F)$}, \\
    \label{eq:ns.bcn}
    \left[ \nu \left(\nabla \otimes \uvec + (\nabla \otimes \uvec)^t \right) - \frac23 \nu \left(\nabla \cdot \uvec\right) \Ivec\right] \cdot \normal = 0, \quad p \normal = \mathbf{g}  
    &\qquad& \text{on $\partial\Omega_N \times (0,t_F)$}, \\
    \label{eq:ns.ic}
    \mathbf{u}(\cdot,t=0) = \mathbf{u}^0
    &\qquad& \text{in $\Omega$}. 
  \end{alignat}
\end{subequations}
where $\mathbf{u} \in \mathbb{R}^{d}$ is the velocity vector, $p$ is the pressure, $\nu>0$ denotes the (constant) viscosity, 
$\mathbf{u}^0$ is the initial condition, $t_F$ is the final simulation time, 
$\normal$ is the unit outward normal to $\partial \Omega$ and $\Ivec = \delta_{ij} \, \mathbf{e}_i \otimes \mathbf{e}_j$, $i,j=1,...,d$, is the identity matrix.
The density has been assumed to be uniform and equal to one and the Stokes hypothesis is used for the definition of viscous stresses.

Regarding boundary conditions in \eqref{eq:ns.bcd}-\eqref{eq:ns.bcn}, 
$\mathbf{f}$ and $\mathbf{g}$ are the boundary data
to be imposed on Dirichlet and Neumann boundaries, 
respectively, such that $\partial \Omega_D \bigcup \partial \Omega_N = \partial \Omega$. 
In the case that $\partial \Omega_D = \partial \Omega$, we also require $\langle p\rangle_\Omega = 0$, 
where $\langle\cdot\rangle_\Omega$ denotes the mean value over $\Omega$.
Note that, while using the divergence free constraint \eqref{eq:ns.mass} we get
$$ \nu \DIV \left[ \left(\nabla \otimes \uvec + (\nabla \otimes \uvec)^t \right) - \frac23 (\nabla \cdot \uvec) \Ivec \right] = \nu \LAP \uvec,$$
this simplification is unsuitable for Neumann boundaries.

Introducing the convective and viscous flux functions 
\begin{equation}
  \begin{array}{lll}
\Fvec^{\nu}\left(\frac{\partial u_i}{ \partial x_j}\right) &= \nu \left(\nabla \otimes \uvec + (\nabla \otimes \uvec)^t \right) - \frac23 \nu (\nabla \cdot \uvec) \Ivec
                                                           &= \displaystyle \nu \left( \frac{\partial u_j }{\partial x_i} + \frac{\partial u_i }{\partial x_j} - \frac23 \frac{\partial u_k}{\partial x_k} \delta_{ij}\right) \, \mathbf{e}_i \otimes \mathbf{e}_j \\
\Fvec^c(u_i,p) &= \mathbf{u} \otimes \mathbf{u} + p \Ivec &= \left(u_i \, u_j + p \delta_{ij} \right) \, \mathbf{e}_i \otimes \mathbf{e}_j
  \end{array}
\end{equation}
Eqs.~\eqref{eq:ns.momentum}-\eqref{eq:ns.mass}
can be compactly rewritten in integral form as
\begin{subequations}
\label{eq:NSf}
  \begin{alignat}{1}
{\intO{\partial_t} \uvec} + \intO{\nabla \cdot \left(\Fvec^c - \Fvec^{\nu}\right)} &= {\bf 0}, \nonumber\\
\intO{\nabla \cdot \uvec} &= 0 \nonumber.
  \end{alignat}
\end{subequations}

In order to define the dG discretization we introduce a triangulation $\Th$ 
of the computational domain $\Omega$, that is the collection of disjoint mesh elements 
$\T \in \Th$ such that $ \bigcup_{\T \in \Th} \overline{\T} = \Omega_h$, where $\Omega_h$ 
is a suitable approximation of $\Omega$.
The mesh skeleton $\Ff$ is the collection of mesh faces $\f$. Internal faces $\f \in \Ffi$ are defined as 
the intersection of the boundary of two neighboring elements: $\f = \partial \T \bigcap \partial \T'$ 
with $\T \neq \T'$. Boundary faces $\f \in \Ffb$ reads $\f = \partial \T \bigcap \partial \Omega_h$.
Clearly $\Ff= \Ffi \cup \Ffb$.

Each component of the velocity vector and the pressure is sought (for $0< t < t_F$)
in the so called \emph{broken polynomial spaces} defined over $\Th$
\begin{equation}
\Poly{d}{k}(\Th) = \left\{ v_h \in L^2(\Omega_h)\; | \; \forall \T \in \Th, v_h|_\T \in \Poly{d}{k}(\T)\right\} 
\end{equation}
where $ \Poly{d}{k}(\T)$ is the space of polynomial functions in $d$ variables and 
total degree $k$ defined over $\T$. 
Since no continuity requirements are enforced at inter-element boundaries, 
$v_h$ admits two-valued traces on the partition of mesh skeleton $\Ffi$.
Accordingly we introduce average and jump operators over internal faces 
\begin{equation}
\mathrm{\textbf{Average}}: \averg{v_h}_{\f} = \frac{1}{2}\left(v_h|_{\T} + v_h|_{\T'} \right),  \qquad \qquad\\
\mathrm{\textbf{Jump}}:    \jump{v_h}_{\f} = \left(v_h|_{\T} - v_h|_{\T'} \right).
\end{equation}
Specific definitions of averages and jumps will be introduced over boundary faces
to take into account Dirichlet and Neumann boundary conditions. 

The dG discretization of the Navier-Stokes equations reads: find 
$(\uVech,p_h) \in [\Poly{}{k}(\Tf)]^{d} \times \Poly{}{k}(\Tf)$ such that,
for all $(\vVech,q_h) \in [\Poly{}{k}(\Tf)]^{d} \times \Poly{}{k}(\Tf)$:
\begin{equation}
\label{eq:NSdiscr}
\begin{array}{llll}
 \sumT \intT{ \partial_t \uVech \cdot \vVech} & - \sumT \intT{\left( \Fvec_h^c - \widetilde{\Fvec}_h^{\nu}\right) : \nabla_h \vVech}
&+\sumF \intB{\sigma}{\normal^{\f} \cdot \left(\widehat{\Fvec}_h^c - \widehat{\Fvec}_h^{\nu}\right) \cdot \jump{\vVech}  } & = 0, \\
& - \sumT \intT{\uVech \cdot \nabla_h q_h} & + \sumF \intB{\sigma}{\normal^{\f}\cdot \widehat{\uvec}_h \; \jump{q_h}} & =0,
\end{array}
\end{equation}
where $\jump{\vVech} = \jump{v_{h,i}}\mathbf{e}_i$ and $\normal^{\f}$ is the normal vector with respect to $\f$.
While obtaining \eqref{eq:NSdiscr} from \eqref{eq:NSf} follows the standard dG FE practice (element-by-element integration by parts 
after having multiplied by a suitable test function), the dG method hinges 
on the definition of suitable numerical viscous $\widetilde{\Fvec}_h^{\nu}$, $\widehat{\Fvec}_h^{\nu}$ and inviscid fluxes $\widehat{\Fvec}_h^c$, $\widehat{\uvec}_h$.

According to the BR1 scheme, proposed in~\cite{BR-jcpns97}, 
$\forall\boldsymbol{\tau}_{h} \in [\Poly{d}{k}(\Th)]^d, v_h \in \Poly{d}{k}(\Th)$, 
the consistent gradient $\mathbf{G}_h(v_h)$ is such that 
\begin{equation}
 \int_{\Omega} \left( \mathbf{G}_h(v_h) {-} \nabla {v_h} \right) \cdot \boldsymbol{\tau}_h {=}
 {-}\sum_{\f \in \Ff} \intF \jump{v_h} \averg{\boldsymbol{\tau}_{h}}\cdot \normal^{\f} {:}{=}
 \sum_{\f\in\Ff} \int_{\Omega} \mathbf{r}^{\f}(\jump{v_h}) \cdot \boldsymbol{\tau}_{h}{=} 
 \sumT \intT{ \mathbf{R}^\T(v_h) \cdot \boldsymbol{\tau}_{h} \nonumber} 
\end{equation}
where $\mathbf{r}^{\f}(\jump{v_h}): \Poly{d}{k}(\f) \rightarrow [\Poly{d}{k}(\Th)]^d$ is the local lifting operator,
$ \displaystyle \mathbf{R}^\T(v_h) := \sum_{\f \in \Ft} \mathbf{r}^{\f}(\jump{v_h})$
is the elemental lifting operator and $\Ft$ is the set of faces belonging to $\partial \T$.
In this work we rely on the BR2 scheme, introduced to reduce the stencil of the BR1 discretization
and analyzed in the context of the Poisson problem by \cite{bmmpr-nmpde} and \cite{Arnold.Brezzi.ea:2002}. 
The BR2 viscous fluxes are functions of elemental spatial derivatives corrected by suitable lifting operator contributions
\begin{equation}
\label{eq:vflux}
\widetilde\Fvec_h^{\nu} = \Fvec^{\nu} \left(\frac{\partial u_{h,i}}{ \partial x_j}- R^\T_j(u_{h,i})\right),\\ \quad \mathrm{and} \quad
\widehat\Fvec_h^{\nu} = \Fvec^{\nu}\left(\averg{\frac{\partial u_{h,i}}{ \partial x_j}- \eta_\f r^\f_j(\jump{u_{h,i}})}\right),
\end{equation}
$\widehat\Fvec^{\nu}$ ensures consistency and stability of the scheme and
$\widetilde\Fvec^{\nu}$ guarantees the symmetry of the formulation.
As proved by Brezzi \etal~\cite{bmmpr-nmpde}, coercivity for the BR2 discretization 
of the Laplace equation holds provided that $\eta_\f$ is greater 
than the maximum number of faces of the elements sharing $\f$.
In order to impose boundary conditions on $\f \in \Ffb$, viscous fluxes reads
\begin{equation}
\begin{array}{lll}
\mathrm{\textbf{Dirichlet}}\vspace{0.5cm}: &  \widehat{\Fvec}_h^\nu = \Fvec\left(\frac{\partial u_{h,i}}{ \partial x_j}- \eta_\f r^\f_j(\jump{u_{h,i}})\right),& \quad \jump{\uVech}_\f = (\uVech|_{\T} - \mathbf{f}). \\
\mathrm{\textbf{Neumann}}:                 &  \normal^\f \cdot {\Fvec}_h^\nu = 0, & \quad  \jump{\uVech}_\F = 0. 
\end{array}
\end{equation}

The inviscid numerical fluxes of the dG discretization 
result from the exact solution of local Riemann problems 
based on an artificial compressibility perturbation of the Euler equations, as proposed in~\cite{Bassi.Crivellini.eq:2006}.
Boundary conditions for inviscid fluxes are enforced weakly 
by properly defining a ghost boundary state $(\uvec_{gb}, p_{gb})$ 
having support on the interface $\f \in \Ffb$ of a ghost neighboring elements $\T_g$.
The ghost boundary state is defined imposing the conservation of Riemann invariants  
based on the hyperbolic nature of the artificial compressibility perturbation of the Euler equation.
Accordingly, both the internal state $(\uvec^{\T}, p^{\T})$ and the boundary data 
($\mathbf{f}$ or $\mathbf{g}$, in case of Dirichlet or Neumann boundary conditions, respectively)
are involved in the definition of $(\uvec_{gb}, p_{gb})$. See~\cite{Bassi.Crivellini.eq:2006, Bassi:Unsteady_incomp, Crivellini:Sphere} for additional details on the method.

\subsection{Implicit time accurate integration \label{sec:time}}
Time integration of can be presented in compact form 
by collecting the velocity vector and the pressure polynomial expansions
in the vector $\wvec_h \eqbydef (u_{h,1},...,u_{h,d},p_h) \in [\Poly{d}{k}(\Th)]^{d+1}$
and identifying the unknown vector at time $t_n$ with $\wvec^n_h$, that is 
$\wvec^n_h = [\uVech(t_n),p_h(t_n)]$.
Moreover, we introduce the flux functions 
$ \widetilde\Fvec_h(\wvec_h) \eqbydef \left[ \Fvec_h^c - \widetilde\Fvec_h^{\nu}, \uVech \right] \in \mathbb{R}^d \otimes \mathbb{R}^{d+1}$ and
$ \widehat\Fvec_h(\wvec_h) \eqbydef \left[ \widehat\Fvec_h^c - \widehat\Fvec_h^{\nu}, \widehat{\uvec}_h \right] \in \mathbb{R}^d \otimes \mathbb{R}^{d+1}$,
collecting the viscous and inviscid flux contributions. 
For all $\wvec_h, \kvec_h \in [\Poly{d}{k}(\Th)]^{d+1}$, 
we define the residual of the dG spatial discretization in \eqref{eq:NSdiscr} as follows
\begin{equation}
\label{eq:NSdiscrRes}
f_h(\wvec_h,\kvec_h) = 
                     -\sumT \intT{ \sum_{i=1}^{d+1} \sum_{p=1}^{d} \widetilde{F}_{p,i}(\wvec) \, \frac{\partial z_i} {\partial x_p}}
                     +\sumF \intF{ \sum_{i=1}^{d+1} \sum_{p=1}^{d} n^{\f}_p \, \widehat{F}_{p,i}(\wvec) \, \jump{z_i}  }, 
\end{equation}
where we dropped the mesh step size subscript when working in index notation.
For all $\wvec_h, \delta \wvec_h, \kvec_h \in [\Poly{d}{k}(\Th)]^{d+1}$, 
the linearization of the residual reads 
\begin{equation}
j_h(\wvec_h,\delta \wvec_h, \kvec_h) = \displaystyle \frac{\partial f_h(\wvec_h,\kvec_h)}{\partial \wvec_h} \delta \wvec_h.
\end{equation}
In particular we distinguish the inviscid $j^{!\nu}_h(\wvec_h,\delta \wvec_h, \kvec_h)$ and the viscous $j^{\nu}_h(\delta \wvec_h, \kvec_h)$ 
contributions
\begin{equation} 
 j^{!\nu}_h(\wvec_h, {\delta} \wvec_h, \kvec_h) = 
   - \sumT \intT{\sum_{i,j=1}^{d+1} \sum_{p=1}^{d} \frac{\partial \widetilde{F}_{p,i}}{\partial{w_j}}(\wvec_h) \, \delta w_j \, \frac{\partial z_{i}}{\partial x_p} } 
   + \sumF \intF{\sum_{i,j=1}^{d+1} \sum_{p=1}^{d} n^{\f}_{p} \frac{\partial \widehat{F}_{p,i}}{\partial{w_j}}(\wvec_h) \, \delta w_j \, \jump{z_{i}} }, \label{eq:jacTrilFnotNU}
\end{equation}
\begin{equation} 
\label{eq:jacTrilnu}
\begin{aligned}[1]
 j^{\nu}_h (\delta \wvec_h,\kvec_h) 
 = - &\sumT \intT \sum_{i,j = 1}^{d+1} \sum_{p,q = 1}^d {\frac{\partial \widetilde{F}_{p,i}}{ \partial \Bigl(\partial {w}_{j} /\partial x_q - {R}_{q}^\T (w_{j})\Bigr)}}
          \left({\frac{\partial (\delta {w}_{j})}{\partial x_q} - {R}_{q}^\T (\delta {w}_{j})}\right) \, \frac{\partial {z}_{i}}{\partial x_p} \; +  \\ 
   + &\sumF \intF \sum_{i,j = 1}^{d+1} \sum_{p,q = 1}^d n_{p}^{\f} \, 
              {\frac{\partial \widehat{F}_{p,i}}{ \partial \Bigl( \partial {w}_{j}/ \partial x_q - \eta_\f {r}_{q}^\f (w_{j}) \Bigr) }} 
        \averg{\frac{\partial (\delta w_j)}{\partial x_q} - {\eta_\f \; {r}_{q}^\f \left(\jump{\delta w_j}\right)}} \, \jump{z_i}. 
\end{aligned}
\end{equation}
Note that, since $\Fvec^\nu$ is a linear function, \eqref{eq:jacTrilnu} is a bilinear form, while, by abuse of notation, \eqref{eq:jacTrilFnotNU} is a 
bilinear (resp. trilinear) when $F^c_{p,i}(\wvec)$ is a linear (resp. non-linear) function of $w_j$. 

In this work time integration is performed via the multi-stage linearly implicit 
(Rosenbrock-type) Runge-Kutta method. 
As an appealing feature the method requires the solution of a linear system 
at each stage $s=\{1,\cdots,n_s\}$, while the Jacobian matrix needs to be assembled only once per time step.
Prior to introducing the formulation for the temporal discretization 
we define the following mass bilinear form: for all $\wvec_h, \kvec_h \in [\Poly{}{k}(\Th)]^{d+1}$ 
\begin{equation}
 m_h(\wvec_h, \kvec_h) = \sumT \intT{ \sum_{i=1}^{d} w_{i} \, z_{i} }. \label{eq:massBil} 
\end{equation}

Given the initial condition $\wvec_h^0=\wvec_h(t=0) \in [\Poly{}{k}(\Th)]^{d+1}$ we define the 
sequence $\wvec_h^{n+1}$ iteratively by means of the Rosenbrock scheme as described Algorithm~\ref{algo:ros}, where $\gamma$, $\mathsf{a}_{ij}$, $\mathsf{c}_{ij}$ and $\mathsf{m}_i$ are real coefficients proper of the Rosenbrock scheme and $\delta \mathbf{w}_h^s$, with $s=\{1,\cdots,n_s\}$, the solutions at each stage of the scheme that are properly combined to compute the solution ${\wvec}_h^{n+1}$ at the next time level. 
\begin{algorithm}[H]
\caption{Multi-stage linearly implicit (Rosenbrock-type) Runge-Kutta method\label{algo:ros}}
\begin{algorithmic}[1]
  \STATE set ${\wvec}_h^{n} = \wvec_h^0$, $n_F = \displaystyle\frac{t_F}{\delta t}$
  \FOR{$n = 0,...,n_F$}
     \FOR{$s = 1,...,n_s$}
     \STATE set $\delta \pvec_h = \mathbf{0}\,\wedge\,\delta \qvec_h = \mathbf{0}$ 
     \FOR{$o = 1,...,s-1$}
       \STATE $\delta \pvec_h \mathrel{+}= \mathsf{a}_{s,o} \, \delta \wvec_h^{o}$ 
       \STATE $\delta \qvec_h \mathrel{+}= \mathsf{c}_{s,o} \, \delta \wvec_h^{o}$
     \ENDFOR
     \STATE{find $\delta \wvec_h^s \in [\Poly{}{k}(\Th)]^{d+1}$ such that, for all $\kvec_h \in [\Poly{}{k}(\Th)]^{d+1}$
       \vspace{-0.3cm}
       \begin{flalign}
       \displaystyle \frac{1}{\mathsf{\gamma} \delta t} m_h( \delta \wvec_h^s, \kvec_h) + 
                                               j_h({\wvec}_h^{n}, \delta \wvec_h^s, \kvec_h) =
         - f_h({\wvec}_h^{n}{+} \delta \pvec_h,\kvec_h) 
        - \frac{1}{\delta t} m_h(\delta \qvec_h,\kvec_h) \label{eq:rossys} &&
       \end{flalign}}
       \vspace{-0.3cm}
     \ENDFOR
     \FOR{$o = 1,...,s$}
       \STATE set ${\wvec}_h^{n+1} = {\wvec}_h^{n}\mathrel{+}\mathsf{m}_o\delta \wvec_h^o$
     \ENDFOR
  \ENDFOR
\end{algorithmic}
\end{algorithm}
%
%
%
The Rosenbrock time marching strategy in Algorithm~\ref{algo:ros} advances 
the solution in time by repeatedly solving the linearized system of equations~\eqref{eq:rossys}, once for each stage of the Runge-Kutta method. Introducing the Jacobian and mass matrix operators 
\begin{equation}
\begin{array}{lll}
(\Jm_{h}\;  \delta \wvec_h, \kvec_h)_{L^2(\Omega)} & =  j_h(\wvec_h, \delta \wvec_{h}, \kvec_{h})                   & \forall \, \wvec_h, \delta \wvec_h, \kvec_h \in [\Poly{d}{\Kl}(\Th)]^{d+1},\\
(\Mm_{h} \;  \delta \wvec_h, \kvec_h)_{L^2(\Omega)} & =  m_h(\delta \wvec_{h}, \kvec_{h})                                     & \forall \,  \delta \wvec_h, \kvec_h \in [\Poly{d}{\Kl}(\Th)]^{d+1},\\
\end{array}
\end{equation}
the equation system \eqref{eq:rossys} can be compactly rewritten as follows:
\begin{equation}
\label{eq:rosSysOp}
\Gm_{h} \; \delta \wvec_{h} = \gv_{h}
\end{equation}
where $\Gm_h = \frac{1}{\gamma \delta t} \Mm_h + \Jm_{h} $ is the global matrix operator,
and $ \delta \wvec_h, \gv_h \in [\Poly{d}{k}(\Th)]^{d+1}$ are the unknown polynomial function 
and the right-hand side arising from the linearly-implicit Runge-Kutta time discretization, respectively.
In this work the four stages, order three (ROSI2PW) scheme of Rang and Angermann~\cite{Lang.Verwer:2001} was employed. 
This scheme preserves its formal accuracy when applied to the system of DAEs arising form the spatial discretization of the INS equations as demonstrated in~\cite{Franciolini2017276}. 

\subsection{Matrix-free iterative solver}\label{sec:matrix-free}
In this work we consider (flexible) GMRES iterative solvers with application to the solution of linear system arising in Rosenbrock-type schemes, see Eq.~\eqref{eq:rossys}. The iterative solver can be implemented matrix-free following the approach of~\cite{D.A.knoll-D.E.keyes}, where the product between the primal Jacobian and the defect vector is approximated by its first order Taylor expansion. Given $\wvec_h, \dvec_h \in [\Poly{d}{\Kl}(\Th)]^{d+1}$, the Jacobian trilinear form can be expressed as 
\begin{eqnarray}\label{eq:matrix-free}
j_h({\wvec}_h^{n}, \dvec_h, {\kvec}_h) =\frac{1}{\Delta}\left(f({\wvec}_h^{n}{+} \Delta\,\dvec_h, {\kvec}_h) - f({\wvec}_h^{n},\kvec_h) \right), \; \forall \kvec_h \in [\Poly{d}{\Kl}(\Th)]^{d+1},
\end{eqnarray}
which involves bilinear form evaluations. According to~\cite{pernice1998nitsol},
\begin{equation}\label{eq:epsdef}
\Delta=\epsilon \dfrac{\sqrt{1+\|{\wvec}_h^{n}\|_{L^2(\Omega)}}}{\|\dvec_h\|_{L^2(\Omega)}},
\end{equation}
with $\epsilon=10^{-9}$ for all the computations~\cite{BassiCrive,crivellini2017matrix,Franciolini2017276}. We remark that the use of~\eqref{eq:matrix-free} does not change the behaviour of the iterative solver for relative tolerances of practical engineering interest, \ie\ when those are greater than the numerical perturbation $\varepsilon$, and does not increase the cost-per-iteration at high order of accuracy, since the algorithm complexity of the residual evaluation scales equally to that of a matrix-vector product with the order of polynomial approximation. See~\cite{Franciolini2017276} for further details. Since the global system matrix is not required for the time integration, the Jacobian matrix needs to be assembled for preconditioning purposes only, and such flexibility can be exploited to reduce the matrix-assembly time and memory footprint, for example by evaluating only parts of the Jacobian blocks and/or reusing those blocks for several successive iterations.

As preconditioners for GMRES iterators we consider the following options:
\begin{enumerate}
\item ASM($i$,ILU($j$)) - Additive Schwarz domain decomposition Method (ASM) preconditioners with $i$ levels of overlap between sub-domains and a block ILU decomposition for each sub-domain matrix with $j$ levels of fill;
\item BJ or ASM(0,ILU(0)) - ASM preconditioner with no overlap between sub-domains and a block ILU decomposition for each sub-domain matrix with same level of fill of the original matrix;
\item EWBJ - Element-wise block Jacobi, a BJ preconditioner neglecting off-diagonal blocks, that is an LU factorization of the diagonal blocks. 
\end{enumerate}
Note that in serial computations ASM(i,ILU(j)) and BJ fall back to ILU(j) and ILU(0), respectively. ASM and BJ performance differ when the computation is performed in parallel, depending on the number of sub-domains. While efficiency of BJ decreases while increasing the number of sub-domains, ASM seeks to heal the convergence degradation
at the expense of an increased memory footprint of the solver as the number of partitions rises, as part of the global matrix non-zeros entries are replicated in neighboring sub-domains. Conversely, EWBJ has optimal scalability properties, involving local to each element operations. It is worth pointing out that, when using a matrix-free iterative solver, only the use of EWBJ leads to a memory saving. In fact, it allows to skip the computation of the off-diagonal contributions, thereby reducing the matrix-assembly computation time. The code relies on the PETSc library to handle linear solvers and parallelism~\cite{petsc-web-page}.

For the sake of compactness, in the remaining of the paper we will denote a solver-preconditioner couple following the convention: $$\mathrm{SOLVER}({\mathrm{MatVecOpt}})[\mathrm{PREC}(\mathrm{Opt})].$$
The MatVecOpt label describes how the matrix-vector products are performed, \ie\ in a matrix-free (MF) or matrix-based (MB) fashion,
while PREC(Opt) describes the type of preconditioning employed.

\subsection{Multigrid preconditioners}
To increase the performance of linear system solutions on stiff space discretizations, we investigate the use of multigrid preconditioning approaches to solve the global equation system \eqref{eq:rosSysOp}. The basic idea is to exploit iterative solvers to smooth-out the high-frequency component of the error with respect to the unknown exact solution. Indeed, being iterative solvers not effective at damping low-frequency error components, the iterative solution of coarser problems is exploited to circumvent this issue, shifting the low-frequency modes towards the upper side of the spectrum. This simple and effective strategy allows to obtain satisfactory rates of convergence all along the iterative process.

As the work aims at obtaining solutions with high order polynomials on rather few and possibly curved mesh elements, and targets the use of the solver on large HPC facilities, we build coarse spaces by reducing the degree of polynomial approximation of the solution of the dG discretization with respect to the original problem of degree $k$, commonly referred in the literature as $p$-multigrid method. The strategy show some advantages over $h$-multigrid approaches on agglomerated mesh elements other than the ease of implementation, as the intergrid transfer operators and the matrix assembly routines are applied in a local to each element fashion, and thus are ideally scalable. We consider $L$ coarse levels spanned by the index $\ell = 1,...,L$ and indicate the fine and coarse levels with $\ell=0$ and $\ell=L$, respectively. The polynomial degree of level $\ell$ is $k_\ell$ and the polynomial degrees of the coarse levels are chosen such that $k_\ell < k_{\ell-1}$, $l = 1,...,L$, with $k_0 = k$. Accordingly the polynomial spaces associated to the coarse levels read $\Poly{d}{\Kl}(\Th)$.
The coarse problems corresponding to \eqref{eq:rosSysOp} are in the form 
\begin{equation}
\label{eq:rosSysOpl}
\Gm_{\ell}\; \delta \wvec_{\ell} = \gv_{\ell}
\end{equation}
where $\Gm_{\ell}$ is the global matrix operator on level $l$ and
$\delta \wvec_\ell, \gv_\ell \in [\Poly{d}{\Kl}(\Th)]^{d+1}$
are the unknown function and the known right-hand side, respectively.

A crucial aspect for the efficiency of the \emph{p}-multigrid iteration is related to the computational cost 
of building coarse grid operators $\Gm_\ell$.
While it is possible to assemble the bilinear and trilinear forms $j_h$, $f_h$ and $m_h$ of Section~\ref{sec:time}
on each level $\ell$ with the corresponding polynomial functions $\wvec_h, \delta \wvec_h, \kvec_h \in \Poly{d}{k_\ell}(\Th)$, significantly better performances are achievable by \emph{restricting} the fine grid operator by means of so called Galerkin projections. 
The former and the latter strategies are named non-inherited and inherited \emph{p}-multigrid, respectively. 
As will be clear in what follows the construction of coarse operators is trivial when 
polynomial expansions are based on hierarchical orthonormal modal basis functions.

\subsubsection{Restriction and prolongation operators}
In this section we describe the prolongation and restriction operators 
required to map polynomial functions on finer and coarser levels, respectively.

Since $ \Poly{d}{\Kl}(\Th) \supset \Poly{d}{\Klpo}(\Th)$, the \textit{prolongation} operator $\IntOp_{\ell+1}^{\ell}: \Poly{d}{\Klpo}(\Th) \rightarrow \Poly{d}{\Kl}(\Th)$, is the injection operator such that	 
\begin{equation}
\sum_{\T \in \Th} \int_{\T} (\IntOp_{\ell+1}^{\ell} w_{h} - w_{h})\; = 0, \;\; \forall w_{h} \in \Poly{d}{\Klpo}(\Th),\nonumber 
\end{equation}
The prolongation operator from level $\ell$ to level $0$ can be recursively defined by the composition of inter-grid prolongation operators: $\IntOp_\ell^0 = \IntOp^0_1 \, \IntOp^1_2 \,... \, \IntOp_\ell^{\ell-1}$.

The ($L^2$ projection) \textit{restriction} operator $\IntOp_\ell^{\ell+1}: \Poly{d}{\Kl}(\Th) {\rightarrow} \Poly{d}{k}(\Th)$, is such that
\begin{equation}
\sum_{\T \in \Th} \int_{\T} (\IntOp_{\ell}^{\ell+1} w_{h} - w_{h})\; z_{h} = 0, \qquad \forall w_{h} \in \Poly{d}{\Kl}(\Th), \; \forall z_{h} \in \Poly{d}{\Klpo}(\Th),
\end{equation}
and the restriction operator from level $0$ to level $\ell$ reads $\IntOp^\ell_0 = \IntOp^\ell_{\ell-1} \,... \,\IntOp^2_1 \,\IntOp^1_0 $.

When applied to vector functions $\wvec_h \in [\Poly{d}{\Klpo}(\Th)]^{d+1}$ the interpolation operators act componentwise,
\eg,~$\IntOpB_{\ell+1}^{\ell} \wvec_h = \sum_{i=1}^{d+1} \IntOp_{\ell+1}^{\ell} w_i$.
It is interesting to remark that using hierarchical orthonormal modal basis functions restriction and prolongation operators are trivial, in particular restriction from $\Poly{d}{\Kl}(\Th)$ into $\Poly{d}{\Klpo}(\Th)$ is as simple as 
keeping the degrees of freedom of the modes of order $k \leq \Klpo$ and discarding the remaining high-frequency modes.
 
\subsection{Fine and coarse grid Jacobian operators}
The non-inherited and the inherited version (denoted with superscript $\IntOp$) of the inviscid and viscous 
Jacobian operators introduced in \eqref{eq:jacTrilFnotNU}-\eqref{eq:jacTrilnu}, can be defined as follows for $\ell=1,...,L$
\begin{equation}
\begin{array}{lll}
(\Jm_{\ell}^{!\nu}\, \delta \wvec_h, \kvec_h)_{L^2(\Omega)}         & = j^{!\nu}_h(\wvec_h, \delta \wvec_{h}, \kvec_{h})                   & \forall \, \wvec_h, \delta \wvec_h, \kvec_h \in [\Poly{d}{\Kl}(\Th)]^{d+1}\\
(\Jm_{\ell}^{\nu}\, \delta \wvec_h, \kvec_h)_{L^2(\Omega)}          & = j^{\nu}_h(\delta \wvec_{h}, \kvec_{h})                               & \forall \,  \delta \wvec_h, \kvec_h \in [\Poly{d}{\Kl}(\Th)]^{d+1}\\
(\Jm^{!\nu, \IntOp}_{\ell} \delta \wvec_h, \kvec_h)_{L^2(\Omega)} & = j^{!\nu}_h(\wvec_h, \IntOpB_\ell^0 \,\delta \wvec_h, \IntOpB_\ell^0 \kvec_{h})  
                                                                                                                                            & \forall \, \wvec_h, \delta \wvec_h, \kvec_h \in [\Poly{d}{\Kl}(\Th)]^{d+1} \\
(\Jm^{\nu, \IntOp}_{\ell} \delta \wvec_h, \kvec_h)_{L^2(\Omega)}  & = j^{\nu}_h(\IntOpB_\ell^0\,\delta \wvec_{h}, \IntOpB_\ell^0 \kvec_h)     
                                                                                                                                            & \forall \,  \delta \wvec_h, \kvec_h \in [\Poly{d}{\Kl}(\Th)]^{d+1}
\end{array}
\end{equation}
The main benefit of inherited algorithms is the possibility to efficiently compute
coarse grid operators by means of the so called Galerkin projection,
avoiding the cost of assembling bilinear and trilinear forms. 
The procedure is described in what follows, focusing on the benefits of using hierarchical orthonormal basis functions. 

The matrix counterpart $\SysM{\ell}^{\IntOp}$ of the operator $\Jm_{\ell}^{\IntOp} = \Jm_{\ell}^{\nu,\IntOp} + \Jm_{\ell}^{!\nu,\IntOp}$ is a sparse block matrix with block dimension $N^{\T}_{\mathrm{dof}}= \mathrm{dim}\left(\Poly{d}{\Kl}(\T)\right)$ and total dimension $\card(\Th) \, N^{\T}_{\mathrm{dof}}\, (d{+}1)$. The matrix is composed of diagonal blocks $\SysM{\T,\T}^{\ell,\IntOp}$ and off-diagonal blocks $\SysM{\T,\T^{'}}^{\ell,\IntOp}$, the latter taking care of the coupling between neighboring elements $\T,\Tpr$ sharing a face $\f$. Once the fine system matrix $\SysM{0}$ is assembled, the diagonal and off-diagonal blocks of the Jacobian matrix of coarse levels can be inherited recursively and matrix-free as follows
\begin{equation}
\SysM{\T,\T}^{\ell+1,\IntOp}     =  \M_{\ell+1,\ell}^\T \; \left(\SysM{\T, \T}^{\ell,\IntOp}\right) \;      \left(\M_{ \ell+1,\ell}^{\T} \right)^t, \qquad \qquad 
\SysM{\T,\T^{'}}^{\ell+1,\IntOp} =  \M_{\ell+1,\ell}^\T \; \left(\SysM{\T, \Tpr}^{\ell,\IntOp}\right) \;  \left(\M_{ \ell+1,\ell}^{\Tpr} \right)^t. \label{eq:GalProj}
\end{equation}
The projection matrices read 
\begin{equation}
\M_{\ell+1,\ell}^\T = {\left(\M_{\ell+1}^\T\right)^{-1}} \int_{\T} {\bf{\varphi}}^{\ell+1} {\otimes} \; {\bf{\varphi}}^{\ell}, 
          \quad \mathrm{where} \quad \M_{\ell+1}^\T = \int_{\T} {\bf{\varphi}}^{\ell+1} {\otimes} \; {\bf{\varphi}}^{\ell+1},
\end{equation}
and $\varphi^\ell$ represents the set of basis functions spanning the space $\Poly{d}{\Kl}(\T)$.
When using hierarchical orthonormal basis functions, $\M_{\ell+1}^\T$ is the unit diagonal elemental mass matrix 
and $\M_{\ell+1,\ell}^\T \in \mathbb{R}^{\mathrm{dim}\left(\Poly{d}{\Klpo}(\T)\right) \times \mathrm{dim}\left(\Poly{d}{\Kl}(\T)\right)}$ 
is a unit diagonal rectangular matrix. 
Accordingly the Galerkin projection in \eqref{eq:GalProj} falls back to a trivial and inexpensive sub-block extraction.

Being $\Poly{d}{k_0}(\Th) \supset \Poly{d}{\Kl}(\Th)$, it can be demonstrated that 
inherited and non-inherited \emph{p}-multigrid algorithms lead to the same inviscid Jacobian operators, that is $\Jm^{!\nu, \IntOp}_{\ell} = \Jm^{!\nu}_{\ell}$.
As opposite inherited and non-inherited viscous Jacobian differ because of the terms involving lifting operators. 
Note that inherited lifting operators act on traces of polynomial functions mapped into $\Poly{d}{k_0}(\Th)$, accordingly
\begin{align} 
\mbox{inherited \emph{p}-multigrid lifting operators,} \quad &\mathbf{r}^{\f}(\jump{z_h}): \Poly{d}{k_0}(\f) \rightarrow [\Poly{d}{k_0}(\Th)]^d, \\
\mbox{non-inherited \emph{p}-multigrid lifting operators,} \quad &\mathbf{r}^{\f}(\jump{z_h}): \Poly{d}{\Kl}(\f) \rightarrow [\Poly{d}{\Kl}(\Th)]^d.
\end{align}

Interestingly, using the definitions of the global and local lifting operators, the bilinear form \label{eq:jacTrilFNU} can be rewritten as follows
\begin{align} 
 j^{\nu}_h (\delta \wvec_h,\kvec_h) 
 = - &\sumT \intT \sum_{i,j = 1}^{d+1} \sum_{p,q = 1}^d {\frac{\partial \widetilde{F}_{p,i}}{ \partial \Bigl(\partial {w}_{j} /\partial x_q - {R}_{q}^\T(w_{j})\Bigr)}}
          \frac{\partial (\delta {w}_{j})}{\partial x_q}  \, \frac{\partial {z}_{i}}{\partial x_p} \; + \nonumber \\ 
   + &\sumF \intF \sum_{i,j = 1}^{d+1} \sum_{p,q = 1}^d n_{q}^{\f} \, 
            {\frac{\partial \widetilde{F}_{p,i}}{ \partial \Bigl(\partial {w}_{j} /\partial x_q - {R}_{q}^\T(w_{j})\Bigr)}}  
        \jump{\delta w_j} \, \averg{\frac{\partial z_i}{\partial x_p}}  + \nonumber \\
   + &\sumF \intF \sum_{i,j = 1}^{d+1} \sum_{p,q = 1}^d n_{p}^{\f} \, 
              {\frac{\partial \widehat{F}_{p,i}}{ \partial \Bigl( \partial {w}_{j}/ \partial x_q - \eta_\f {r}_{q}^\f({w_{j}) \Bigr) }}} 
        \averg{\frac{\partial (\delta w_j)}{\partial x_q}} \, \jump{z_i} +\nonumber \\
   - &\sumF \intF \sum_{i,j = 1}^{d+1} \sum_{p,q = 1}^d \eta_\f
              {\frac{\partial \widehat{F}_{p,i}}{ \partial \Bigl( \partial {w}_{j}/ \partial x_q - \eta_\f {r}_{q}^\f({w_{j}) \Bigr) }}} 
               {r}^\f_{q}\left(\jump{\delta w_j}\right) \, {r}^\f_{p}\left(\jump{z_i}\right) \label{eq:jacTrilFNURef}
\end{align}
showing that only the last term, \textit{i.e.}, the stabilization term, cannot be reformulated lifting-free. In particular, it can be demonstrated that the inherited stabilization term introduces an excessive amount of stabilization with respect to its non-inherited counterpart, see~\ref{app:stab} for the theoretical estimates. In the context of \emph{h}-multigrid solution strategies, this showed to be detrimental for multigrid algorithm performance, see~\cite{botti2017h}, where the authors consider dG discretizations of the INS equations, and~\cite{AntoniettiDiosBrenner}, where preconditioners for weakly over-penalized symmetric interior penalty dG discretization of elliptic problems are devised. In those works, the use of rescaled-inherited coarse space operators was proposed in order to recover the correct amount of stabilization of the viscous operator and the optimal multigrid efficiency. In the numerical results, the use of rescaled inherited coarse grid operators extended to the $p$-version of the multigrid linear solver is also assessed, following the theoretical estimates of~\ref{app:stab}. A rather limited benefits was observed on practical three dimensional simulations involving convection-dominated regimes, which justifies the use of standard inherited approaches for production runs involving the under resolved direct numerical simulations of turbulent flows.

\subsubsection{The \emph{p}-multigrid iteration}
In this section we provide an overlook of the sequence of operations involved in \emph{p}-multigrid iterations. 
The recursive \emph{p}-multigrid $\mathcal{V}$-cycle and full \emph{p}-multigrid $\mathcal{V}$-cycle
for the problem $\Gm_{\ell} \; \delta \wvec_{\ell} = \gv_{\ell}$ on level $\ell$ reads:

\begin{minipage}[t]{6cm}
  \null
\begin{algorithm}[H]
\caption{$\overline{\wvec}_{\ell} = \mathrm{MG}_{\mathcal{V}}(l,\gv_{\ell},\wvec_{\ell})$ \label{algo:vcic}}
 \begin{algorithmic}[0]
 \IF {$(\ell = L)$}
 \STATE{$\overline{\wvec}_{\ell} = \mathrm{SOLVE} (\Gm_{\ell}, \gv_{\ell}, 0)$}
 \ENDIF
 \IF {$(\ell < L)$}
   \STATE{\underline{\emph{Pre-smoothing:}}}
   \STATE{$\overline{\wvec}_{\ell} = \mathrm{SMOOTH}(\Gm_{\ell}, \gv_{\ell}, \wvec_{\ell})$} 
   \vspace{0.2cm}
   \STATE{\underline{\emph{Coarse grid correction:}}}
   \STATE $\dvec_{\ell} = \gv_{\ell} - \Gm_{\ell} \overline{\wvec}_{\ell}$
   \STATE $\dvec_{\ell+1} = \mathcal{I}_{\ell}^{\ell+1} \dvec_{\ell}$
   \STATE $\evec_{\ell+1} = \mathrm{MG}_{\mathcal{V}}(\ell+1,\dvec_{\ell+1},0)$ 
   \STATE $\widehat{\wvec}_{\ell} = \overline{\wvec}_{\ell} + \mathcal{I}_{\ell+1}^{\ell} \evec_{\ell+1}$
   \vspace{0.2cm} 
   \STATE{\underline{\emph{Post-smoothing:}}}
   \STATE{$\overline{\wvec}_{\ell} = \mathrm{SMOOTH}(\Gm_{\ell}, \gv_{\ell}, \widehat{\wvec}_{\ell})$} 
 \ENDIF
 \STATE{return $\overline{\wvec}_{\ell}$}
\end{algorithmic}
\end{algorithm}
\end{minipage}
\hspace{0.5cm}
\begin{minipage}[t]{6cm}
  \null
\begin{algorithm}[H]
\caption{$\overline{\wvec}_{\ell} = \mathrm{MG}_{\mathrm{full}}(l,\gv_{\ell},\wvec_{\ell})$ \label{algo:fullvcic}}
 \begin{algorithmic}[0]
 \IF {$(\ell = L)$}
 \STATE{$\overline{\wvec}_{\ell} = \mathrm{SOLVE} (\Gm_{\ell}, \gv_{\ell}, 0)$}
 \ENDIF
 \IF {$(\ell < L)$}
   \STATE $\gv_{\ell+1} = \mathcal{I}_{\ell}^{\ell+1} \gv_{\ell}$
   \STATE $\widehat{\wvec}_{\ell+1} = \mathrm{MG}_{\mathrm{full}}(\ell+1,\gv_{\ell+1},0)$ 
   \STATE{\underline{\emph{$\mathcal{V}$-cycle correction:}}}
   \STATE $\widehat{\wvec}_{\ell} = \mathcal{I}_{\ell+1}^{\ell} \widehat{\wvec}_{\ell+1}$
   \STATE $\dvec_{\ell} = \gv_{\ell} - \Gm_{\ell} \widehat{\wvec}_{\ell}$
   \STATE $\evec_{\ell} = \mathrm{MG}_{\mathcal{V}}(\ell,\dvec_{\ell},0)$ 
   \STATE $\overline{\wvec}_{\ell} = \widehat{\wvec}_{\ell} + {\evec_{\ell}}$
 \ENDIF
 \STATE{return $\overline{\wvec}_{\ell}$}
\end{algorithmic}
\end{algorithm}
\end{minipage}

To obtain an application of the \emph{p}-multigrid preconditioner the multilevel iteration is 
invoked on the problem $\Gm_{h} \; \delta \wvec_{h} = \gv_{h}$.
While one $\mathrm{MG}_{\mathcal{V}}$ iteration requires two applications of the smoother on the finest level (one pre- and one post-smoothing) and one application of the coarse level smoother independently from the number of levels, one $\mathrm{MG}_{\mathrm{full}}$ iteration requires one application of the finest level smoother and $L$ applications of the coarse level smoother. 

In this work the \emph{p}-multigrid Full-$\mathcal{V}$ cycle iteration will be applied for the numerical solution of linearized equations systems arising in Rosenbrock time marching strategies for dG discretizations of incompressible flow problems. In the context of such problems we seek for the best performance employing full \emph{p}-multigrid and tuning preconditioners and smoothing options. We remark that, in this work, all the smoothers of the multigrid strategy are chosen to be GMRES, and thus two nested Krylov iterative solvers are envoked. In this setting, the outer solver follows the flexible GMRES implementation~\cite{saad1993flexible}, since the action of the multigrid linear solver needs to be stored at each iteration. It is worth noticing that, in order to reduce the overall memory footprint, a matrix-free implementation of the $p$-multigrid linear solver needs to be employed. To this end, we recall what reported in Section~\ref{sec:MemSav} regarding matrix-free iterative strategies, and we remark that only if EWBJ preconditioner is coupled with a matrix-free GMRES smoother on the finest space, an overall reduction of memory footprint of the solver can be achieved. As demonstrated in results section, such strategy can be conveniently used for practical simulations without spoiling the convergence rates of the multigrid iteration. Another important aspect regarding the choice of the smoothers involves the coarse space. To this end, the use of more powerful preconditioners like BJ or ASM for the smoothers on the coarsest space is typically suggested to ensure a satisfactory convergence rate not polluted by the domain decomposition. It is worth noting that, while matrix-free iterative solvers can be employed at no additional cost at high order, they are more expensive for low order polynomials, and thus the use of matrix-based methods would speed-up the solution process. In this configuration, the off-diagonal blocks needed by the coarse level operators can be computed at the coarsest polynomial degree for the sake of efficiency.
In the rest of the paper several combinations of preconditioners are investigated with particular attention to the quantification of the parallel performance and of the memory savings. We point out that, since the number of non-zeros of the primal jacobian matrix scales as $k^{2d}$, the memory footprint of a coarse space matrix can be less than 1\% that of the EWBJ used on the fine space, and thus the overall memory saving is not compromised.


\subsection{Memory footprint considerations}\label{sec:MemSav}
\begin{figure}[htbp!]
\centering
\subfigure[$d=2$]{\includegraphics[angle=0,width=0.75\textwidth]{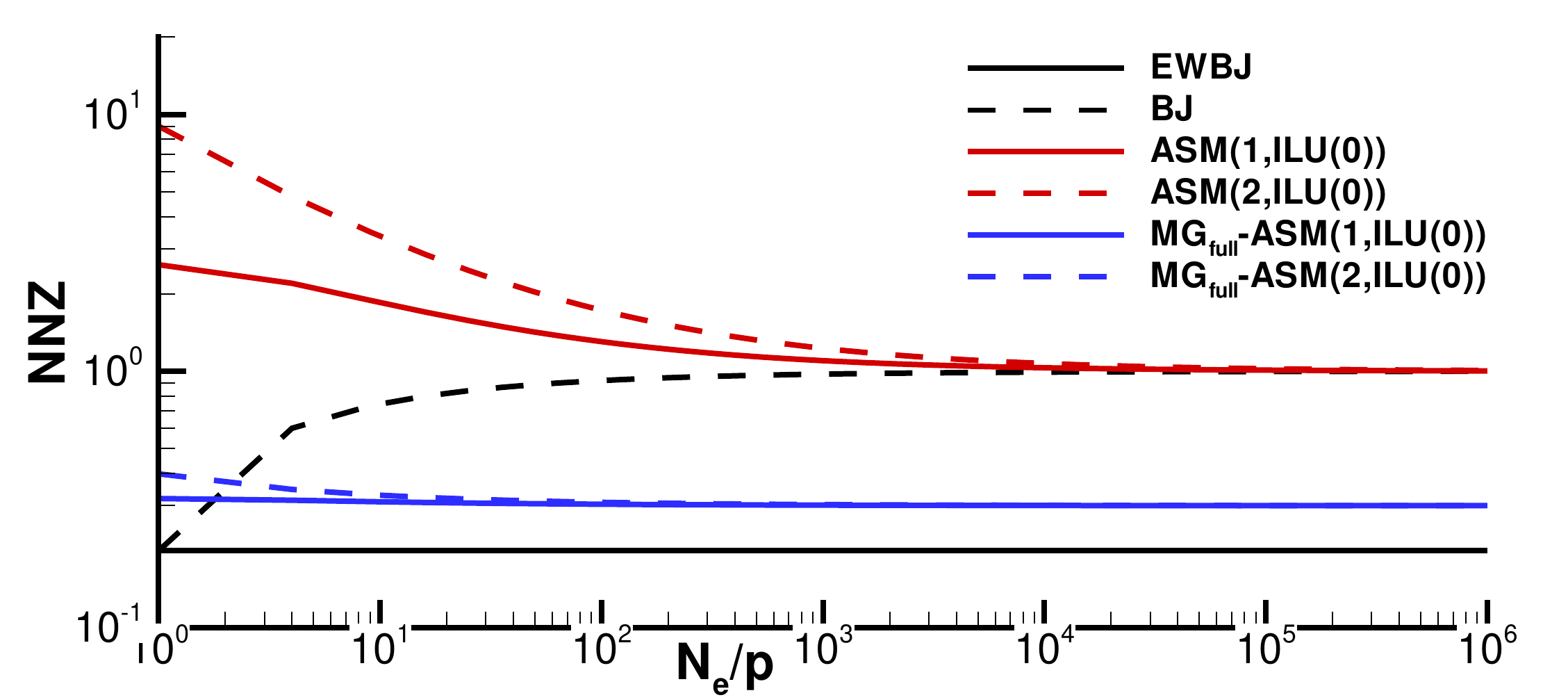} \label{fig:Mem2DCyl}}
\subfigure[$d=3$]{\includegraphics[angle=0,width=0.75\textwidth]{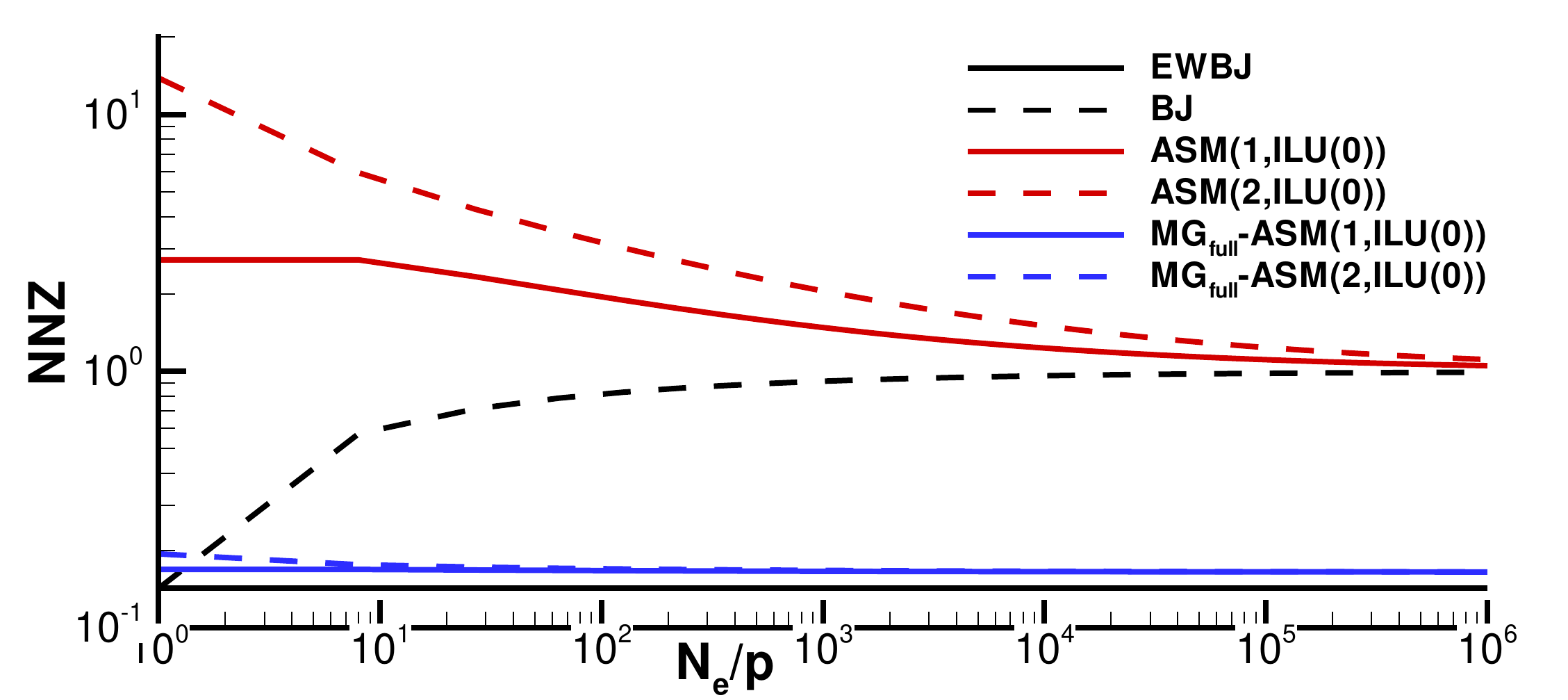} \label{fig:Mem3DSph}}
\caption{Estimated relative number of non-zeros (NNZ) of the preconditioner with respect to the non-zeros of the Jacobian as a function of the number of elements per partition for a two dimensional ($d=2$) and three-dimensional case ($d=3$). See text for details. \label{fig:MemSolver}}
\end{figure}

In this section an estimate of the memory footprint of all the strategies employed in this paper is devised to fully appreciate the memory savings achievable trough a matrix-free solver preconditioned with \emph{p}-multigrid. We observe that the memory footprint of the global block matrix scales as $$\card(\Th) \; (\overline{\card}(\Ft)+1) \, ((d+1) \, \dim(\Poly{d}{k}))^2,$$ where $\overline{\card}(\Ft)$ is the average number of element's faces, $d+1$ is the number of variables in $d$ space dimensions and $((d+1) \, \dim(\Poly{d}{k}))^2$ is the number of non-zeros in each matrix block. While for a matrix-based implementation we assume that both the global system matrix and the preconditioner are stored in memory, for a matrix-free approach only the preconditioner is stored. The preconditioner's memory footprint is carefully estimated: 
\begin{enumerate}
    \item for EWBJ, we consider only the non-zero entries of a block-diagonal matrix. 
    \item For BJ we consider the storage of ILU(0) factorization applied to the domain-wise portion of the iteration matrix, which neglects the off diagonal blocks related to faces residing on a partition boundary.
    \item For ASM($q$,ILU(0)), we assume that the preconditioner applies the ILU(0) decomposition to a larger matrix, bigger than the sub-domain matrix of the BJ precondtioner. The exact number of additional non-zero blocks is difficult to estimate for general unstructured grids since it depends on mesh topology, element types (in case of hybrid grids) and the partitioning strategy. Nevertheless, an estimation can be done based on the following simplifications: we assume a square and cubical domain discretized by uniformly distributed quadrilateral and hexahedral elements in $2d$ and $3d$, respectively, and we consider periodic boundary conditions on the domain boundaries. Accordingly the number of non-zero blocks in each sub-domain matrix can be estimated as follows
\begin{equation}
\displaystyle
 \left(2d{+}1 \right) \left( \frac{N_e}{p}{+}2dq \left( \frac{N_e}{p} \right)^{\frac{d{-}1}{d}} +  2^{d{-}1} d \sum_{i=1}^{q} \left( i{-}1 \right) \right)   
-2d \left(  \left( \left( \frac{N_e}{p} \right)^{\frac{1}{d}}{+}2q \right)^{d{-}1}{-}2^{d{-}1}  \left( d{-}2 \right)  \sum_{i=0}^{q} \left( q{-}i \right)   \right) 
\label{eq:mem_estimation}
\end{equation}
where $q$ is the number (or depth) of overlapping layers, $p$ is the number of processes, $N_e = \card(\Th)$ is the number of mesh elements and $\overline{\card}(\Ft)=2d$. In Eq.~\eqref{eq:mem_estimation}, the first term takes into account that each element of a partition, which is widened with the overlapping elements, contributes to the Jacobian matrix with $(1+2d)$ blocks, being $2d$ the number of faces of an element, while the second term subtracts the blocks not considered by the preconditioner, 
being they due to connection between elements at the boundary faces of the augmented partition. For $q=\{0,1,2\}$ we get an estimation of the number of non-zero blocks corresponding to BJ, ASM(1,ILU(0)) and ASM(2,ILU(0)) preconditioners, respectively. The number of non-zero entries of the global matrix can be obtained multiplying by the number of non-zeros in each block. 
\end{enumerate}  
Figures~\ref{fig:Mem2DCyl} and~\ref{fig:Mem3DSph} report the ratio between the estimated number of non-zeros of the preconditioner and the system matrix with respect to the number of sub-domains. For $(N_{e}/p)=1$, corresponding to one element per partition, BJ reduces to EWBJ, which provides a $1/(2d+1)$ decrease of the number of non-zeros. On the other hand, for both ASM(1,ILU(0)) and ASM(2,ILU(0)) the number of non-zero entries grows as $(N_{e}/p)$ approaches one. 
In the same manner the memory footprint of \emph{p}-multigrid preconditioners can be estimated.

We consider as reference the three-level \emph{p}-multigrid strategy whose specs, also reported on top of Table~\ref{tab:3dmg}, reads: $k=6$, FGMRES(MF) outer solver, GMRES(MB)[ASM(1,ILU(0)] smoothing on the coarsest level ($k=1$), GMRES(MF)[EWBJ] on the finest level ($k=6$) and GMRES(MB)[EWBJ] on the intermedite level ($k=2$). We remark that the memory allocation of coarse levels preconditioners has as a small impact on the total number of non-zeros, due the reduction of block size. For instance, in three space dimensions, $\mathbf{G}_{2}$ and $\mathbf{G}_{1}$ have $440$ and $70$ time less non-zeros that the $\mathbf{G}_0$ matrix, respectively. 

\section{Numerical results}\label{sec:INSResults}

%
In this section the performance of the \emph{p}-multigrid matrix-free preconditioner is compared to state-of-the-art single-grid strategies in the context of unsteady flow simulations. 
Three incompressible flow problems of increasing complexity are considered:
\begin{inparaenum}[i)]\item the two-dimensional laminar flow around a circular cylinder at $Re=200$; \item the three-dimensional laminar flow around a sphere at $Re=300$; \item the implicit LES of the transitional flow on a flat plate with semi-circular leading edge at $Re=3450$, namely the T3L1 test case of the ERCOFTAC (European Research Community on Flow, Turbulence and Combustion) test case suite; \item the implicit LES of the Boeing Rudimentary Landing Gear test case at $Re=10^6$.
\end{inparaenum}
This latter two test cases are representative of the target applications for the solution strategy here proposed.

\subsection{Laminar flow past a two-dimensional circular cylinder at $Re=200$}
\label{par:2dcyl}

\begin{figure}[b!]
\centering
\subfigure[Velocity magnitude iso-contours] {\includegraphics[angle=0,width=0.75\textwidth]{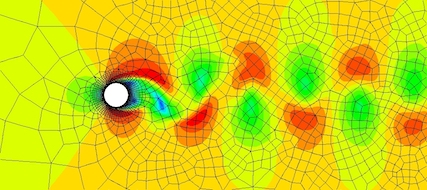}  \label{fig:2dViewCylinder_0}}
\subfigure[$C_d$ and $C_l$ coefficients history] {\includegraphics[angle=0,width=0.75\textwidth]{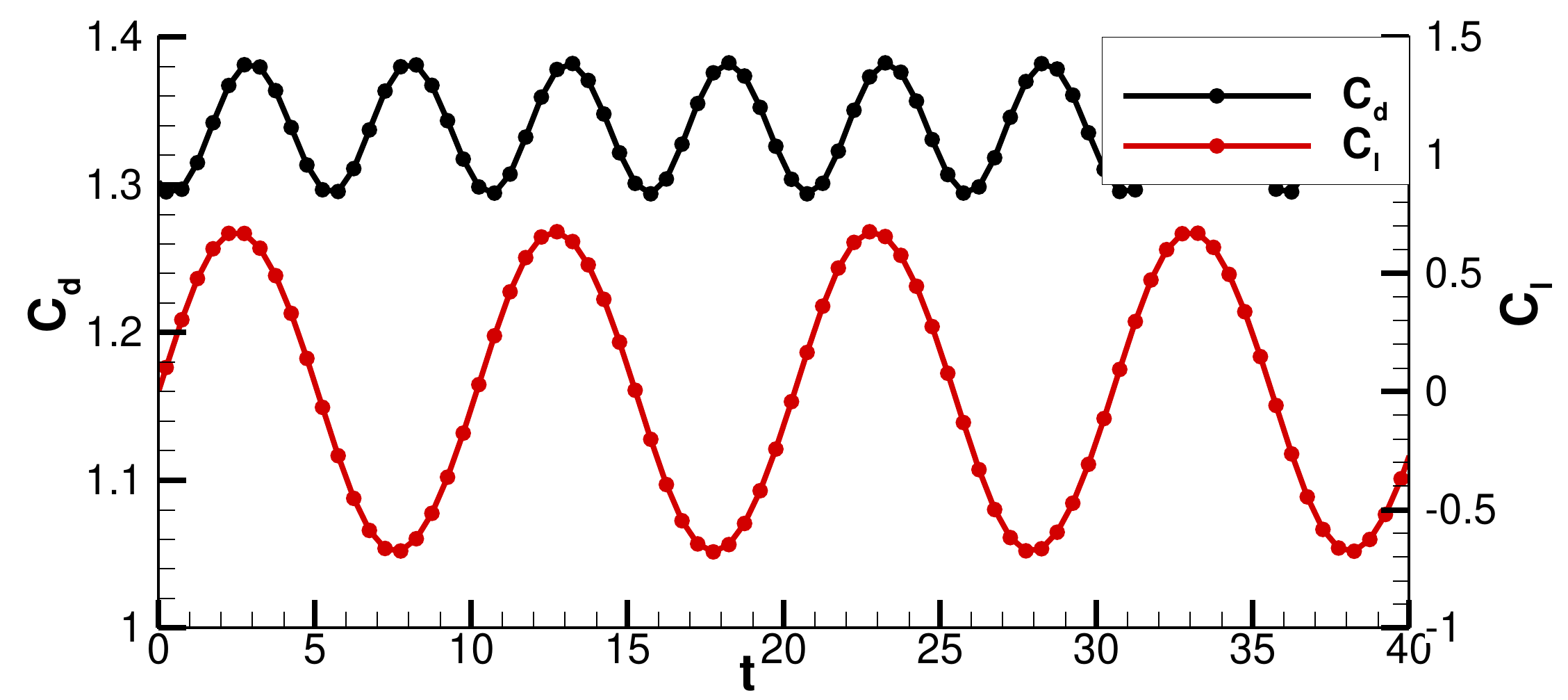}  \label{fig:2dViewCylinder_1}}
\caption{ Laminar flow around a circular cylinder at $Re=200$. Velocity magnitude iso-contour. \label{fig:2dViewCylinder}}
\end{figure}
The laminar flow around a circular cylinder at $Re=200$ is solved with $k=6$ on a computational grid made of 4710 elements with curved edges represented by cubic Lagrange polynomials. 
The domain extension is $[-50,100]{\times}[-50,50]$ in terms of non-dimensional units. We remark that the grid was deliberately generated with a severe grid refinement in the wake region, as well as large elements at the far-field, in order to challenge the solution strategy on a stiff space discretization. Dirichlet and Neumann boundary conditions are imposed at the inflow and outflow boundaries, respectively, while symmetry flow conditions are employed at the top and bottom boundaries. The no-slip Dirichlet boundary condition is imposed on the cylinder wall. A snapshot of the mesh and the velocity magnitude contours is shown in Fig.~\ref{fig:2dViewCylinder_0}.
Time marching is performed by means of the linearly-implicit four-stage, order three ROSI2PW Runge-Kutta method~\cite{Lang.Verwer:2001}. The scheme is specifically designed to accurately deal with PDAEs of index 2, like the INS equations. A non-dimensional fixed time step $\Delta t=0.25$, corresponding to 1/20 of the shedding period, is found to be adequate to describe the flow physics and large enough tostress the solution strategy. Fig.~\ref{fig:2dViewCylinder_1} reports lift and drag coefficients history. The drag coefficient $C_d=1.335$ and the Strouhal number $\text{St}=0.1959$ are in good agreement with~\cite{meneghini2001numerical} and references therein. Even if, for the sake of efficiency, it is possible to adaptively estimate the relative defect drop $\text{rTol}$ that the linear solver should attain~\cite{Franciolini2017276}, we set the fix value $\text{rTol} = 10^{-5}$ to compare different preconditioners with similar solution accuracy. The resulting time discretization error, estimated using the embedded Runge--Kutta scheme, is small enough not to affect the overall solution accuracy. Moreover, the defect tolerance is large enough to ensure that the matrix-free approximation error does not affect GMRES convergence, see~\cite{Franciolini2017276}.
The right preconditioning approach is employed throughout all the numerical experiments of this paper to reach convergence levels independent of the preconditioner.



\paragraph{Performance assessment}
The \emph{p}-multigrid preconditioner approach seeks to minimize the number of GMRES iterations on the fine grid by means of a full \emph{p}-multigrid solution strategy. A full $\mathcal{V}$-cycle \emph{p}-multigrid iteration (see Algorithm~\ref{algo:fullvcic}) has many parameters to tune in order to get the best performance, among the others we mention the following: 
\begin{inparaenum}[i)]
\item the choice of the smoother and its preconditioner on each level, 
\item the number of smoothing iteration on each level,
\item the forcing term (controlling the exit condition based on the relative defect drop) of the coarse solver and its maximum number of iterations.
\end{inparaenum}
Accordingly the combination of these parameters lead to a multitude of different configurations that is hard to explore comprehensively. Nevertheless, we provide some useful indications that can be directly applied to the simulation of realistic flow problems. In general, with respect to the linear test cases, such as that presented in~\ref{sec:PoissResults}, the use of a full \emph{p}-multigrid strategy together with an increased number of smoothing iterations proved to deal more efficiently with the non-linearity of the governing equations.
The experiments are devoted to show the benefits of \begin{inparaenum}[i)] \item the use of a memory saving smoother on the fine space to reduce the memory footprint and increase the computational efficiency; \item the use of a rescaled-inherited approach for the coarse space operators to improve the convergence rates of the iterative solver; \end{inparaenum}
Code profiling is applied for 10 time steps starting from a fully developed flow solution obtained with the same polynomial degree, same time step size and solving linear system up to the same tolerance. In practice, the performance of the preconditioners is averaged on the solution of 40 linear systems. The efficiency of each setting is monitored in parallel, to assess the behavior of different preconditioners in a realistic setup for this kind of computations. The runs are performed on a computational node made by two sixteen-core AMD Opteron CPUs.
 \begin{table}[b!]
 \centering
 \footnotesize
 \begin{tabular}{c | cc | cc | cc | cc | cc }
\multicolumn{1}{c}{Solver} &\multicolumn{2}{c}{GMRES(MB)}&\multicolumn{2}{c}{GMRES(MB)} &\multicolumn{2}{c}{GMRES(MF)}  &\multicolumn{2}{c}{GMRES(MF)}&\multicolumn{2}{c}{GMRES(MF)} \\
\multicolumn{1}{c}{Prec} &\multicolumn{2}{c}{BJ}&\multicolumn{2}{c}{ASM(1,ILU(0))} &\multicolumn{2}{c}{EWBJ}  &\multicolumn{2}{c}{BJ}&\multicolumn{2}{c}{ASM(1,ILU(0))} \\
 \hline
 nProcs & ITs & TotTime & ITs & TotTime & ITs & TotTime & ITs & TotTime &ITs & TotTime \\
         1 & 123.5 &  3805 &  123.5 &  3805 & 542.8 & 36700 & 112.2 &   9860 & 112.2 & 9860 \\
         2 & 108.0 &  1756 &   121.3 &   1917 & 538.7 & 17980 & 102.4 &   4547 & 109.9 & 4782 \\
         4 & 105.3 &    859 &  123.9 &  982 & 543.9 & 9281 & 103.5 &  2325 & 111.2 & 2426 \\
         8 & 138.0 &   543 & 120.4 &  515 & 542.2 & 4615 & 121.4 & 1333 & 111.2 & 1253 \\
       16 & 199.7 &   497 & 134.7 &  378 & 554.4 & 2934 & 171.3 & 995 & 122.8 & 750\\
 \hline
 \end{tabular}
 \caption{Two-dimensional cylinder test case. Single-grid parallel performances, matrix-based and matrix-free implementations. Comparison of the average number of GMRES iterations (ITs) and the whole elapsed CPU time (solution plus assembly) TotTime.
 }
 \label{tab:cyl2d.sg}
 \end{table}
First, we report in Table~\ref{tab:cyl2d.sg} the results obtained using single-grid preconditioners. As expected, the numerical experiments show a sub-optimal parallel efficiency for the BJ preconditioner, indeed the average number of linear iterations increases while increasing the number of sub-domains. The iterations number increase tops at 62\% when comparing the simulation on 16 cores against the serial one. The ASM(1,ILU(0)) preconditioner partially heals the performance degradation providing a 10\% increase of the iterations number at the expenses of a higher memory requirement, as explained in Section~\ref{sec:MemSav}. 
The matrix-based and the matrix-free versions of GMRES provide a similar number of iterations with a CPU time that is in favour of the former. This can be explained by the high quadrature cost associated to non-affine mesh elements, see~\eg \cite{BottiAppPoly}. In fact, while Franciolini \etal \cite{Franciolini2017276} demonstrated that the residual computation and a matrix-vector product has similar costs when dealing with high-order discretizations on affine elements, in this case the numerical quadrature expense has a higher relative cost on residual evaluation.
It is worth pointing out that no attempt was made to optimize numerical quadrature, in particular elements located far from curved boundaries are still treated as high-order non-affine elements for the sake of simplicity. Even if matrix-free iterations can be further improved in term of efficiency in realistic applications, this is beyond the scope of the present work.
\begin{table}[b!]
\centering 
\begin{tabular}{ c | c c | c c | c c | c c }
\multicolumn{1}{c}{Solver} & $\ell$ & \multicolumn{1}{c}{$\Kl$} & \multicolumn{1}{c}{Tol} & \multicolumn{1}{c}{ITs} & \multicolumn{2}{c}{Smoother} & \multicolumn{2}{c}{Prec}\\
\cline{1-9}
\multicolumn{1}{c}{\multirow{2}{*}{FGMRES[MG$_{\text{full}}$]}} &\multicolumn{1}{c}{0,1} &\multicolumn{1}{c}{$6,2$} & --  & \multicolumn{1}{c}{$\ast$} & \multicolumn{2}{c}{GMRES(MB)} & \multicolumn{2}{c}{$\ddagger$}\\
\multicolumn{1}{c}{}  &\multicolumn{1}{c}{2} & \multicolumn{1}{c}{$1$} & -- & \multicolumn{1}{c}{40} & \multicolumn{2}{c}{GMRES(MB)} & \multicolumn{2}{c}{ASM(1,ILU(0))}  \\ \hline
\multicolumn{1}{c|}{\multirow{2}{*}{scaling off}} & \multicolumn{2}{c|}{$^\ast$3} & \multicolumn{2}{c|}{$^\ast$3} & \multicolumn{2}{c|}{$^\ast$8} & \multicolumn{2}{c}{$^\ast$8} \\
 & \multicolumn{2}{c|}{$^\ddagger$BJ} & \multicolumn{2}{c|}{$^\ddagger$ASM(1,ILU(0))} & \multicolumn{2}{c|}{$^\ddagger$BJ} & \multicolumn{2}{c}{$^\ddagger$EWBJ} \\\hline
nProcs & ITs & SU$_{MB}$ & ITs & SU$_{MB}$ & ITs & SU$_{MB}$ & ITs & SU$_{MB}$\\
1 & 4.10 & 2.02 & 4.10 & 1.98 & 2.98 & 1.76 & 5.48 & 1.56 \\
2 & 4.63 & 1.82 & 4.05 & 1.90 & 3.10 & 1.65 & 5.48 & 1.49 \\
4 & 5.73 & 1.65 & 4.05 & 1.90 & 3.63 & 1.53 & 5.63 & 1.48 \\
8 & 7.20 & 1.39 & 4.35 & 1.74 & 3.85 & 1.40 & 5.63 & 1.40 \\
16 & 8.63 & 1.37 & 5.38 & 1.71 & 5.05 & 1.28 & 5.70 & 1.53 \\
\hline
\multicolumn{1}{c|}{\multirow{2}{*}{scaling on}} & \multicolumn{2}{c|}{$^\ast$3} & \multicolumn{2}{c|}{$^\ast$3} & \multicolumn{2}{c|}{$^\ast$8} & \multicolumn{2}{c}{$^\ast$8} \\
 & \multicolumn{2}{c|}{$^\ddagger$BJ} & \multicolumn{2}{c|}{$^\ddagger$ASM(1,ILU(0))} & \multicolumn{2}{c|}{$^\ddagger$BJ} & \multicolumn{2}{c}{$^\ddagger$EWBJ} \\\hline
nProcs & ITs & SU$_{MB}$ & ITs & SU$_{MB}$ & ITs & SU$_{MB}$ & ITs & SU$_{MB}$\\
1 & 3.43 & 2.11 & 3.43 & 2.11 & 2.48 & 1.88 & 3.50 & 1.92 \\
2 & 3.68 & 1.99 & 3.55 & 1.97 & 2.80 & 1.72 & 3.53 & 1.84 \\
4 & 4.78 & 1.79 & 3.60 & 1.95 & 2.55 & 1.80 & 3.83 & 1.79 \\
8 & 5.13 & 1.51 & 3.60 & 1.74 & 2.58 & 1.57 & 3.68 & 1.60 \\
16 & 7.65 & 1.20 & 4.10 & 1.67 & 2.85 & 1.49 & 3.50 & 1.68 \\
\end{tabular}
\caption{Two-dimensional cylinder test case. Effects of the smoother type and the rescaled-inherited coarse spaces on parallel performance. Comparison of the average number of FGMRES iterations (ITs) and the speed-up (SU$_{\mathrm{MB}}$) of the \emph{p}-multigrid preconditioner with respect to the best performing single-grid preconditioner. The asterisk and the double dagger symbols in the solver specs row are placeholders for the number of iterations (ITs) and coarse solver type of each column, respectively.}
\label{tab:cyl2d.mgfine}
\end{table}
Table~\ref{tab:cyl2d.mgfine} allows to evaluate the impact of the fine smoother preconditioner on the computational efficiency. We report two parameters of interest, the average number of FGMRES iterations (ITs) and the speed-up with respect to the best performing single-grid preconditioner, $\text{SU}_{\text{MB}}=\text{TotTime}/\text{TotTime}_\text{ref}$, where $\text{TotTime}_\text{ref}$ is the total CPU time of the GMRES(MB)[ASM(1,ILU(0))] approach. The specs of the \emph{p}-multigrid iteration setup are reported in the top of the table. We also compare the standard-inherited approach (\emph{scaling off}) with the rescaled-inherited one (\emph{scaling on}).

FGMRES[MG$_\text{full}$] with 3 GMRES(MB)[BJ] smoothing iterations provides a speed-up of about 2 in serial computations with respect to the reference strategy. Although the solver is still faster than the reference one, the parallel performance is not satisfactory being an increase in the number of iterations observed. As expected a better scalability can be obtained with 3 GMRES(MB)[ASM(1,ILU(0))] smoothing iterations. The numerical experiment revealed that to increase the number of iterations from 3 to 8 is mandatory to maintain the smoothing efficiency of GMRES(MB)[EWBJ], and to achieve a satisfactory performance in parallel. Indeed, despite being less performing in serial runs, the number of iterations is almost independent from the number of processes. It is worth noting that increasing the number of iterations of GMRES(MB)[BJ] does not pay off in terms of speedup. 
The number of FGMRES iterations is significantly reduced in all the cases when considering rescaled-inherited coarse grid operators. However, it can be seen that the strategy does not always pay off in terms of speedup, as an increased cost for the matrix assembly is required to compute the rescaled stabilization terms. It is worth noticing that the GMRES(MB)[EWBJ] smoother, despite being less powerful per-iteration, is competitive with more expensive preconditioners in parallel computations. Interestingly, the EWBJ preconditioner is also the cheapest from the memory footprint viewpoint.

\begin{table}[b!]
\centering
\begin{tabular}{ c | c c | c c | c c }
\multicolumn{1}{c}{Solver} & $\ell$ & \multicolumn{1}{c}{$\Kl$} & \multicolumn{1}{c}{Tol} & \multicolumn{1}{c}{ITs} & \multicolumn{1}{c}{Smoother} & \multicolumn{1}{c}{Prec}\\
\cline{1-7}
\multicolumn{1}{c}{\multirow{3}{*}{FGMRES[MG$_\text{full}$]}} &\multicolumn{1}{c}{0} &\multicolumn{1}{c}{$6$} & --  & \multicolumn{1}{c}{8} & $\ast$ & EWBJ\\
\multicolumn{1}{c}{} &\multicolumn{1}{c}{1} &\multicolumn{1}{c}{$2$} & --  & \multicolumn{1}{c}{8} & GMRES(MB) & EWBJ \\
\multicolumn{1}{c}{}  &\multicolumn{1}{c}{2} & \multicolumn{1}{c}{1} & -- & \multicolumn{1}{c}{40} & GMRES(MB) & $\ddagger$  \\ \hline
\multirow{2}{*}{scaling off} & \multicolumn{2}{c|}{$^\ast$GMRES(MB)} & \multicolumn{2}{c|}{$^\ast$GMRES(MB)} & \multicolumn{2}{c}{$^\ast$GMRES(MB)} \\
 & \multicolumn{2}{c|}{$^\ddagger$BJ} & \multicolumn{2}{c|}{$^\ddagger$ASM(1,ILU(0))} & \multicolumn{2}{c}{$^\ddagger$ASM(1,ILU(1))} \\\hline
nProcs & ITs & SU$_{MB}$ & ITs & SU$_{MB}$ & ITs & SU$_{MB}$ \\
1 & 5.48 & 1.56 & 5.48 & 1.56 & 4.73 & 1.64 \\
2 & 5.63 & 1.47 & 5.48 & 1.49 & 4.73 & 1.56  \\
4 & 5.43 & 1.52 & 5.63 & 1.48 & 4.73 & 1.57 \\
8 & 5.90 & 1.38 & 5.63 & 1.40 & 4.75 & 1.47  \\
16 & 6.85 & 1.37 & 5.70 & 1.53 & 4.95 & 1.62 \\\hline
\multirow{2}{*}{scaling on} & \multicolumn{2}{c|}{$^\ast$GMRES(MB)} & \multicolumn{2}{c|}{$^\ast$GMRES(MB)} & \multicolumn{2}{c}{$^\ast$GMRES(MB)} \\
 & \multicolumn{2}{c|}{$^\ddagger$BJ} & \multicolumn{2}{c|}{$^\ddagger$ASM(1,ILU(0))} & \multicolumn{2}{c}{$^\ddagger$ASM(1,ILU(1))} \\\hline
nProcs & ITs & SU$_{MB}$ & ITs & SU$_{MB}$ & ITs & SU$_{MB}$ \\
1 & 3.50 & 1.92 & 3.50 & 1.92 & 3.15 & 1.96 \\
2 & 3.55 & 1.85 & 3.53 & 1.84 &  3.15 & 1.88 \\
4 & 3.35 & 1.91 & 3.83 & 1.79 &  3.15 & 1.89 \\
8 & 3.33 & 1.70 & 3.68 & 1.60 & 3.15 & 1.67 \\
16 & 3.48 & 1.70 & 3.50 & 1.68 & 3.18 & 1.73 \\\hline\hline
\multirow{2}{*}{scaling off} & \multicolumn{2}{c|}{$^\ast$GMRES(MF)} & \multicolumn{2}{c|}{$^\ast$GMRES(MF)} & \multicolumn{2}{c}{$^\ast$GMRES(MF)} \\
 & \multicolumn{2}{c|}{$^\ddagger$BJ} & \multicolumn{2}{c|}{$^\ddagger$ASM(1,ILU(0))} & \multicolumn{2}{c}{$^\ddagger$ASM(1,ILU(1))} \\\hline
nProcs & SU$_{MB}$ & SU$_{MF}$ & SU$_{MB}$ & SU$_{MF}$ & SU$_{MB}$ & SU$_{MF}$ \\
1 & 0.61 & 1.59 & 0.61 & 1.59 & 0.69 & 1.80 \\
2 & 0.60 & 1.41 & 0.61 & 1.45 & 0.69 & 1.63 \\
4 & 0.60 & 1.42 & 0.59 & 1.41 & 0.69 & 1.63 \\
8 & 0.55 & 1.43 & 0.57 & 1.48 & 0.66 & 1.70 \\
16 & 0.61 & 1.61 & 0.73 & 1.92 & 0.77 & 2.03 \\\hline
\multirow{2}{*}{scaling on} & \multicolumn{2}{c|}{$^\ast$GMRES(MF)} & \multicolumn{2}{c|}{$^\ast$GMRES(MF)} & \multicolumn{2}{c}{$^\ast$GMRES(MF)} \\
 & \multicolumn{2}{c|}{$^\ddagger$ BJ} & \multicolumn{2}{c|}{$^\ddagger$ASM(1,ILU(0))} & \multicolumn{2}{c}{$^\ddagger$ASM(1,ILU(1))} \\\hline
nProcs & SU$_{MB}$ & SU$_{MF}$ & SU$_{MB}$ & SU$_{MF}$ & SU$_{MB}$ & SU$_{MF}$ \\
1 & 0.89 & 2.31 & 0.89 & 2.30 & 0.98 & 2.54 \\
2 & 0.92 & 2.18 & 0.89 & 2.11 & 0.98 & 2.32 \\
4 & 0.93 & 2.21 & 0.87 & 2.06 & 0.97 & 2.30 \\
8 & 0.86 & 2.23 & 0.85 & 2.20 & 0.92 & 2.39 \\
16 & 1.06 & 2.80 & 1.07 & 2.81 & 1.16 & 3.05 \\
\end{tabular}
\caption{Two-dimensional cylinder test case. Effects of the coarse level solver on parallel performance.
Comparison of the average number of FGMRES iterations (ITs) and the speed-up of the \emph{p}-multigrid preconditioner with respect to the best performing single-grid preconditioner in its matrix-based and matrix-free implementation (SU$_{\mathrm{MB}}$ and SU$_{\mathrm{MF}}$, respectively). The asterisk and the double dagger symbols in the solver specs row are placeholders for the smoother and coarse solver types of each column, respectively.}
\label{tab:cyl2d.mgcoarse}
\end{table}
Table~\ref{tab:cyl2d.mgcoarse} compares the computational efficiency when varying the preconditioner on the coarsest level. We fix the GMRES(MB)[EWBJ] smoother on all the other levels, with the idea of exploiting its performance on parallel runs. The top and bottom of the table include results for a matrix-based and a matrix-free approach, respectively. Note that only on the finest level the matrix-vector products are performed matrix-free, both within the outer FGMRES iteration and the fine GMRES smoother. Indeed, since the coarse levels operators are fairly inexpensive to store in memory, the moderate memory savings of a matrix-free implementation would not justify the increased computational costs at low polynomial orders. The results highlight that a further improvement in computational efficiency is achieved by means of a [ASM(1,ILU(1))] preconditioner for the coarsest smoother: the number of FGMRES iterations decreases while maintain optimal scalability. In addition, the speedup values for the matrix-free approach increase at large number of cores. We remark that, due to low polynomial degree of the coarsest level, the additional level of fill of the ILU factorization is not significant from the memory footprint viewpoint. Interestingly, the increased robustness of the rescaled-inherited \emph{p}-multigrid approach results in similar speedups for all the coarse level solver options, but also shows that with a powerful smoother on the coarsest level the standard inherited approach shows competitive speedup values as well.

Table~\ref{tab:cyl2d.mgcoarse} also reports numerical experiments using the matrix-free solver on the fine space. Computational efficiency is assessed using two different speedup values 
\begin{inparaenum}[i)] 
\item SU$_{\text{MB}}$ considers as a reference the GMRES(MB)[ASM(1,ILU(0))] solver, that is the best performing matrix-based single-grid strategy,
\item SU$_{\text{MF}}$ considers as a reference the GMRES(MF)[BJ] solver, that is the best performing matrix free single-grid strategy.
\end{inparaenum}
We disregard the inefficient single-grid GMRES(MF)[EWBJ] solver, despite requiring the lowest memory footprint. The results show that, since for this test case the cost-per-iteration of the matrix-free solver is higher than that of a matrix-based, reducing the number of FGMRES iterations does pay off. Accordingly, the benefits of rescaled-inherited coarse grid operators are more evident: the total execution time is comparable with GMRES(MB)[ASM(1,(ILU(0))] and almost three times faster than GMRES(MF)[BJ]. However, as already mentioned, further optimizations on the quadrature rules would reduce considerably this penalty.

To conclude, Table~\ref{tab:cyl2d.mgmf} compares three- and four-levels \emph{p}-multigrid preconditioners. The additional level significantly reduces the number of FMGRES iterations at the expense of storing a fourth degree coarse grid operator. The use of a rescaled-inherited approach reduce the number of iterations further, but the CPU time compares similarly for a matrix-based solution strategy. On the other hand, the benefits in terms of speedup are most significant in the matrix-free framework, where solution times dominates assembly times.
\begin{table}[b!]
\centering
\begin{tabular}{ c | c c | c c | c c | c c }

 \multicolumn{1}{c}{Solver} & \multicolumn{1}{c}{$\ell$} & \multicolumn{1}{c}{$\Kl$} & \multicolumn{1}{c}{rTol} & \multicolumn{1}{c}{ITs} & \multicolumn{2}{c}{Smoother} & \multicolumn{2}{c}{Prec} \\
 \cline{1-9}
 \multicolumn{1}{c}{\multirow{3}{*}{FGMRES[MG$_{\text{full}}$]}} &\multicolumn{1}{c}{0}&\multicolumn{1}{c}{$6$} &- &\multicolumn{1}{c}{8} & \multicolumn{2}{c}{$\ast$} & \multicolumn{2}{c}{EWBJ}\\
 \multicolumn{1}{c}{} & \multicolumn{1}{c}{1,...,L-1}&\multicolumn{1}{c}{$\ddagger$} &- &\multicolumn{1}{c}{8} & \multicolumn{2}{c}{GMRES(MB)} & \multicolumn{2}{c}{EWBJ}\\
 \multicolumn{1}{c}{}  &\multicolumn{1}{c}{L} & \multicolumn{1}{c}{1} & -  & \multicolumn{1}{c}{40} & \multicolumn{2}{c}{GMRES(MB)} & \multicolumn{2}{c}{ASM(1,ILU(1))}\\ \hline
%
%
%

\multirow{3}{*}{scaling off} & \multicolumn{2}{c|}{GMRES(MB)$^\ast$} & \multicolumn{2}{c|}{GMRES(MB)$^\ast$} & \multicolumn{2}{c|}{GMRES(MF)$^\ast$} & \multicolumn{2}{c}{GMRES(MF)$^\ast$} \\
\multicolumn{1}{c|}{}&  \multicolumn{2}{c|}{2$^\ddagger$ (L=2)} &  \multicolumn{2}{c|}{4,2$^\ddagger$ (L=3)} &  \multicolumn{2}{c|}{2$^\ddagger$ (L=2)}  & \multicolumn{2}{c}{4,2$^\ddagger$ (L=3)} \\\hline
nProcs & ITs & SU$_{MB}$ & ITs & SU$_{MB}$ & SU$_{MB}$ & SU$_{MF}$ & SU$_{MB}$ & SU$_{MF}$ \\
1 & 4.73 & 1.64 & 3.00 & 1.62 & 0.69 & 1.80 & 0.88 & 2.29 \\
2 & 4.73 & 1.56 & 3.00 & 1.53 & 0.69 & 1.63 & 0.87 & 2.06 \\
4 & 4.73 & 1.57 & 3.00 & 1.54 & 0.69 & 1.63 & 0.87 & 2.05 \\
8 & 4.75 & 1.47 & 3.10 & 1.43 & 0.66 & 1.70 & 0.78 & 2.02 \\
16 & 4.95 & 1.62 & 3.33 & 1.55 & 0.77 & 2.03 & 0.97 & 2.54 \\\hline
\multirow{3}{*}{scaling on} & \multicolumn{2}{c|}{GMRES(MB)$^\ast$} & \multicolumn{2}{c|}{GMRES(MB)$^\ast$} & \multicolumn{2}{c|}{GMRES(MF)$^\ast$} & \multicolumn{2}{c}{GMRES(MF)$^\ast$} \\
\multicolumn{1}{c|}{}&  \multicolumn{2}{c|}{2$^\ddagger$ (L=2)} &  \multicolumn{2}{c|}{4,2$^\ddagger$ (L=3)} &  \multicolumn{2}{c|}{2$^\ddagger$ (L=2)}  & \multicolumn{2}{c}{4,2$^\ddagger$ (L=3)} \\\hline
nProcs & ITs & SU$_{MB}$ & ITs & SU$_{MB}$ & SU$_{MB}$ & SU$_{MF}$ & SU$_{MB}$ & SU$_{MF}$  \\
1 & 3.15 & 1.96 & 2.00 & 1.91 & 0.98 & 2.54 & 1.19 & 3.09 \\
2 & 3.15 & 1.88 & 2.00 & 1.82 & 0.98 & 2.32 & 1.18 & 2.80 \\
4 & 3.15 & 1.89 & 2.00 & 1.83 & 0.97 & 2.30 & 1.19 & 2.82 \\
8 & 3.15 & 1.67 & 2.00 & 1.63 & 0.92 & 2.39 & 1.11 & 2.87 \\
16 & 3.18 & 1.73 & 2.00 & 1.72 & 1.16 & 3.05 & 1.33 & 3.50 \\
 \end{tabular}
 \caption{Two-dimensional cylinder test case. Comparison of a three-level and four-level \emph{p}-multigrid strategy in optimal settings for matrix-based and matrix-free fine level options in terms of SU MB and SU MF, respectively). The asterisk and the double dagger symbols in the solver specs row are placeholders for the smoother type and (number, order) of the coarse levels, respectively.}
 \label{tab:cyl2d.mgmf}
 \end{table}
 
In conclusion, it has been demonstrated how the use of cheap preconditioners like EWBJ on the fine space smoothers of multigrid cycle, coupled with a matrix-free implementation of the iterative solver, can be used to devise an efficient and memory saving solution strategy. The optimal scalability properties of the EWBJ, which involve local-to-each element operations, can be conveniently coupled with a more powerful preconditioning strategy on the coarse space smoothers, for example an Additive Schwarz method, with helps to maintain an optimal solution of the coarse space problem even in the context of highly parallel runs. In the next few sections, test cases of increasing complexity will be presented to further assess the performace of the devised strategy. We remark that similar specs to those reported herein will be employed for the multigrid iteration, which proved to be optimal in the context of the solution of incompressible Navier--Stokes equations.

\subsection{Three-dimensional laminar flow past a sphere at $Re=300$}

\begin{figure}[b!]
\centering
\subfigure[$C_p$ iso-contours]{\includegraphics[angle=0,width=0.75\textwidth]{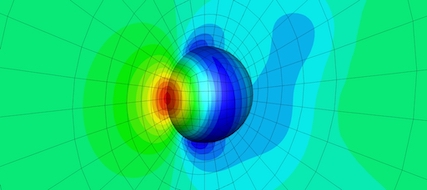}  \label{fig:3dViewSphere_0}}
\subfigure[$C_D$]{\includegraphics[angle=0,width=0.75\textwidth]{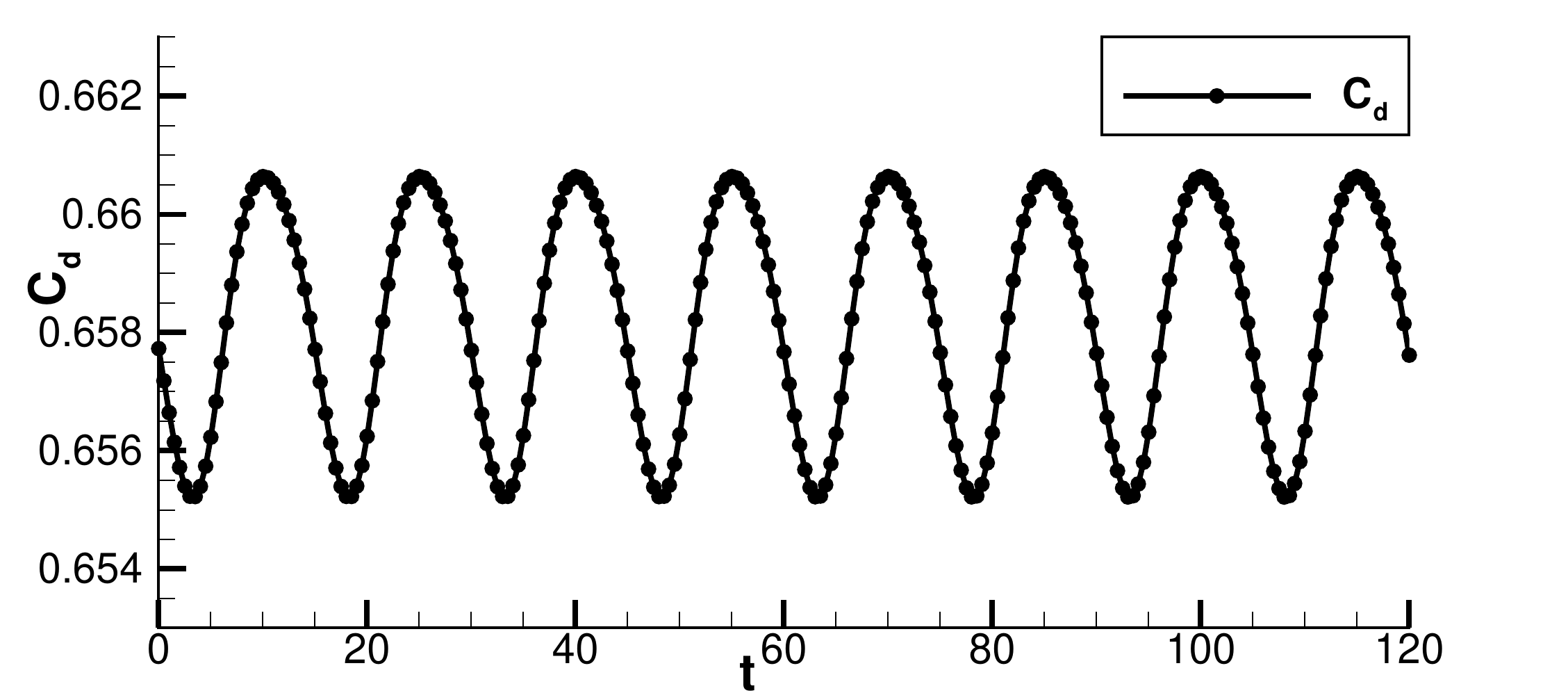}  \label{fig:3dViewSphere_1}}
\caption{Laminar flow around a Sphere at $Re=300$. Pressure coefficient iso-contours (top) and drag coefficient history (bottom).}
\end{figure}

As a three-dimensional validation test case we computed the unsteady laminar flow past a sphere at $Re=300$. The solution is characterised by a perfectly periodic behaviour, with the flow maintaining a plane of symmetry~\cite{johnson1999flow, tomboulides2000numerical, ploumhans2002vortex, Crivellini:Sphere}. In our computations, the symmetry plane was enforced by defining an appropriate boundary condition.
The mesh is made of 3560 elements with a bi-quadratic geometrical representation of the wall boundary, see Fig.~\ref{fig:3dViewSphere_0}. The computational domain is obtained via extrusion of the wall surface discretization. While the no-slip condition is set at the wall, velocity inflow and pressure outflow boundary conditions are imposed on the spherical farfield located at $50$ diameters. A $k=6$ representation of the solution was employed for all computations presented hereafter. We remark that the small number of mesh elements together with the lack of a refined region in the wake of the sphere reduce the stiffness of the problem. The parallel performance is evaluated running on the Marconi-A1 HPC platform hosted by CINECA, the italian supercomputing center. Scalability is assessed on a single-node base, as the CPU time of the serial computation exceeded the maximum wall-clock time allowed by CINECA. The number of mesh elements is optimised to ensure that all the solution strategies fit the memory of a single node (118 GB). Despite the small size of the problem the following numerical experiments aim at providing reliable indications on the parallel performance that can be extended to real-size production runs.

The solution is integrated in time with a fixed non-dimensional time step $\delta t=0.5$ and a relative tolerance on the linear system defect drop of $\text{rTol}=10^{-5}$. The drag coefficient time history is shown in Fig.~\ref{fig:3dViewSphere_1}, its mean value reads $0.659$, and the Strouhal number is $\text{St}=0.133$, in agreement with the published literature, see~\cite{Crivellini:Sphere}. Despite the geometry being represented with second degree polynomial spaces, the degree of exactness of quadrature rules does not consider the degree of mappings from reference to physical mesh elements. Accordingly, bilinear forms are exactly integrated only over affine mesh elements, located far away from the sphere boundaries. We numerically verified that, for this test case, this practice does not compromise accuracy while significantly improving the matrix-free computational time, see~\cite{Franciolini2017276}. For the sake of efficiency of parallel runs, the mesh has been partitioned using the \emph{local} two-level partitioning strategy described in~\cite{Crivellini:MPI-OMP}. The first-level decomposition is performed according to the number of nodes, and the second-level decomposition acts over each node-local partition according to the number of cores per node, such that the extra-node MPI communications are minimized.
%
%

\paragraph{Performance assessment}

 \begin{table}[b!]
 \centering
 \begin{tabular}{c | cc | cc | cc | cc }
\multicolumn{1}{c}{Solver} &\multicolumn{2}{c}{GMRES(MB)}&\multicolumn{2}{c}{GMRES(MB)}  &\multicolumn{2}{c}{GMRES(MF)}&\multicolumn{2}{c}{GMRES(MF)} \\
\multicolumn{1}{c}{Prec} &\multicolumn{2}{c}{BJ}&\multicolumn{2}{c}{ASM(1,ILU(0))} &\multicolumn{2}{c}{BJ}&\multicolumn{2}{c}{ASM(1,ILU(0))} \\\hline
 nProcs & ITs & TotTime & ITs & TotTime & ITs & TotTime & ITs & TotTime \\
36 & 78.5 & 1448.1 & 34.5 & 1245.9 & 77.1 & 1365.9 & 34.7 & 1220.7 \\
72 & 86.7 & 774.0 & 35.0 & 675.2 & 87.0 & 743.9 & 35.0 & 671.9\\
144 & 84.9 & 380.1 & 38.2 & 386.7 & 85.2 & 370.3 & 38.2 & 381.9\\
288 & 102.7 & 226.7 & 40.2 & 233.1 & 102.4 & 221.3 & 40.3 & 228.5\\
576 & 111.8 & 126.0 & 41.3 & 150.6 & 113.6 & 129.2 & 41.3 & 133.8 \\
 \hline
 \end{tabular}
 \caption{Three dimensional incompressible flow around a sphere.
  Single-grid parallel performances, matrix-based and matrix-free implementations. Comparison of the average number of GMRES iterations (ITs) and the whole elapsed CPU time (solution plus assembly) TotTime. Computations performed on Marconi-A1{@}CINECA.}
\label{tab:3dsg}
 \end{table}
 
Table~\ref{tab:3dsg} reports the parallel performance of the single-grid matrix-based and matrix-free solvers running in parallel up to 576 cores, \ie\ the domain is decomposed using 6 elements per partition on average. Increasing the number of sub-domains from 36 to 576 leads to an increased number of GMRES iterations: 42\% and 20\% up when employing a BJ and an ASM(1,ILU(0)) preconditioner, respectively. 
Thanks to the use of quadrature rules suited for affine meshes, the CPU time of the matrix-free solver is similar to the matrix-based one.

\begin{table}[b!] 
\centering
\begin{tabular}{ c | c c | c c | c c | c c }
 \multicolumn{1}{c}{Solver} & \multicolumn{1}{c}{$\ell$} & \multicolumn{1}{c}{$\Kl$} & \multicolumn{1}{c}{rTol} & \multicolumn{1}{c}{ITs} & \multicolumn{2}{c}{Smoother} & \multicolumn{2}{c}{Prec} \\
 \cline{1-9}
 \multicolumn{1}{c}{\multirow{3}{*}{FGMRES[MG${}_\text{full}$]}} &\multicolumn{1}{c}{0}&\multicolumn{1}{c}{$6$} &- &\multicolumn{1}{c}{8} & \multicolumn{2}{c}{$\ast$} & \multicolumn{2}{c}{EWBJ}\\
 \multicolumn{1}{c}{} & \multicolumn{1}{c}{1,...,L-1}&\multicolumn{1}{c}{$\ddagger$} &- &\multicolumn{1}{c}{8} & \multicolumn{2}{c}{GMRES(MB)} & \multicolumn{2}{c}{EWBJ}\\
 \multicolumn{1}{c}{}  &\multicolumn{1}{c}{L} & \multicolumn{1}{c}{1} & -  & \multicolumn{1}{c}{40} & \multicolumn{2}{c}{GMRES(MB)} & \multicolumn{2}{c}{ASM(1,ILU(0))}\\ \hline
%
%
%
\multirow{3}{*}{scaling off} & \multicolumn{2}{c|}{$^\ast$GMRES(MB)} & \multicolumn{2}{c|}{$^\ast$GMRES(MB)} & \multicolumn{2}{c|}{$^\ast$GMRES(MF)} & \multicolumn{2}{c}{$^\ast$GMRES(MF)} \\
\multicolumn{1}{c|}{}&  \multicolumn{2}{c|}{$^\ddagger$2 (L=2)} &  \multicolumn{2}{c|}{$^\ddagger$4,2 (L=3)} &  \multicolumn{2}{c|}{$^\ddagger$2 (L=2)}  & \multicolumn{2}{c}{$^\ddagger$4,2 (L=3)} \\\hline
nProcs & ITs & SU$_{MB}$ & ITs & SU$_{MB}$ & SU$_{MB}$ & SU$_{MF}$ & SU$_{MB}$ & SU$_{MF}$ \\
36 & 4.00 & 1.29 & 2.00 & 1.61 & 1.37 & 1.29 & 2.28 & 2.15 \\
72 & 4.00 & 1.37 & 2.00 & 1.68 & 1.52 & 1.46 & 2.38 & 2.29 \\
144 & 4.00 & 1.29 & 2.00 & 1.43 & 1.45 & 1.41 & 2.07 & 2.02 \\
288 & 4.00 & 1.37 & 2.00 & 1.67 & 1.56 & 1.52 & 2.14 & 2.09 \\
576 & 4.00 & 1.28 & 2.00 & 1.55 & 1.46 & 1.50 & 1.55 & 1.59 \\\hline
\multirow{3}{*}{scaling on} & \multicolumn{2}{c|}{GMRES(MB)$^\ast$} & \multicolumn{2}{c|}{$^\ast$GMRES(MB)} & \multicolumn{2}{c|}{$^\ast$GMRES(MF)} & \multicolumn{2}{c}{$^\ast$GMRES(MF)} \\
\multicolumn{1}{c|}{}&  \multicolumn{2}{c|}{$^\ddagger$2 (L=2)} &  \multicolumn{2}{c|}{$^\ddagger$4,2 (L=3)} &  \multicolumn{2}{c|}{$^\ddagger$2 (L=2)}  & \multicolumn{2}{c}{$^\ddagger$4,2 (L=3)} \\\hline
nProcs & ITs & SU$_{MB}$ & ITs & SU$_{MB}$ & SU$_{MB}$ & SU$_{MF}$ & SU$_{MB}$ & SU$_{MF}$  \\
36 & 3.0 & 1.43 & 2.00 & 1.51 & 1.53 & 1.44 & 2.09 & 1.97 \\
72 & 3.0 & 1.52 & 2.00 & 1.61 & 1.62 & 1.56 & 2.18 & 2.09 \\
144 & 3.0 & 1.36 & 2.00 & 1.50 & 1.55 & 1.51 & 1.93 & 1.88 \\
288 & 3.0 & 1.56 & 2.00 & 1.60 & 1.71 & 1.67 & 2.02 & 1.97 \\
576 & 3.0 & 1.45 & 2.00 & 1.43 & 1.63 & 1.67 & 1.73 & 1.78 \\
\end{tabular}
\caption{Three dimensional incompressible flow around a Sphere. Efficiency of a three and four level \emph{p}-multigrid strategy varying the fine level smoother. Comparison of the average number of FGMRES iterations (ITs) and the speed-up of the \emph{p}-multigrid preconditioner with respect to the best performing single-grid preconditioner in its matrix-based and matrix-free implementation (SU$_{\mathrm{MB}}$and SU$_\mathrm{MF}$,respectively). The asterisk and the double dagger symbols in the solver specs row are placeholders for the smoother and coarse solver types of each column, respectively. Computations performed on Marconi-A1{@}CINECA.}
\label{tab:3dmg}
\end{table}
Results reported in Table~\ref{tab:3dmg} for a three- and four-levels \emph{p}-multigrid strategy and the exact same setup of two-dimensional computations confirm the efficacy of the multigrid preconditioner: the number of iterations stays the same up to 576 cores and the speedup are maintained in this largely-parallelized scenario. In addition, it is worth noticing how the use of a matrix-free implementation reduce the CPU time over its matrix-based counterpart, since the matrix assembly time is reduced as discussed in Section~\ref{sec:matrix-free}. It is worth noting that the stabilization scaling provides only slight improvements in terms of FGMRES iterations, while the speedup values looks almost similar to those obtained using standard-inheritance, especially for the most parallelized cases. We finally remark that the four-level \emph{p}-multigrid preconditioner, with the same settings on the fine/coarse space smoothers, is almost two times faster than the best single-grid setup.


\section{Application to under resolved simulations of incompressible turbulent flows}
In this section the devised $p$-multigrid matrix-free implementation is applied to the implicit LES of two test cases. The first is the transitional flow around a flat plate with a semi-circular leading edge of diameter $d$ at $Re_d=3450$ and a low level of free-stream turbulence intensity ($\text{Tu}=0.2\%$). The second is the turbulent flow around the Boeing Rudimentary Landing Gear test case at $Re=10^6$.

\subsection{ERCOFTAC T3L1 test case}
\label{crivellini_sec:results}
This test case, named T3L1, is part of the ERCOFTAC test case suite. The solution exhibits at leading edge a laminar separation bubble and, downstream the transition, an attached turbulent boundary layer. Those complex flow features are perfectly suited to evaluate the efficiency of the solver and highlight the advantages of using a dG-based ILES approach. ILES naturally resolves all the flow scales (in a DNS-fashion) if the numerical resolution is enough to do so, while the numerical dissipation plays the role of a sub-grid scale model for the spatially under-resolved regions of the domain. In this test case, the laminar region is fully resolved, while in the turbulent region the dissipation of the numerical scheme dumps the under-resolved scales. We remark that in all the computations the same settings of Table~\ref{tab:3dmg} are employed for the $p$-multigrid iteration.

\begin{figure}[b!]
\centering
\includegraphics[angle=0,width=0.75\textwidth]{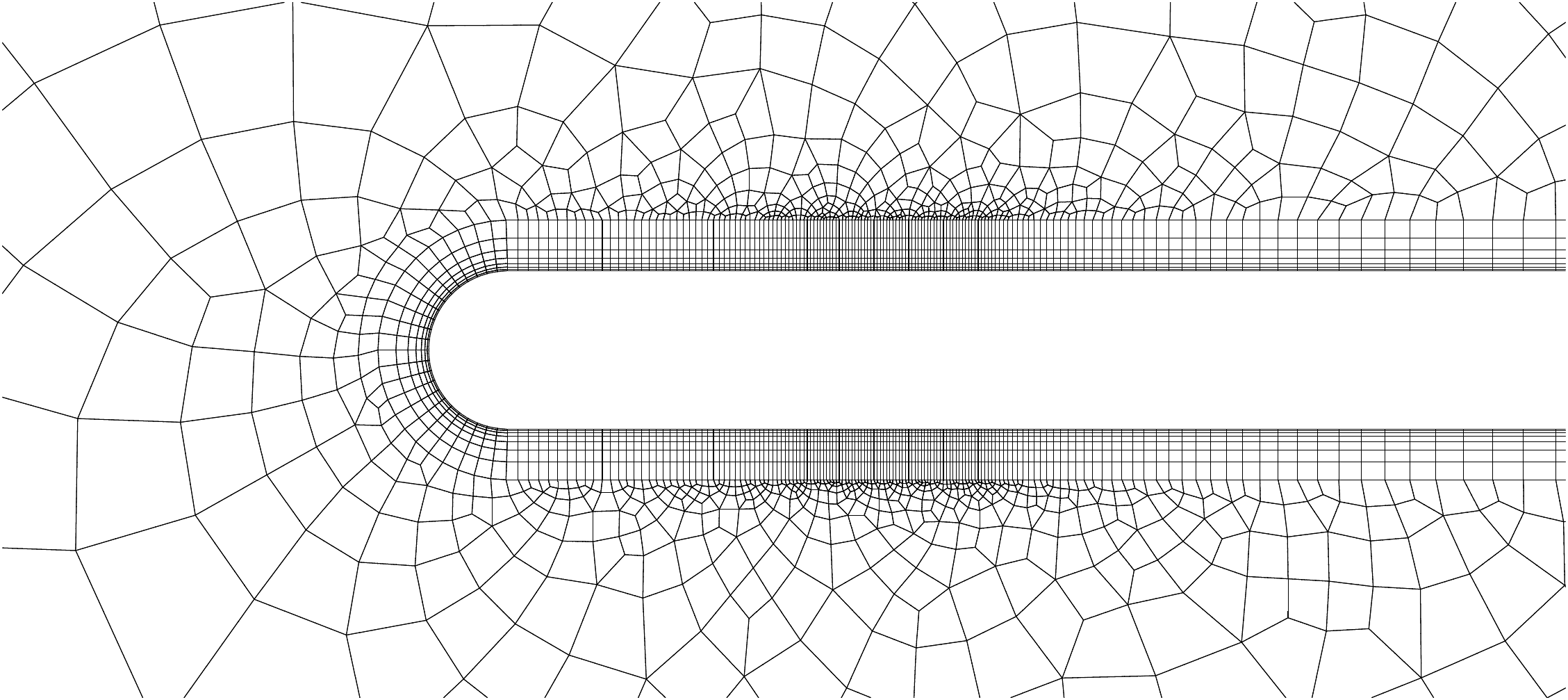}
\caption{T3L1 test case. Near-wall detail of the computational grid.\label{fig:t3l-grid}}
\end{figure}

The simulations were performed in parallel using $540$ cores on a hybrid mesh of $38320$ elements with curved edges. The unstructured grid is strongly coarsened moving away from the plate, while a structured-like boundary layer is used at the wall. The first cell height is $10^{-2}d$ and the mesh is refined near the reattachment region, where the minimum dimension along $x$ axis is $2\cdot 10^{-2}d$, see Fig.~\ref{fig:t3l-grid}. The domain extension on the $x-y$ plane is taken from ~\cite{Langari.Yang.2013}, \emph{i.e.}, $28d \times 17d$, and it is extruded using $10$ elements along the span-wise direction $z$ for a length of $2d$, as in~\cite{yang_voke_2001}. To our best knowledge, only those two works report a LES simulation of this ERCOFTAC test case. In both the cases, the numerical method was based on a standard second-order scheme and a dynamic subgrid scale model. 
A direct comparison of the present computations with previously published works in terms of DoFs is not trivial due to the differences of the computational domains. In \cite{yang_voke_2001} the DoFs count per variable is on the order of $1.88\cdot 10^6$ and the domain extension in the $x-y$ plane is $1.9$ times smaller. In ~\cite{Langari.Yang.2013} the Dofs count is on the order of $4.39\cdot 10^6$ and the domain is $4$ times larger in the span-wise direction. The result with the lowest resolution presented in this paper has about $1.39\cdot 10^6$ DOFs and takes advantage of the unstructured nature of the grid by increasing the mesh density during the turbulent transition and reattachment, that is for $x/d \gtrapprox 4.5$ up to the outflow.

\paragraph{Physical discussion}
This type of flow problem is reported to be very sensitive to the free-stream turbulence at the inlet ($\text{Tu}$). 
The free-stream flow was carefully manipulated to reproduce those reported by ERCOFTAC as well as previous numerical computations~\cite{yang_voke_2001,Langari.Yang.2013}. In those works, a white-noise random perturbation was added at the inflow velocity to mimic the low experimental turbulence level, \textit{i.e.}, $\text{Tu}<0.2$\%. 
In the present work, due to an aggressive mesh coarsening in the far-field regions, the generation of a free-stream turbulence at inlet is unfeasible. In fact, the coarse spatial discretization at far-field would rapidly damp any random perturbation introduced upstream. 
Accordingly, the turbulent fluctuations were synthetically injected, via a spatially-supported random forcing term, in those regions of the domain where the mesh density is enough not to dissipate small scales. The random forcing analytic expression assumes a Gaussian distribution in the $x$ direction and is homogeneous in $y{-}z$ 
\begin{equation}
f_i= A e^{\left(-\frac{x_1-\overline{x}_1}{2\mu}\right)} r_i
\end{equation}
where $A$, $\overline{x}_1$, $\mu$ and $r_i$ are, respectively, the amplitude coefficient, the location of the forcing plane, the amplitude of the Gaussian support and a normalised random vector component such that $\sqrt{r_i r_i} = 1$. The Gaussian function was centered in $\overline{x}_1/D=-3$, and the constants $A$, $\mu$ were adjusted, via trial and error approach, to meet the experimental $\text{Tu}$ levels. We avoid a fine control algorithm of the turbulent length-scale since the reattachment length is pretty insensitive to this value, see \cite{HILLIER198149,nakamura_ozono_1987,YANG20091026}. In the present configuration the expected turbulence intensity is met setting $A=0.06$ and $\mu=0.01$. We remark that perturbing the velocity field through a forcing term in the momentum equations guarantees a divergence-free perturbation.

\begin{figure}[htbp!]
\centering
\subfigure[$\text{Tu}=0\%$]{\includegraphics[angle=0,width=0.75\textwidth]{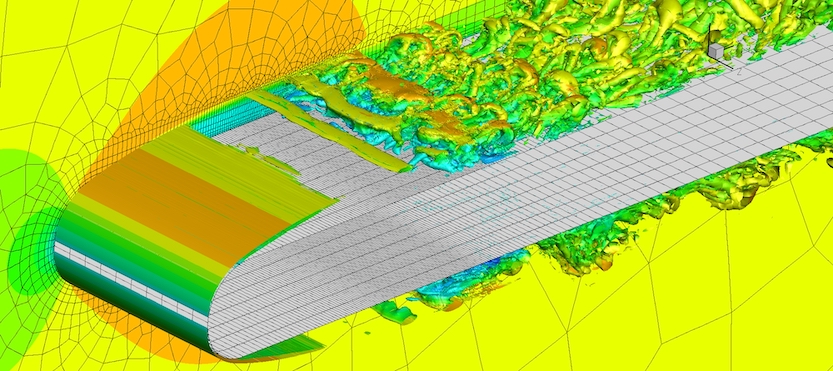} \label{fig:t3l_plot_0}}
\subfigure[$\text{Tu}=0.2\%$]{\includegraphics[angle=0,width=0.75\textwidth]{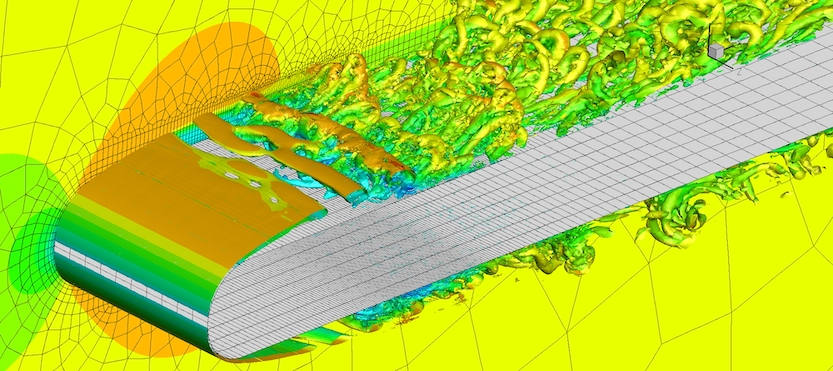} \label{fig:t3l_plot_02}}
\caption{T3L1 test case, $k=6$ solutions for different $\text{Tu}$ levels. $\lambda_2=-1$ iso-contour and periodic plane coloured by the streamwise velocity. \label{fig:t3l_plot}}
\end{figure}
Fig.~\ref{fig:t3l_plot} depicts the instantaneous flow fields computed with $\text{Tu}=0$\% and $\text{Tu}=0.2$\%. In both cases, the quasi two-dimensional Kelvin-Helmholtz instabilities in the shear-layer region above the separation bubble and their convection downstream are observed. As expected, the low free-stream turbulence intensity value promotes the instability of the quasi two-dimensional structures arising from the upstream flow separation. For both the conditions, hairpin vortices developing after flow reattachment and the breakdown to turbulence are similar. Distortion along the spanwise direction is anticipated upstream in the $\text{Tu}=0.2$\% case. 

\begin{figure}[htbp!]
\centering
\includegraphics[angle=0,width=0.75\textwidth]{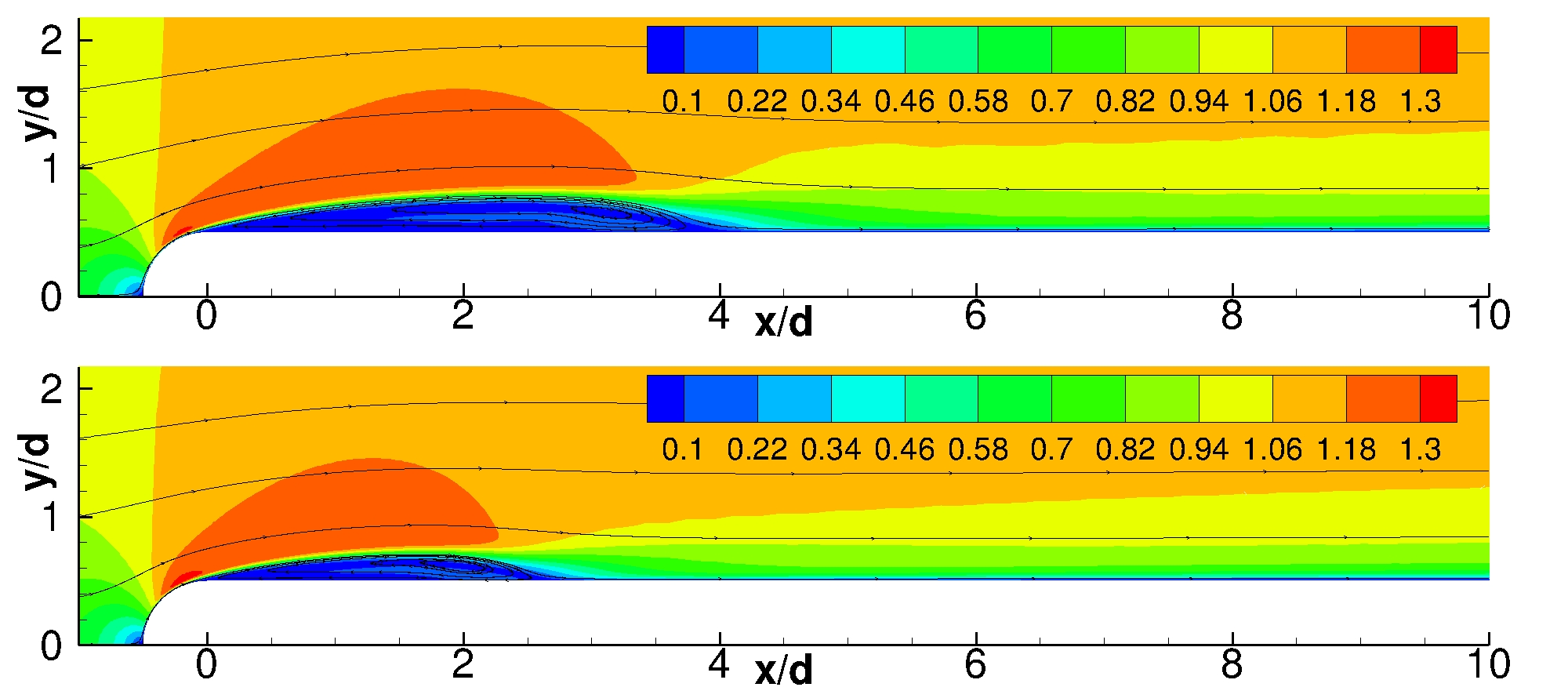}
\caption{T3L1 test case. Effect of the $\text{Tu}$ level. Average velocity magnitude iso-contours, $k=6$ solutions. Top: $\text{Tu}=0.0\%$; Bottom: $\text{Tu}=0.2\%$. \label{fig:t3l_um}}
\end{figure}

\begin{figure}[htbp!]
\centering
\includegraphics[angle=0,width=0.75\textwidth]{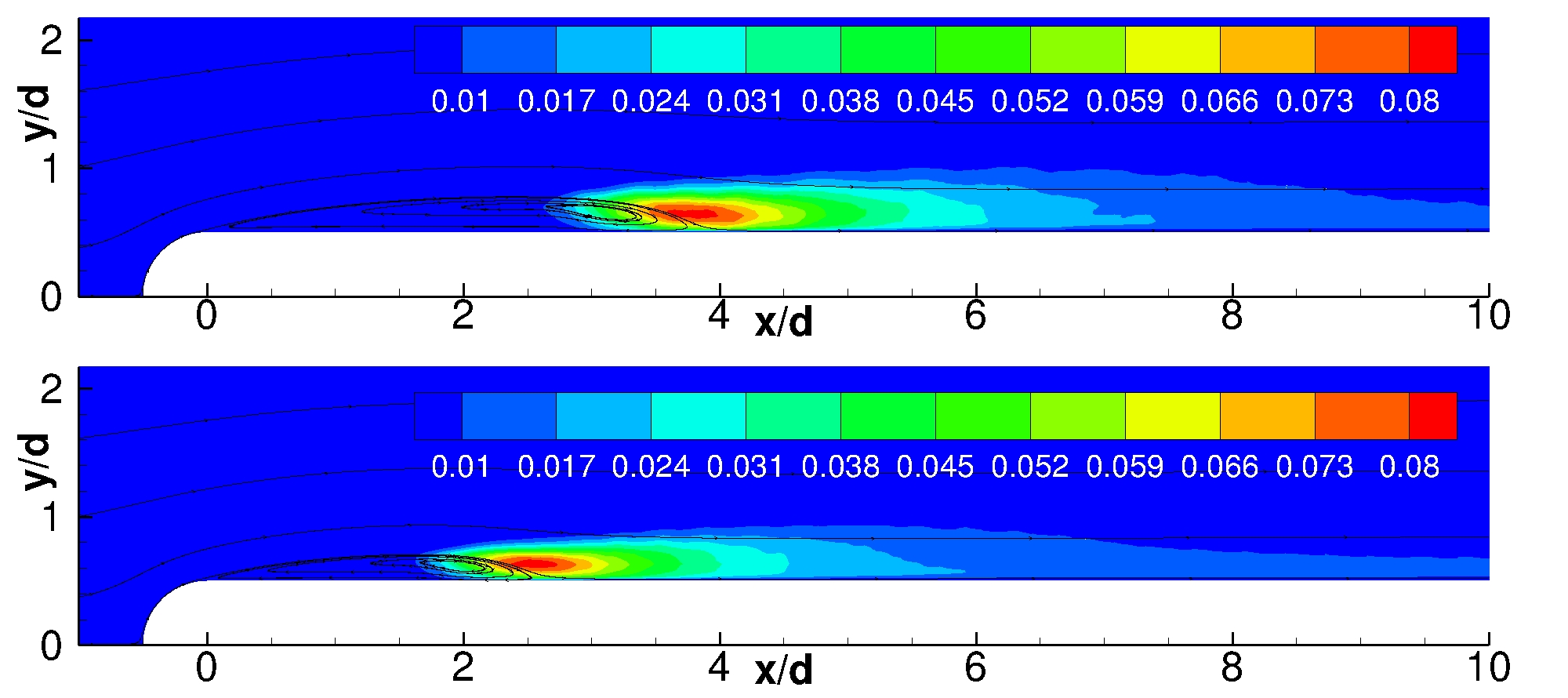}
\caption{T3L1 test case. Effect of the $\text{Tu}$ level. Turbulent kinetic energy iso-contours, $k=6$ solutions. Top: $\text{Tu}=0.0\%$; Bottom: $\text{Tu}=0.2\%$. \label{fig:t3l-k}}
\end{figure}

\begin{figure}[htbp!]
\centering
\includegraphics[angle=0,width=0.7\textwidth]{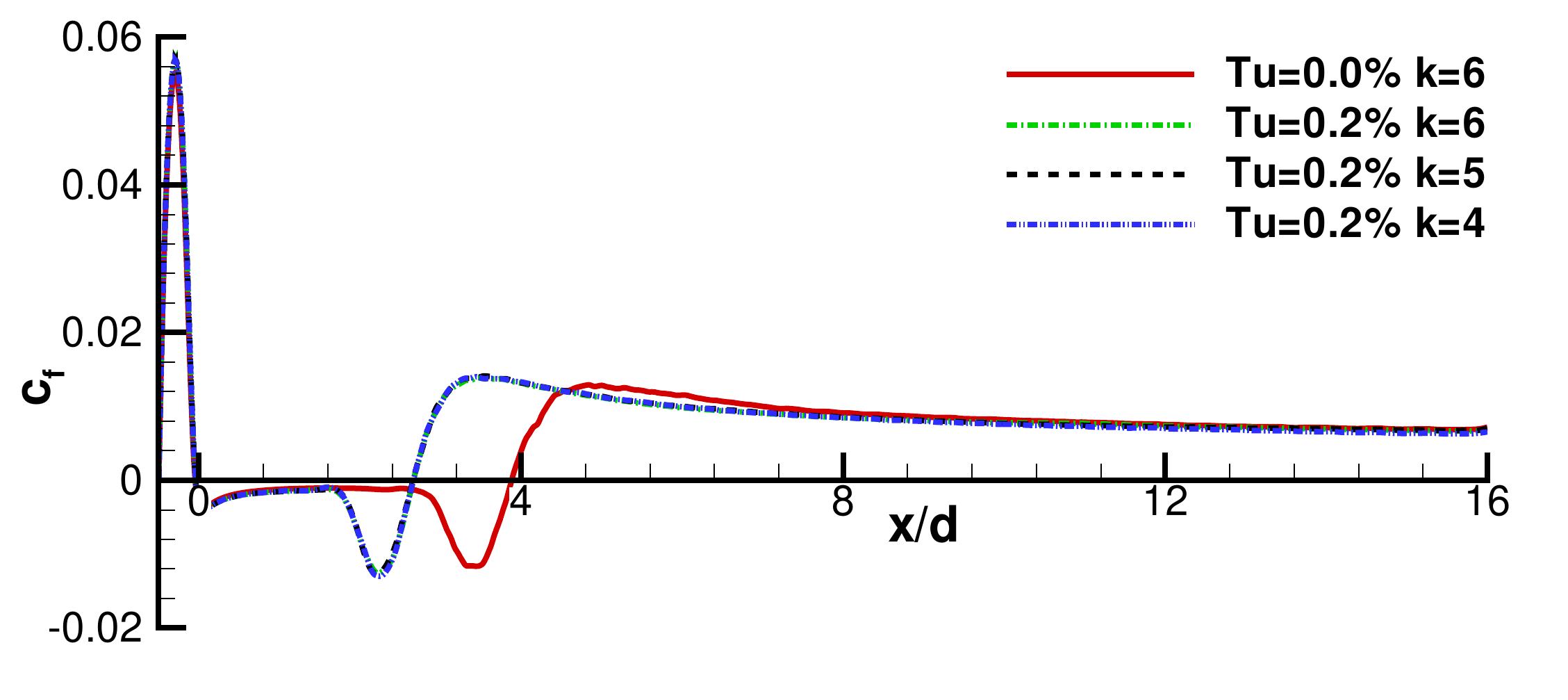}
\caption{T3L1 test case. Effect of the $\text{Tu}$ level on the skin friction coefficient $c_f$.\label{fig:t3l_cf}}
\end{figure}

\begin{figure}[htbp!]
\centering
\includegraphics[angle=0,width=0.75\textwidth]{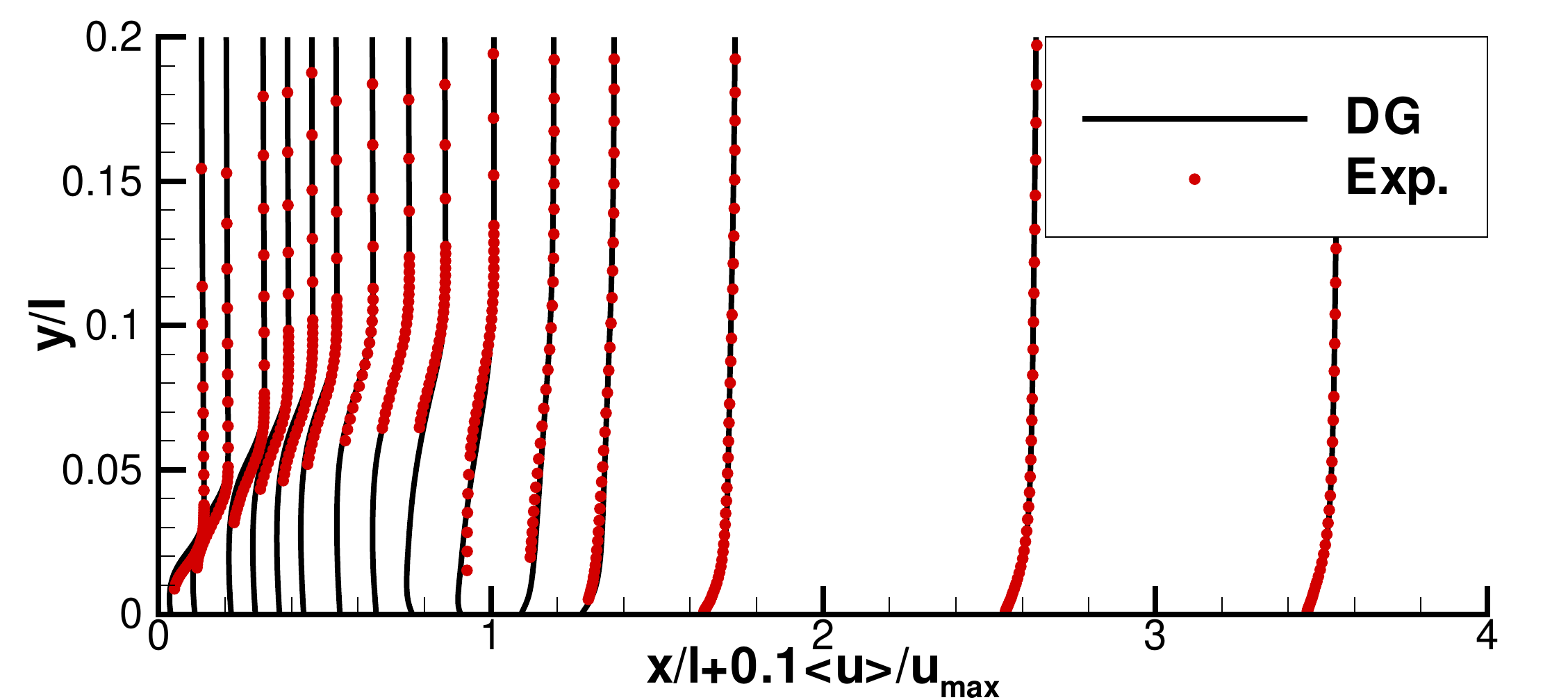} 
\includegraphics[angle=0,width=0.75\textwidth]{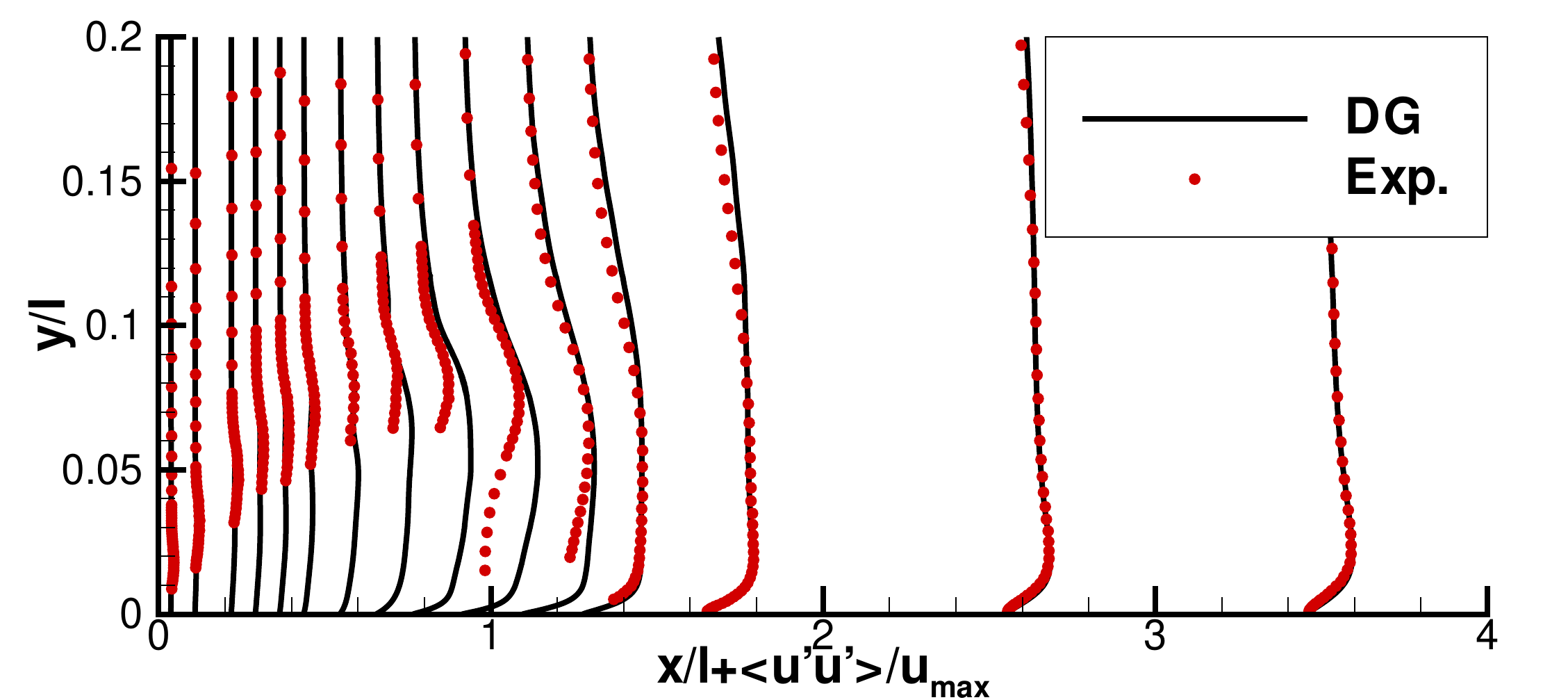} 
\caption{T3L1 test case, $k=6$ solution for $Tu=0.2$\%. Time- and spanwise-averaged velocity profiles in comparison with experimental data~\cite{cutrone2008predicting}. Mean (top) and RMS (bottom) flow velocity. Abscissas are non-dimensionalized using the experimental reattachment length $l/d=2.75$. \label{fig:P6ave}}
\end{figure}

\begin{figure}[htbp!]
\centering
\includegraphics[angle=0,width=0.75\textwidth]{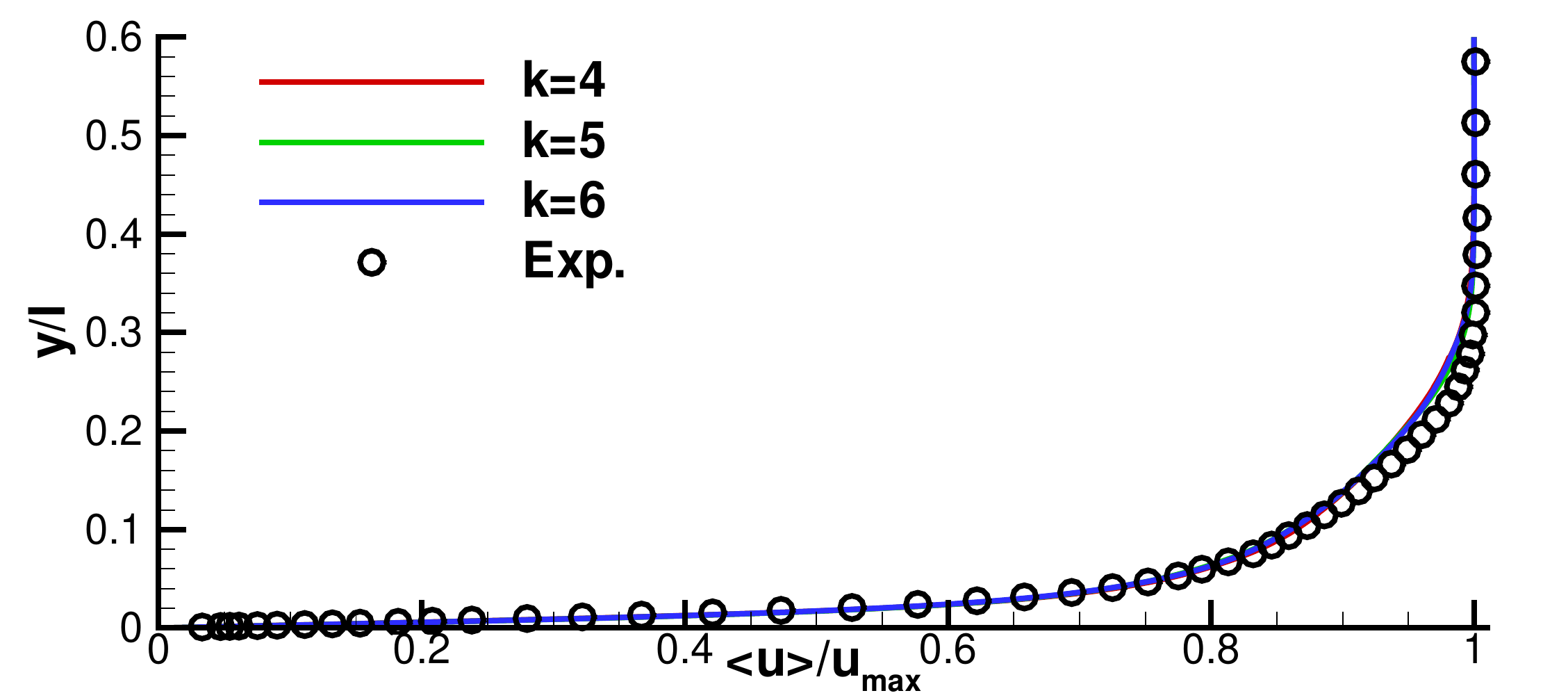} 
\includegraphics[angle=0,width=0.75\textwidth]{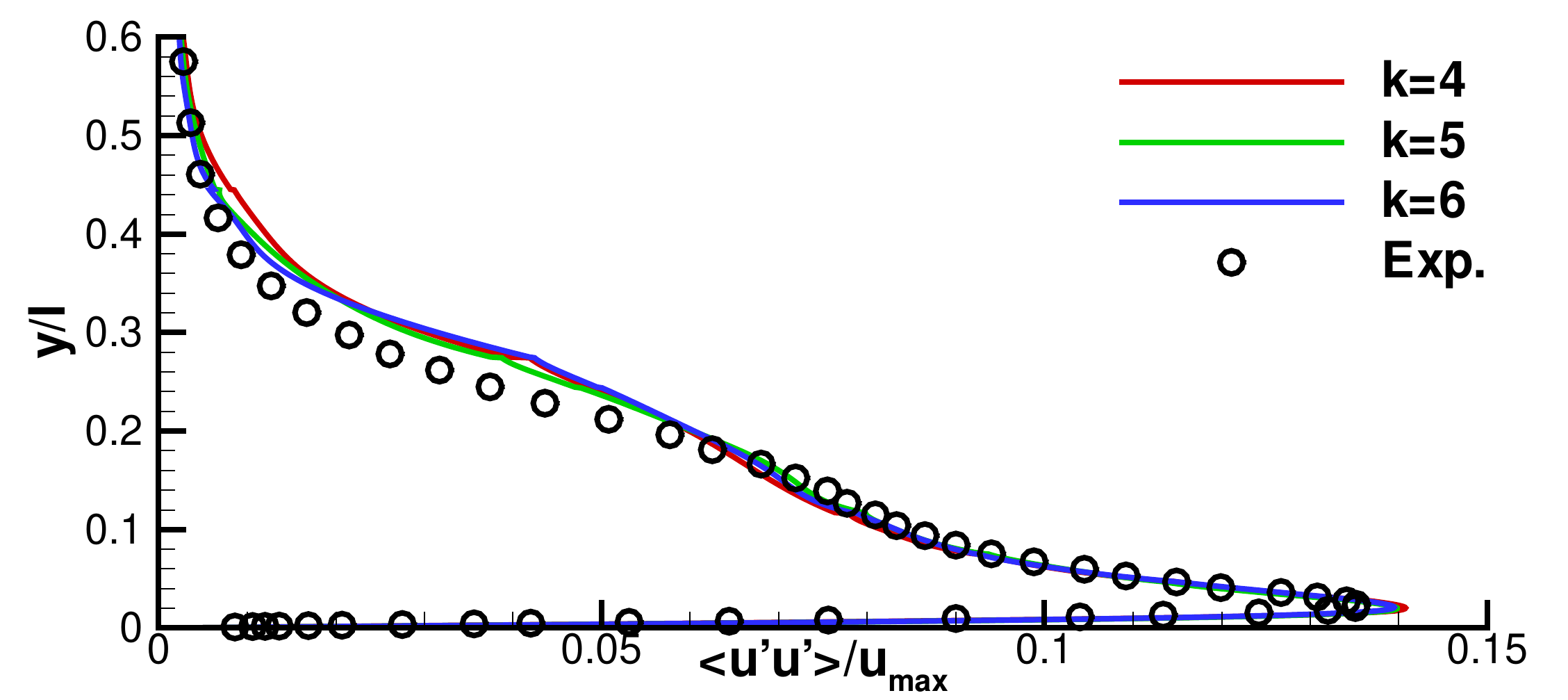} 
\caption{T3L1 test case, $k=6$ solution for $Tu=0.2$\%. Time- and spanwise-averaged velocity profiles in comparison with experimental data~\cite{cutrone2008predicting} at $x/l=3.45$. Mean (top) and RMS (bottom) velocity. \label{fig:t3l_last}}
\end{figure}

\begin{figure}[htbp!]
\centering
\includegraphics[angle=0,width=0.75\textwidth]{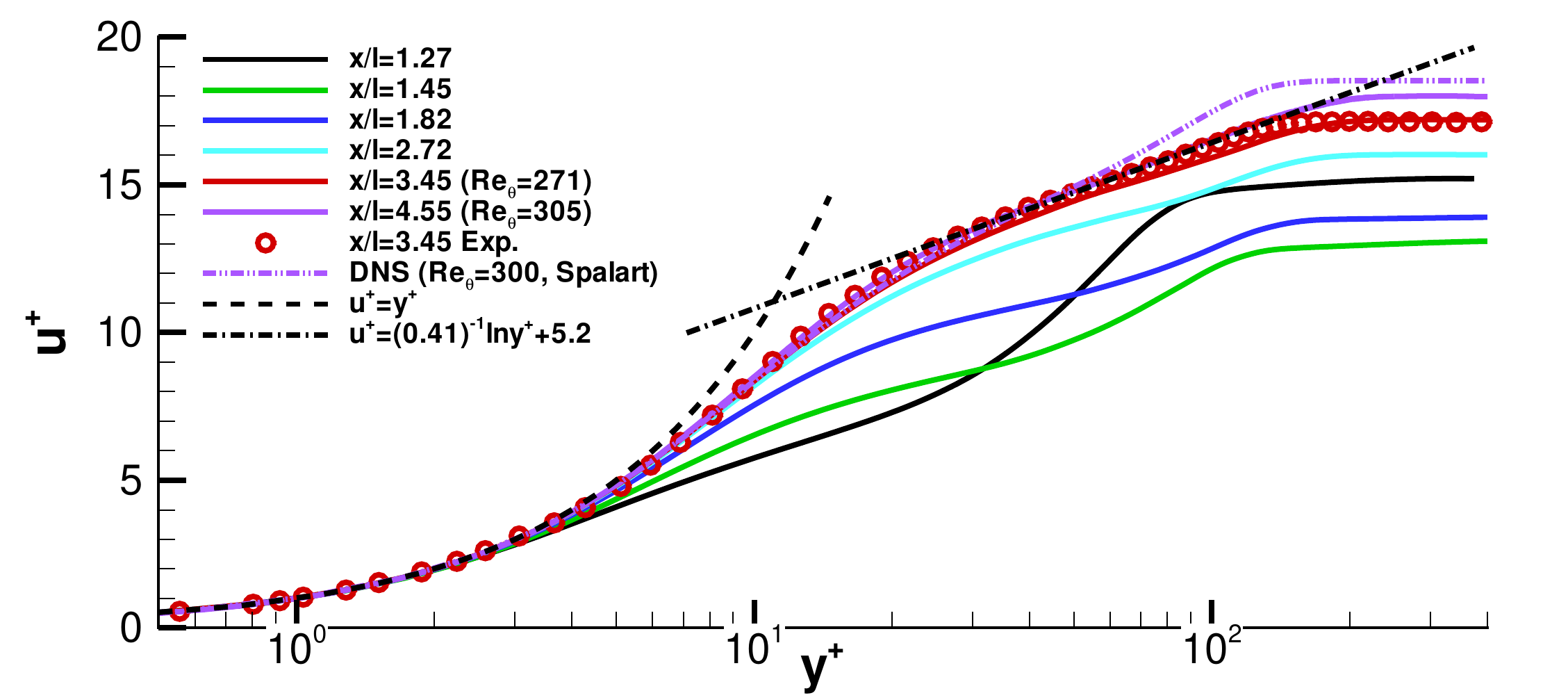}
\caption{T3L1 test case, $k=6$ solution for $Tu=0.2$\%. Non dimensional stream-wise velocity profile for different values of $x/l$ after reattachment in comparison to the theoretical law of the wall and DNS data~\cite{spalart_1988}. \label{fig:t3l-parete}}
\end{figure}

As reported in previous studies, the bubble length is found to be very sensitive to the inlet turbulence intensity, see for example~\cite{lamballais2010direct}. In particular, when increasing the $\text{Tu}$ from $0$\% to $0.2$\% the bubble length reduces from $x_R/d=3.90$ to $2.69$, as shown by the statistically-converged time and spanwise averaged velocity contours in Figs.~\ref{fig:t3l_um} and \ref{fig:t3l-k}. The length predicted for the $\text{Tu}=0.2$ case is in a better agreement with the experimental data ($l/d=2.75$) than other numerical computations~\cite{yang_voke_2001,Langari.Yang.2013} ($2.59$ and $3.00$, respectively). Moreover, we verified by lowering the polynomial degree of the dG discretization that our statistical average $x_R/d$ is almost converged with respect to the spatial resolution: for $k=5$ and $k=4$ we obtain $2.70$ and $2.73$, respectively. Convergence of the statistics is also confirmed by polynomial degree independence observed for the skin friction coefficients, see Fig.~\ref{fig:t3l_cf}. As opposite to the behavior documented in ~\cite{yang_voke_2001}, no hysteresis effects are observed in our computations. Accordingly, if the random source term generating small turbulent perturbations is suddenly suppressed in the fully developed flow field at $\text{Tu}=0.2$\%, the solution of a zero free-stream turbulence case is quickly recovered.
%
%
Figure~\ref{fig:P6ave} compares velocity profiles with the experimental ones. We consider the mean stream-wise velocity ${<}{u}{>}$, and the velocity fluctuation (or velocity RMS), ${<}{u'}{u'}{>}$, as a function of the normal direction for different stations. Velocity is normalized by the local maximum velocity $u_{max}$, computed independently for each of the stations. The random forcing efficacy is demonstrated by the very good agreement with experimental data close to the plate stagnation point. We point out that for $x_1/l<1.64$ improvements with respect to previous computational investigations are difficult to evaluate. As opposite, for $x_1/l>1.64$ our results still compare favourably with the experimental data, while the matching is less evident in~\cite{Langari.Yang.2013}. We stress that our velocity fluctuations compare favourably with the experiments up to $3.45l$, which was omitted in previous works, see Figures ~\ref{fig:P6ave} and ~\ref{fig:t3l_last}.
This supports the claim that present computations provide a larger fully resolved region, all the polynomial degrees here considered. Note that some jumps at inter-element boundaries are still noticeable, especially for $k=4$.
Fig.~\ref{fig:t3l-parete} reports the computed averaged velocty profiles in wall units, for different stations located downstream to the reattachment region. For $x/l \ge 3.45$ the profile approaches the turbulent law of the wall, showing some discrepancies with respect to the equilibrium boundary layer in the outer layer. For the sake of comparison, the zero pressure gradient flat plate DNS result at $\Reynolds_\theta=300$ of Spalart~\cite{spalart_1988} is reported together with the numerical solution at $x/l=4.55$, which shows almost the same $\Reynolds_\theta$. We remark that the station at $x/l=3.45$ is compared to the experimental data, $\Reynolds_\theta\approx270$.     

\paragraph{Performance assessment}

\begin{table}[b!]
\footnotesize 
\begin{tabular}{lllllll}
\hline\noalign{\smallskip}
  Solver & GMRES(MB) & GMRES(MB) & GMRES(MF) & FGMRES(MB) & FGMRES(MF) & FGMRES(MF)  \\
  Prec & BJ & ASM(1,ILU(0)) & BJ & MG$_\text{full}$ & MG$_\text{full}$ &  MG$_\text{full}$ (LAG=3)  \\
\noalign{\smallskip}\hline\noalign{\smallskip}
CPU Ratio & 1 & 1.11 & 0.95 & 0.50 & 0.47 & 0.31 \\
Memory Ratio & 1& 1.6 &  0.6 & 0.65 & 0.15 & 0.15 \\
ITs & 115 & 72 & 115 & 3.0 & 3.0 & 3.31 \\
\noalign{\smallskip}\hline\noalign{\smallskip}
\end{tabular}
\caption{Performance comparison of the solver on the T3L1 test case. Computational time, total memory footprint non-dimensionalized with the GMRES(MB)[BJ] solver, and average number of GMRES iterations per time step, for the BJ, ASM(1,ILU(0)) and \emph{p}-multigrid preconditioners (see text for settings details). Results obtained on 540 Intel Xeon CPUs of Marconi-A1{@}CINECA.}
\label{crivellini_tab:perf3d}
\end{table}
Table~\ref{crivellini_tab:perf3d} reports the computational performances of the different solution strategies obtained for a dG approximation with $k=6$, the same polynomial degree of reference employed in previous sections. The Table aims at comparing the performance of the solution strategies accounting for CPU time, memory footprint as well as average number of GMRES iterations. The time step size of the third-order accurate linearly-implicit Rosenbrock-type time integration scheme was $16$ and $8$ times larger than those used in~\cite{Langari.Yang.2013} and \cite{yang_voke_2001}, respectively. The relative defect tolerance for the linear solver reads $\text{rTol} =10^{-5}$. Once again we consider GMRES(MB)[BJ] as the reference solution strategy. 

Note that, being the number of curved elements small when compared to the number of affine elements, the degrees of exactness of quadrature rules neglects the second degree geometrical representation of cells close to leading edge. We point out that, since the boundary layer is still laminar at the leading edge, this under-integration does not affect the stability of the scheme as well as the accuracy of the numerical results.

Comparing the matrix-based (GMRES(MB)[BJ]) to the matrix-free solver (GMRES(MF)[BJ]), we observe the same computational efficiency but 40\% less memory usage. On the other hand, the use of Additive Schwarz preconditioned GMRES (GMRES[ASM(1,ILU(0))]) decreases the overall parallel efficiency of the method: the CPU time increases by the 11\% and the memory requirements raises by 60\%, due to the increased number of overlapping block elements employed. The \emph{p}-multigrid precondioned solver specs are as follows: FGMRES(MB or MF) as outer solver, a full \emph{p}-multigrid interation with $L=3$, $\Kl = 6,2,1$, GMRES(MB or MF)[EWBJ] smoother for $\ell=0$ (8 iterations), GMRES(MB)[EWBJ] smoother for $\ell=1$ (8 iterations) and GMRES(MB)[ASM(1,ILU(0))] smoother on the coarsest level ($\ell=2$, 40 iterations). Note that for the finest level outer solver and smoother we consider both matrix-based and matrix-free implementations for the sake of comparison. The computational efficiency of the method improves considerably with respect to the reference (first column of Tab.~\ref{crivellini_tab:perf3d}): 
\begin{inparaenum}[i)]
\item in a matrix-based framework a 50\% decrease of the CPU time is observed and the memory requirements reduce by 35\%, 
\item in a matrix-free framework the CPU times gains are unaltered and the memory savings reach 85\%.
\end{inparaenum}
Such significant memory footprint reductions are mainly due the finest level strategy: only a block diagonal matrix is allocated and a small number of Krylov subspaces is employed for the GMRES algorithm. We remark that in this case the actual memory footprint has been computed through the PETSc library, and it is in line with the values that can be estimated a-priori using the model of Section~\ref{sec:MemSav}.

As a further optimization of the matrix-free approach we consider the possibility to lag the computation of the system matrix employed for preconditioning purposes, this means that the preconditioner is ``freezed'' for several time steps. Clearly this strategy reduces assembly times but degrades convergence rates if the discrepancy between the matrix and the preconditioner gets too severe. In the present computation optimal performance are achieved by lagging the operators evaluation for $3$ time steps. By doing so, the CPU time is further reduced to the $0.31$ of the baseline, see Tab.~\ref{crivellini_tab:perf3d}, which corresponds to a speed-up of $3.22$. As a side effect, the average number of GMRES iterations slightly increases, from $3.0$ to $3.31$, due to a loss of efficiency of the multigrid preconditioner. We remark that, in a matrix-free framework, lagging the preconditioners only acts on the coarse space multigrid operators, while the finest space still gets updated thanks to the matrix-free approach.


\subsection{Boeing Rudimentary Landing Gear test case}

The final validation case reported in this work deals with the implicit LES of the incompressible flow around the Boeing Rudimentary Landing Gear (RLG). The purpose of the test is to demonstrate the applicability of the solution strategies proposed in this work within an industrially relevant test case.

The RLG was designed by Spalart \emph{et al.}~\cite{spalart2010initial}, and experimentally studied in~\cite{venkatakrishnan2012experimental}, to become a benchmark for testing turbulence modelling approaches. The test case was also included within the test case suite of the ATAAC EU-funded project~\cite{pengST12}. The flow conditions involve a Reynolds number of $10^6$, based on the freestream velocity $V_{\infty}=40~m/s$ and on the wheel diameter $D=0.406~m$. The Mach number of the experiments was $M=0.12$, which resemble an incompressible flow problem. While the structural elements of the landing gear are rectangular to fix the location of the separation points, the boundary layers on the wheels are tripped, see~\cite{spalart2010initial}. The structure of the unsteady flow around the landing gear is mainly characterized by large separated and recirculating regions on the wheels and axles, as well as by the front-rear wheel interaction, which make the use of unsteady scale-resolving simulation mandatory.

\begin{figure}[b!]
\centering
\includegraphics[angle=0,width=0.985\textwidth]{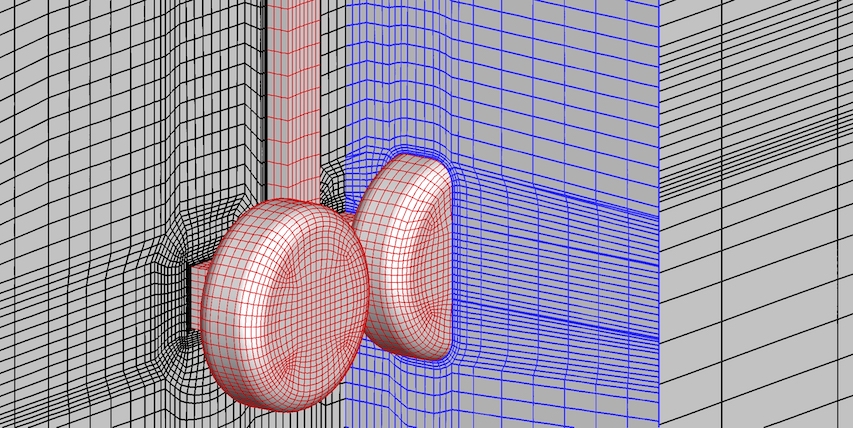}
\caption{Boeing Rudimentary Landing Gear test case. Details of the multiblock structured grid provided by DLR under the ATAAC and TILDA European projects. \label{fig:GridOverall}}
\end{figure}

The computational domain is delimited by four symmetry planes, one inflow, one outflow, and the landing gear wall surface. A snapshot of the grid showing the wall surface (red), the symmetry plane discretization (black) and an internal slice (blue) is reported in Figure~\ref{fig:GridOverall}.
The mesh employed takes advantage of the symmetry of the problem and it discretises only the half of the domain. The grid was made by $115\cdot{10}^3$ hexa elements with second-order geometrical representation of the curved boundaries. It shows a severe wall refinement to accommodate a suitable wall resolution given the high Reynolds number of the case. The first cell height is $\delta y=6.054\cdot 10^{-5} D$, which provides an equivalent wall normal resolution of $1.851\cdot 10^{-5}$, obtained by dividing for $(n_v)^{1/3}$. Exploiting the maximum value of the skin friction coefficient over the entire wall boundary obtained during the post-processing phase, the grid allows a maximum wall normal resolution of $y^+=2.8$. It is worth noting that the estimated miminum aspect ratio of the cell is of the order of $10^4$, which increases considerably the condition number of the iteration matrix.

The solution has been obtained by using $\Pol_4$ polynomials, providing a total of $4.025\cdot{10}^6$ degrees of freedom, while the four-stage, order-three ROSI2PW scheme was employed for time integration. Using as reference quantities the free stream velocity and the wheel diameter, the non-dimensional time step size was $\Delta t=0.001$. To compute the average fields, the solution was advanced in time for roughly 50 convective times, some of them performed using a lower-order space discretization. The average process lasted roughly $T=23$ convective times, which may be not enough to have converged first-order statistics. However, it has been verified that the average quantities did not change sensibly from $T>17.5$. 

\paragraph{Physical discussion}
\begin{figure}[t!]
\centering
\includegraphics[angle=0,width=0.475\textwidth]{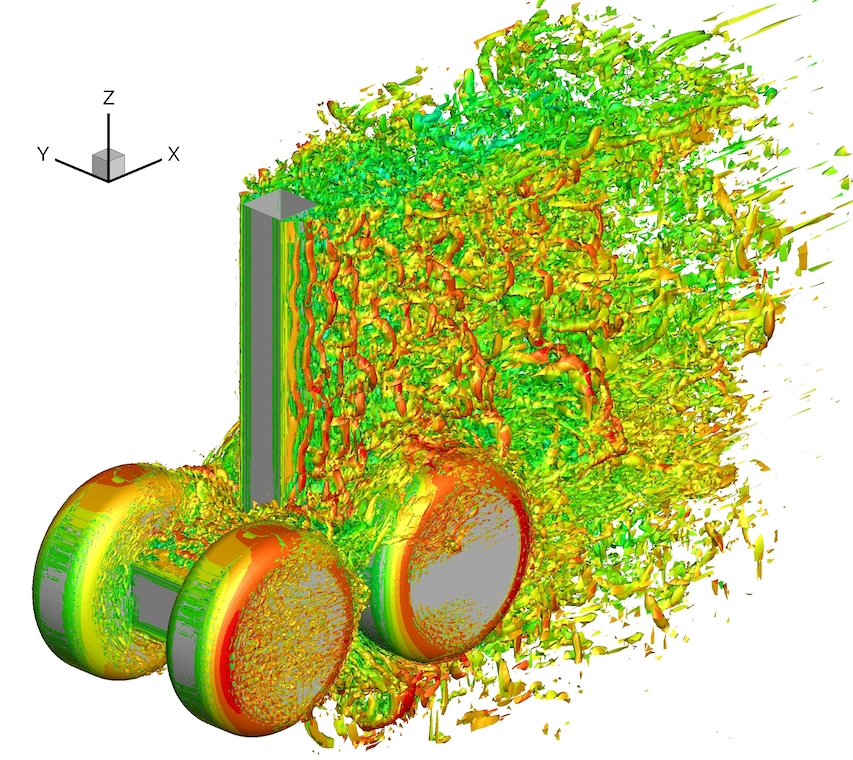} 
\quad
\includegraphics[angle=0,width=0.475\textwidth]{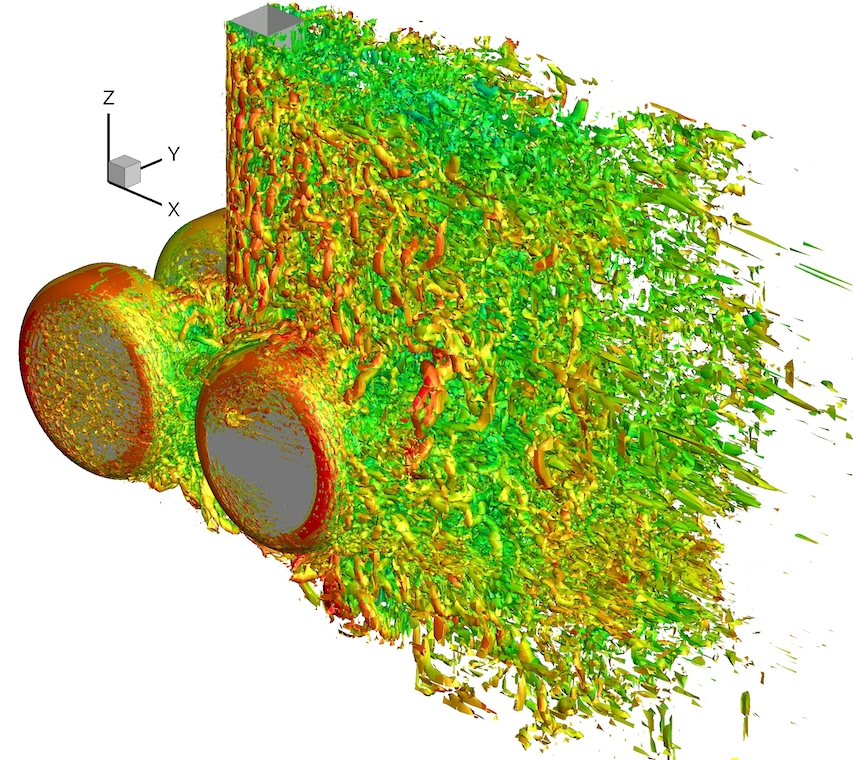} 
\caption{RLG test case at $Re=10^6$. Incompressible flow solution using $k=4$ polynomials. $\lambda_2$ iso-contour coloured by stream-wise velocity magnitude. Front view (left) and rear view (right). \label{fig:rlg_l2}}
\end{figure}
Figure~\ref{fig:rlg_l2} shows the features of the flow field through the istantaneous $\lambda_2$ iso-contour plot coloured by the stream-wise velocity magnitude. The shape of the iso-contours suggests that the flow is mainly attached to the wheels, although a very small laminar separation can be observed on the fore side of the wheel.

\begin{figure}[t!]
\centering
\includegraphics[angle=0,width=0.475\textwidth]{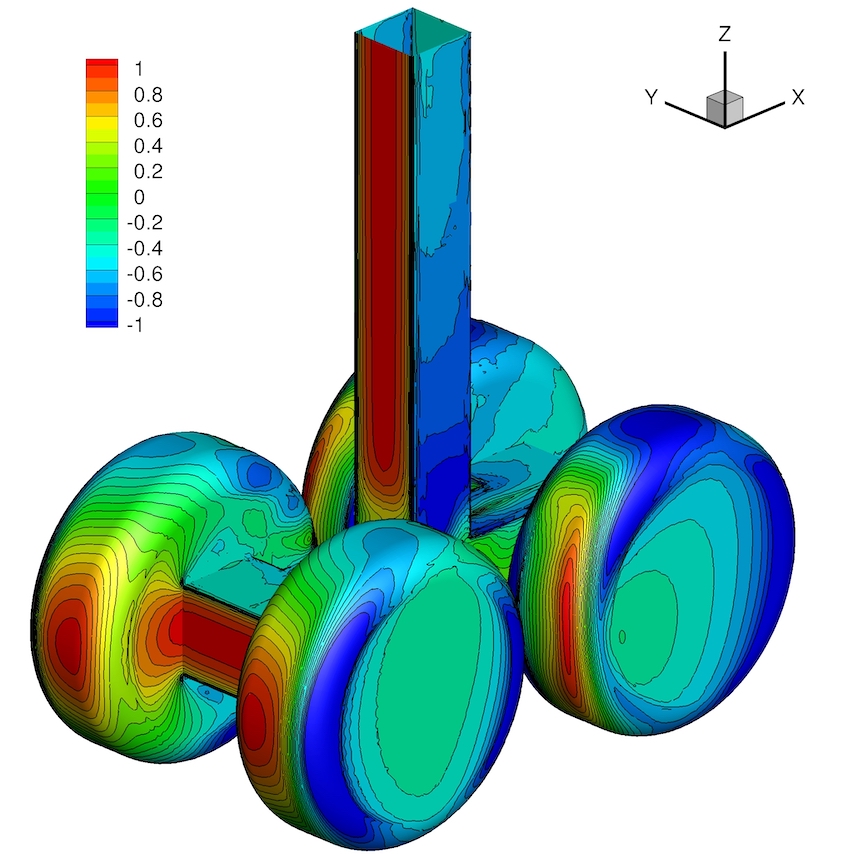} 
\quad
\includegraphics[angle=0,width=0.475\textwidth]{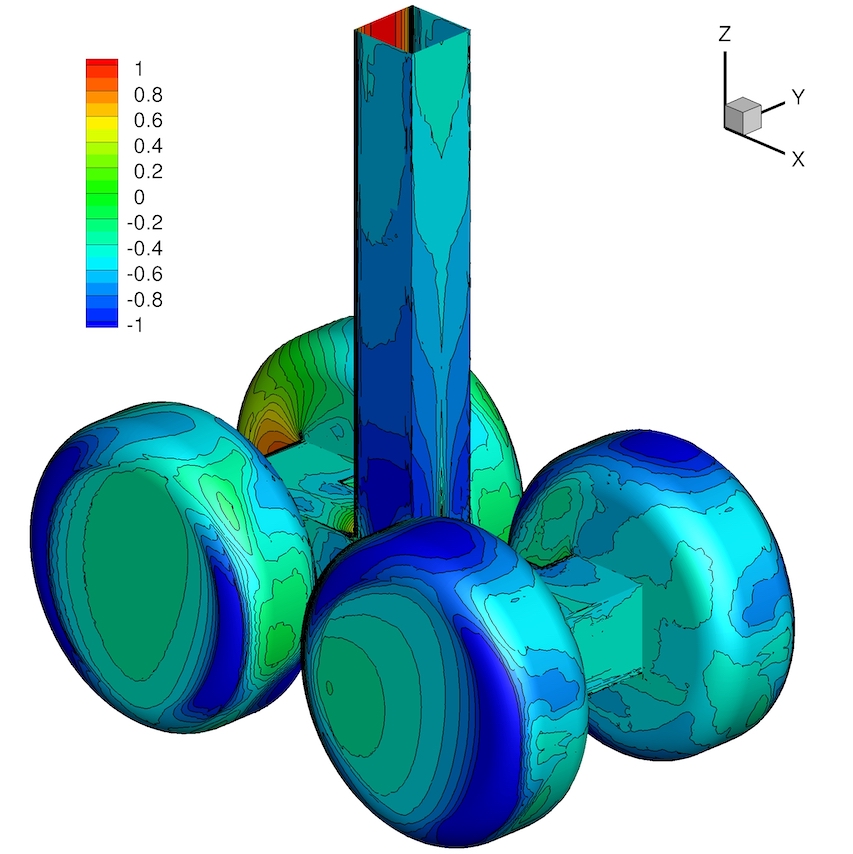} 
\caption{RLG test case at $Re=10^6$. Incompressible flow solution using $k=4$ polynomials. Mean pressure coefficient $C_p$ contours on the wall surface. Front view (left) and rear view (right). \label{fig:rlg_cp}}
\end{figure}
\begin{figure}[t!]
\centering
\includegraphics[angle=0,width=0.475\textwidth]{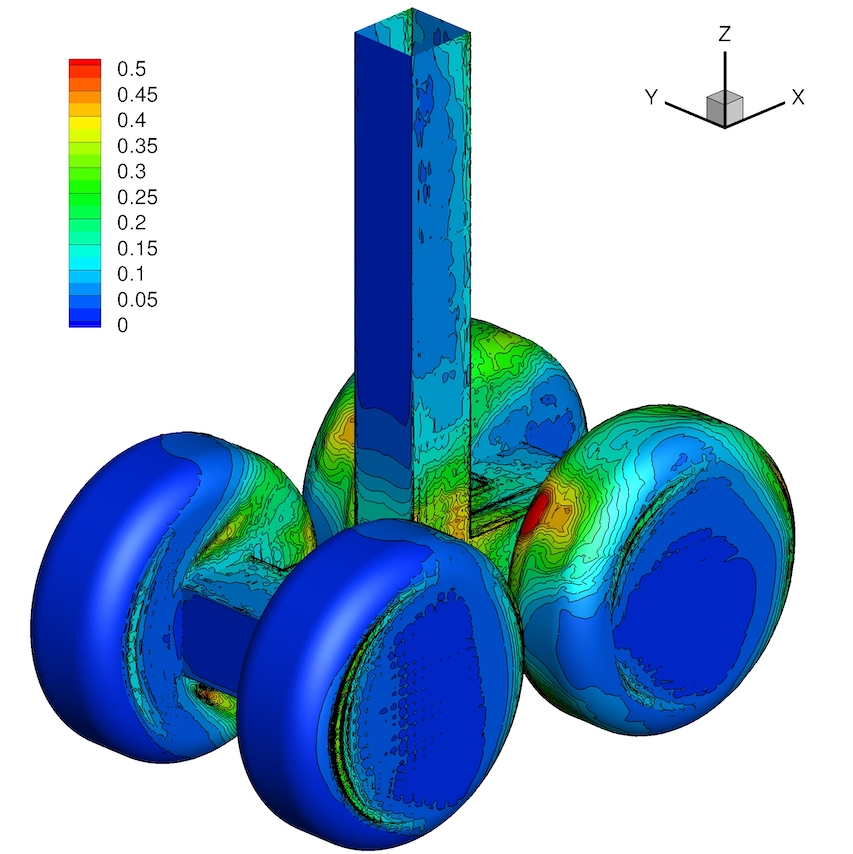} 
\quad
\includegraphics[angle=0,width=0.475\textwidth]{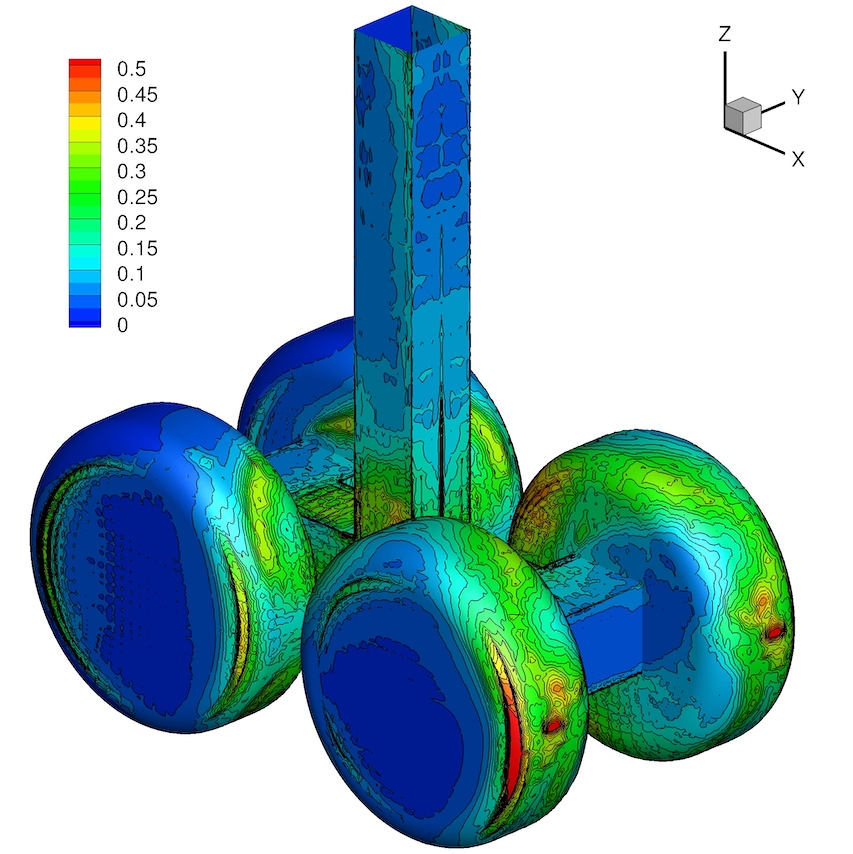} 
\caption{RLG test case at $Re=10^6$. Incompressible flow solution using $k=4$ polynomials. Pressure coefficient RMS, $C_p^{RMS}$ contours on the wall surface. Front view (left) and rear view (right). \label{fig:rlg_cpp}}
\end{figure}
\begin{figure}[t!]
\centering
\includegraphics[angle=0,width=0.475\textwidth]{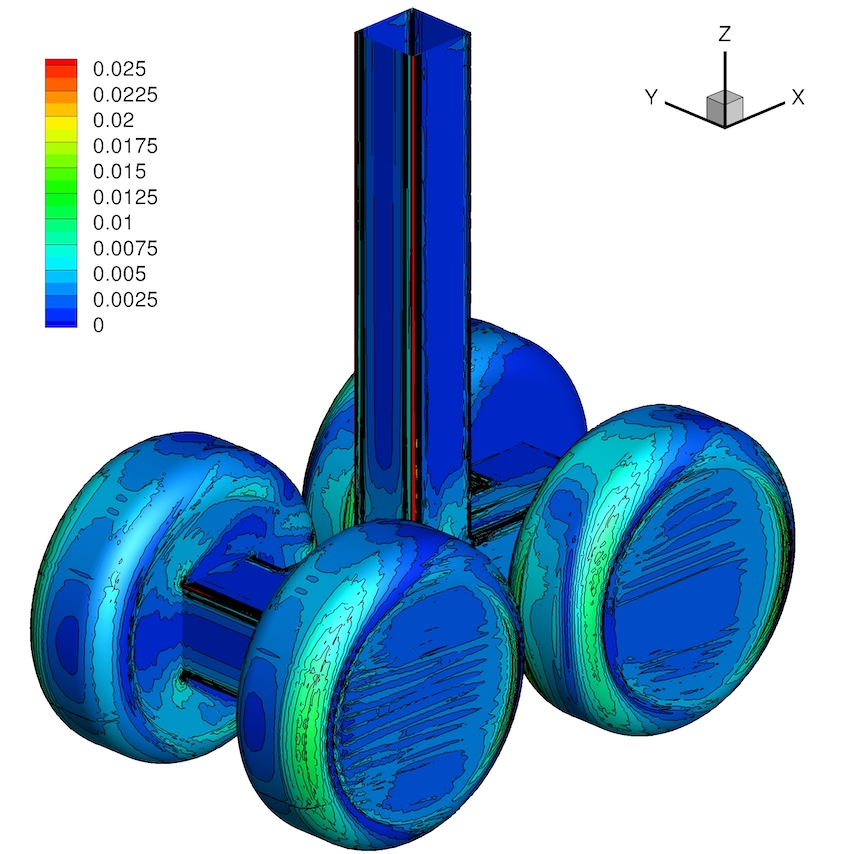} 
\quad
\includegraphics[angle=0,width=0.475\textwidth]{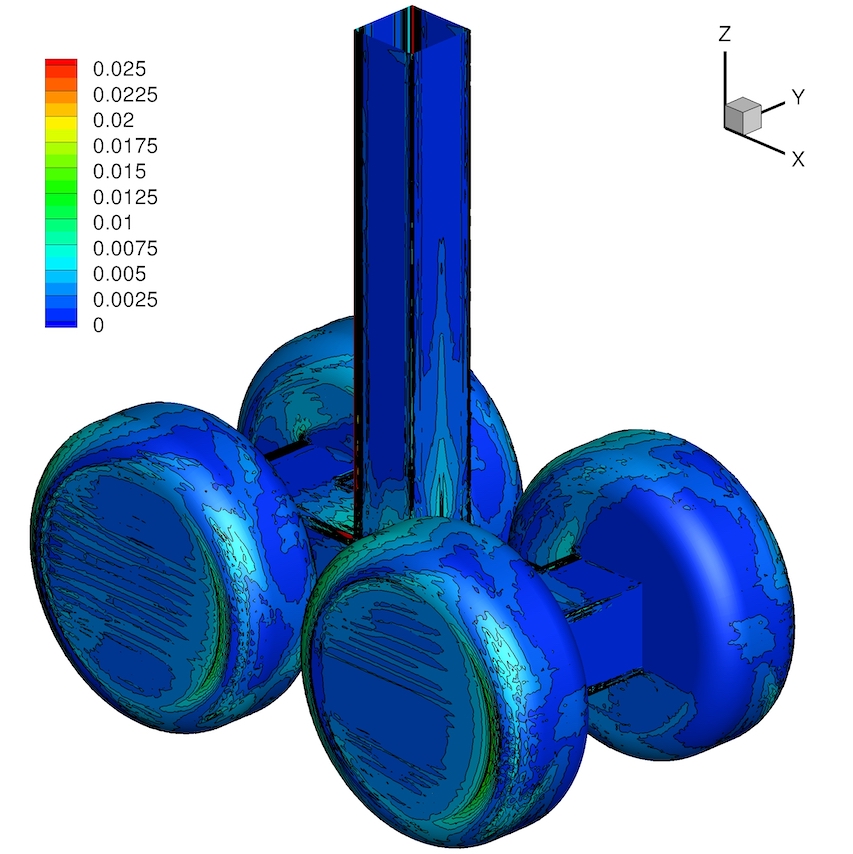} 
\caption{RLG test case at $Re=10^6$. Incompressible flow solution using $k=4$ polynomials. Mean skin friction coefficient $C_f$ contours on the wall surface. Front view (left) and rear view (right). \label{fig:rlg_cf}}
\end{figure}

The averaged fields in terms of pressure coefficient $C_p$, the root mean square value of the pressure coefficient $C_p^{RMS}$ on the landing gear have are reported in Figure~\ref{fig:rlg_cp}, \ref{fig:rlg_cpp} and~\ref{fig:rlg_cf}. A qualitative agreement with the surface plots reported in~\cite{spalart2010initial,dong2018numerical}, obtained through a hybrid RANS/LES approach, can be observed especially as regards the front views. On the other hand, the rear view highlights the presence of pressure oscillations that suggest a very coarse space resolution. Those oscillations are even more evident if first order statistics (\eg\, $C_p^{RMS}$) or the skin friction coefficient (which involve state derivatives) are considered.
\begin{figure}[t!]
\centering
\includegraphics[angle=0,width=0.475\textwidth]{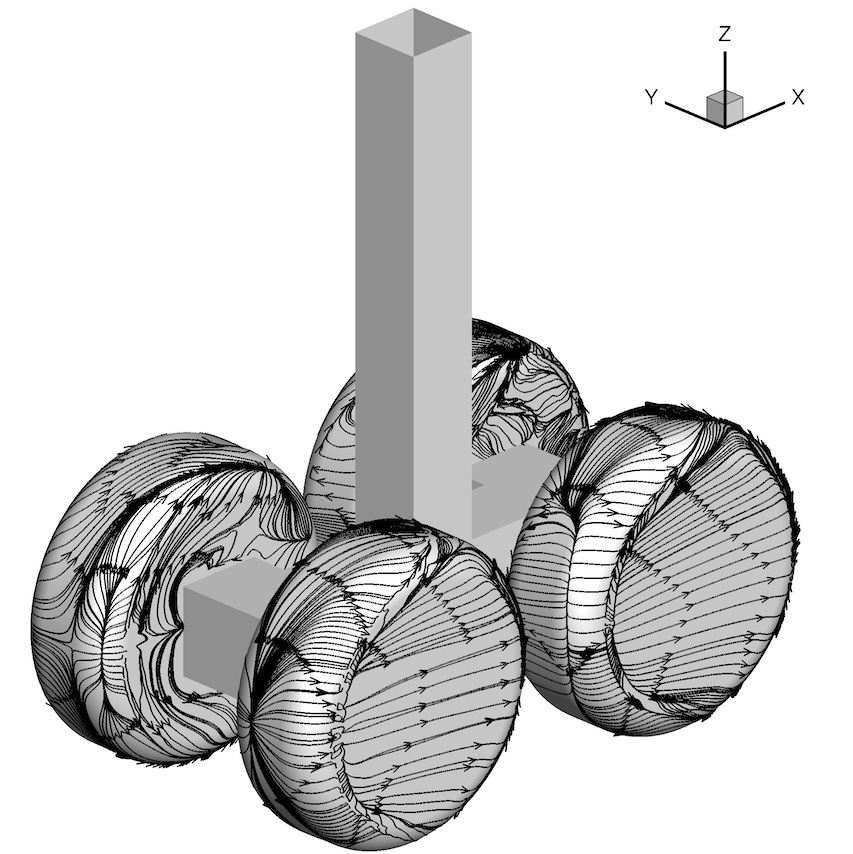} 
\quad
\includegraphics[angle=0,width=0.475\textwidth]{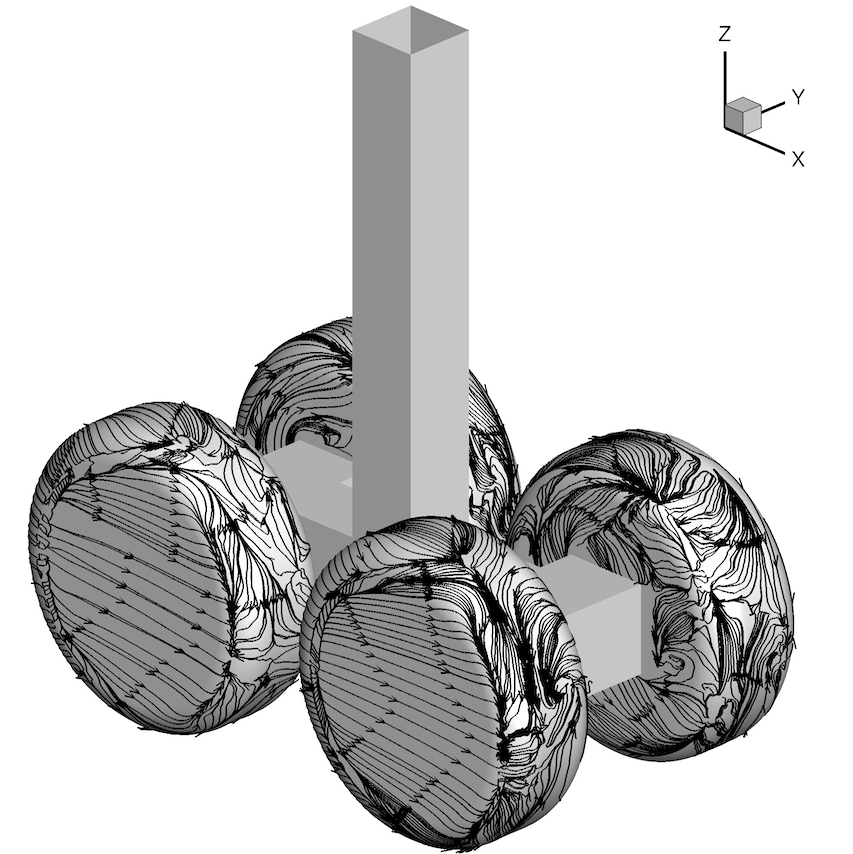} 
\caption{RLG test case at $Re=10^6$. Incompressible flow solution using $k=4$ polynomials. Average wall streamline path. Front view (left) and rear view (right). \label{fig:rlg_cf_line}}
\end{figure}
Figure~\ref{fig:rlg_cf} reports the surface plot of the skin friction coefficient $C_f=2\tau_w/\rho V_{\infty}^2$ and looks qualitatively similar to those reported in previous numerical simulations~\cite{spalart2010initial,dong2018numerical}.

Figure~\ref{fig:rlg_cf_line} show the streamline patterns on the wheels. The patterns show the bifurcation line of separation and reattachment on the front and rear wheels as reported in~\cite{dong2018numerical}. However, the simulation shows on both sides of the wheels a region with separated flow and reversed streamline patterns. Such a region seems to be different to that reported by the experiments~\cite{venkatakrishnan2012experimental} as well as previous numerical simulations. It is worth to point out that no transition tripping was employed in the current simulation, differently to what has been done for the experiments and previous numerical simulations based on RANS and hybrid RANS/LES modeling. 

\begin{figure}[t!]
\centering
\subfigure[$C_p$, front wheel]{\includegraphics[angle=0,width=0.485\textwidth]{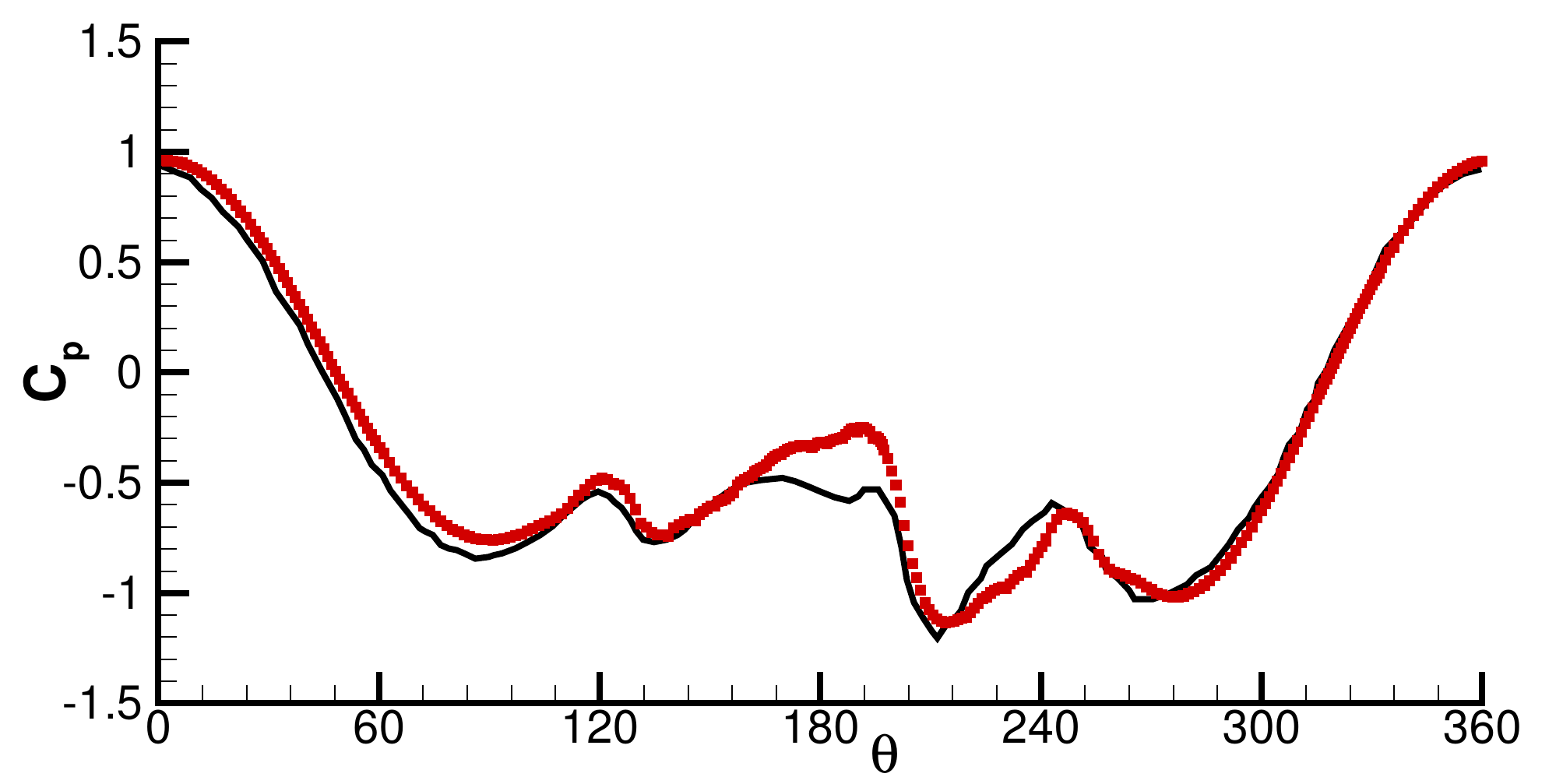} \label{fig:rlg-cp-midfront}}
\subfigure[$C_p$, rear wheel]{\includegraphics[angle=0,width=0.485\textwidth]{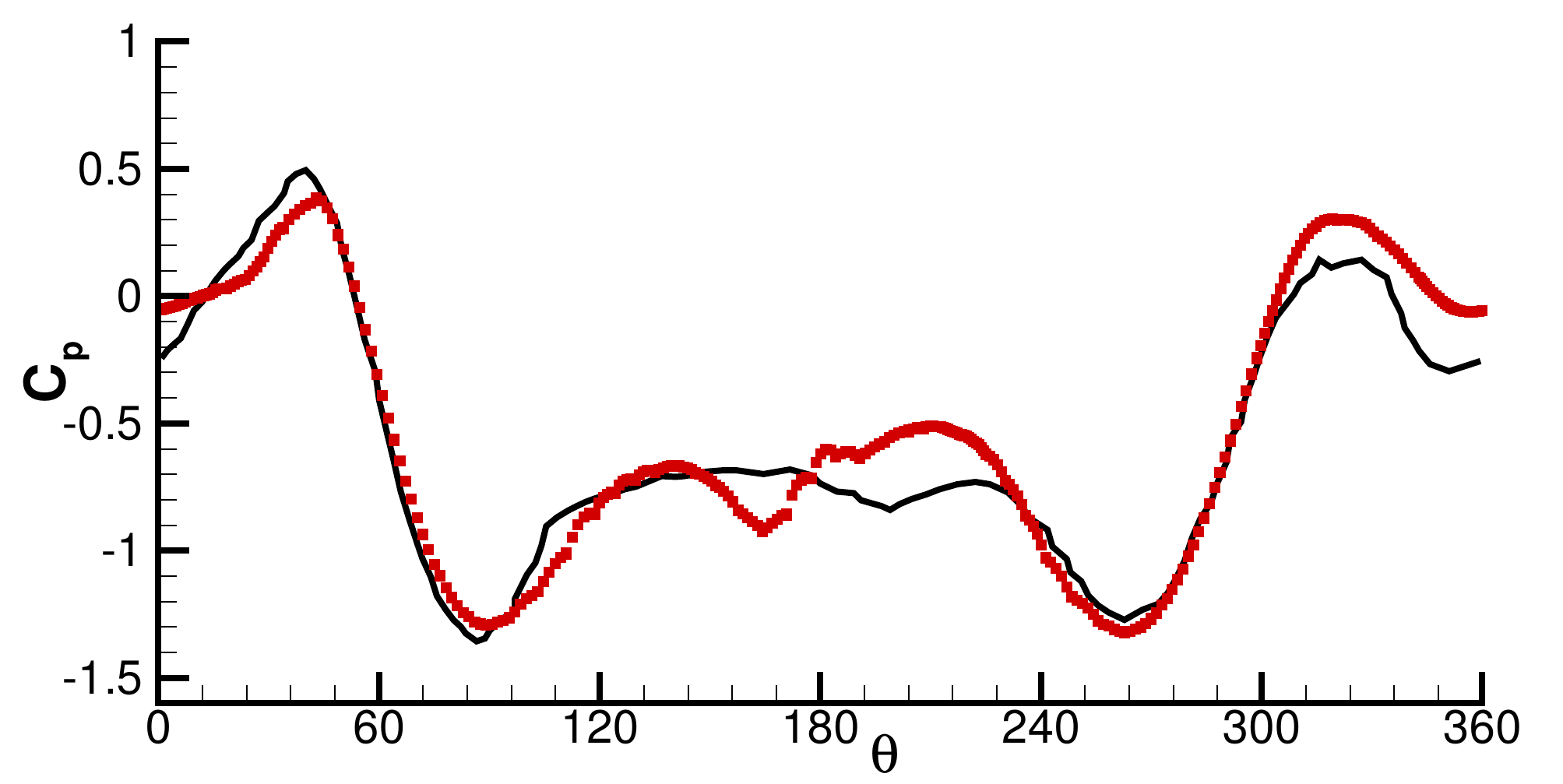} \label{fig:rlg-cp-midrear}}
\subfigure[$C_p^{RMS}$, front wheel]{\includegraphics[angle=0,width=0.485\textwidth]{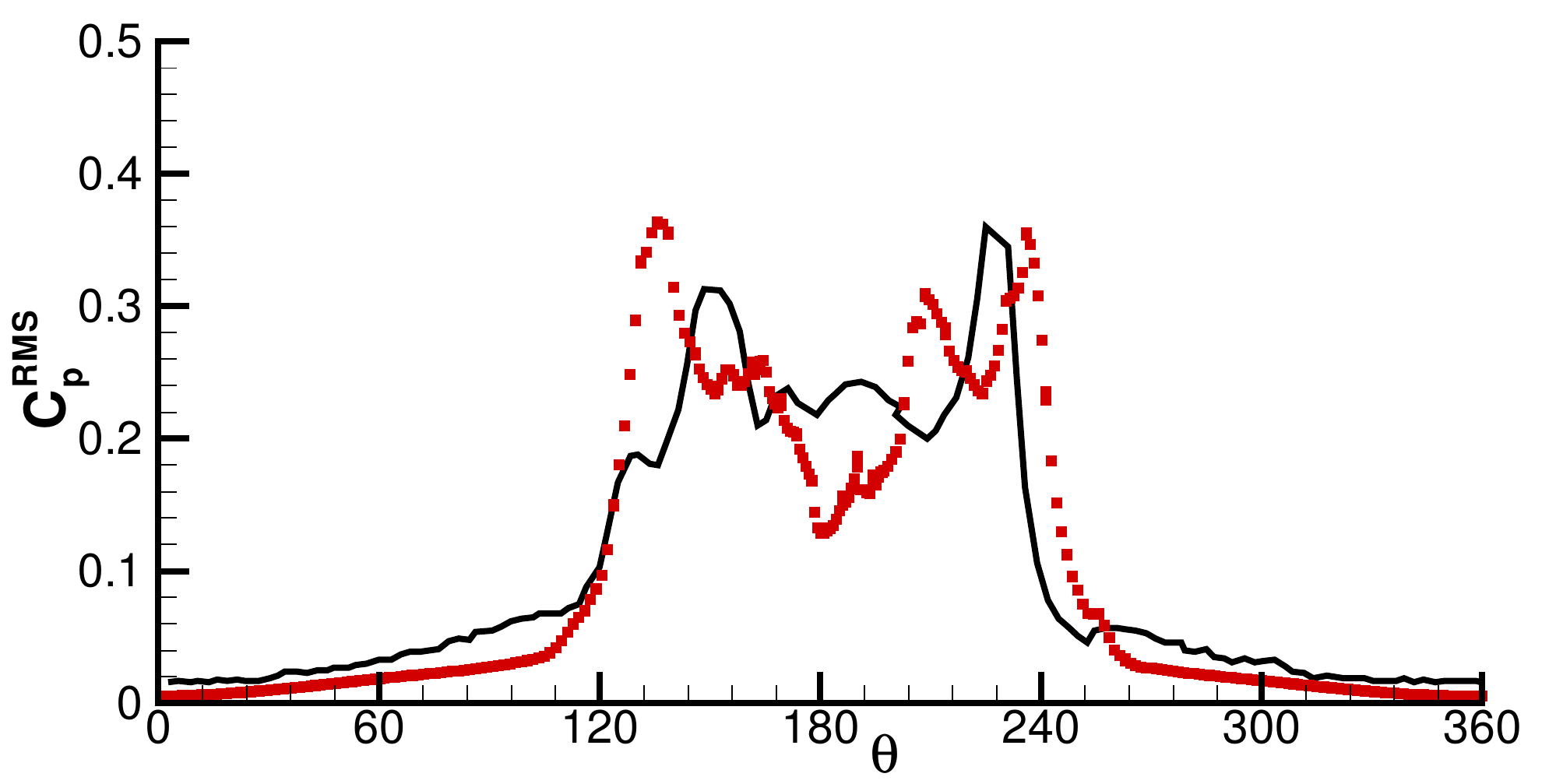} \label{fig:rlg-cpp-midfront}}
\subfigure[$C_p^{RMS}$, rear wheel]{\includegraphics[angle=0,width=0.485\textwidth]{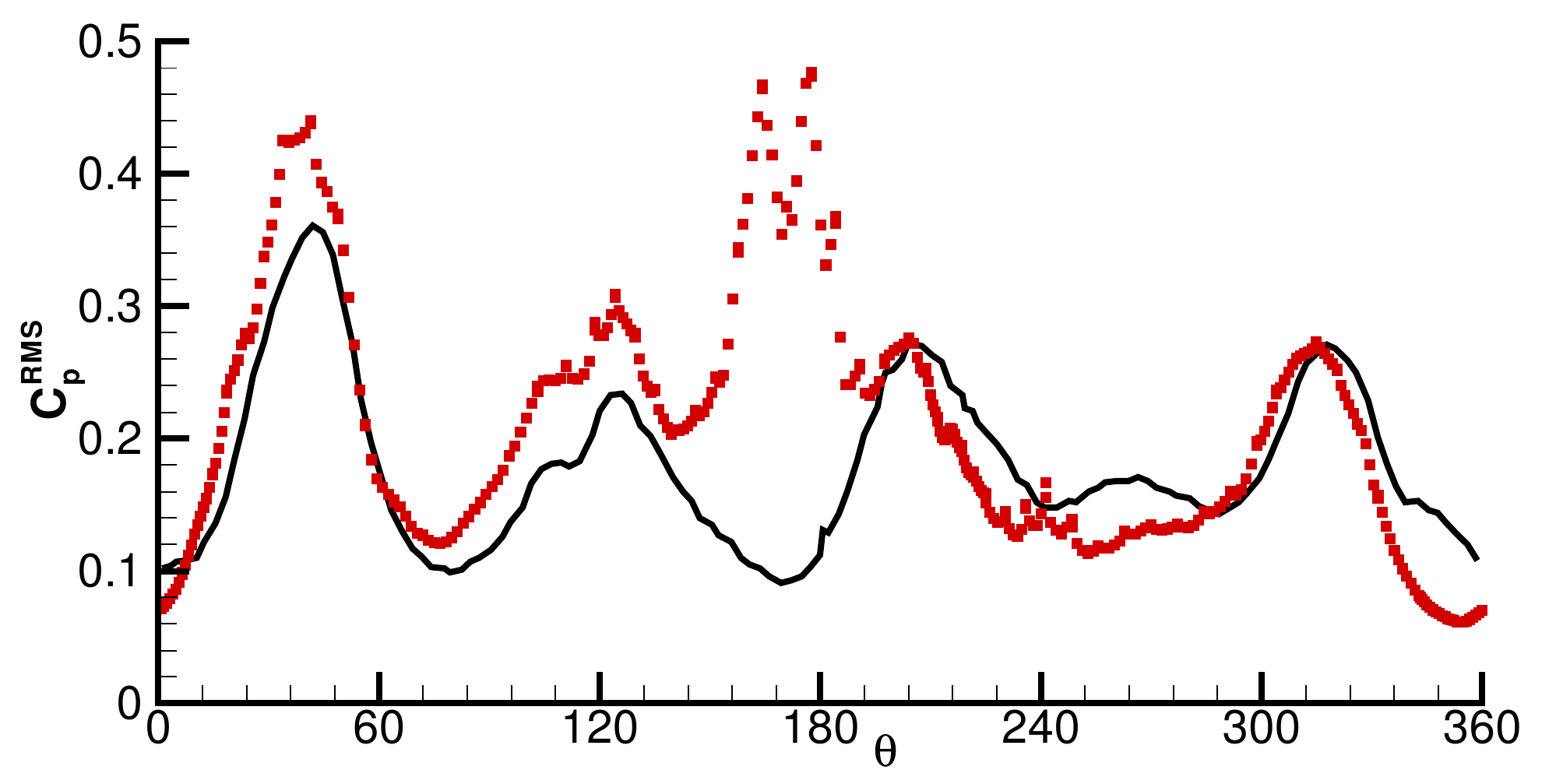} \label{fig:rlg-cpp-midrear}}
   \caption{RLG test case at $Re=10^6$. Incompressible flow solution using $k=4$ polynomials. Average pressure coefficient $C_p$ and RMS $C_p^{RMS}$ distribution on the mid-line of the fore and back wheels versus the azimuthal angle $\theta$. Numerical simulations (red dots) compared to experimental data (black solid lines). \label{fig:cp-mid}}
\end{figure}
Figure~\ref{fig:cp-mid} reports the average pressure coefficient $C_p$ and its root mean square value $C_p^{RMS}$ on the mid-line of the wheels, versus the azimuthal angle $\theta$, in comparison with experimental data from NAL~\cite{venkatakrishnan2012experimental}. While the pressure coefficient compare favourably with experimental data, its root mean square value shows a more oscillatory behaviour, originating from low spatial resolution and possibly a too short averaging time. However, the locations of the peaks of fluctuation as well as its value is pretty well captured.

\paragraph{Performance assessment}
Linear systems arising from the time integration were solved using FGMRES preconditioned with a $p$-multigrid strategy with similar settings to that of Table~\ref{tab:3dmg}, employing an additive Schwarz preconditioned smoothers on the coarsest level of the multigrid iteration, and an element-wise block Jacobi method on the other levels to maximise the scalability of the algorithm. Despite working with an average of 19 elements per partition in such a complex test case, the iterative solver converged using an average of 3.5 iterations per stage, confirming the efficacy of this preconditioning approach. It is worth to remark that the present computation has been performed on roughly $6\cdot{10^3}$ cores of the Marconi A1 cluster hosted by CINECA, using a total wall clock time of 94 hours. The average wall time per convective time unit was roughly 4 hours.

\clearpage
\section{Conclusion}

The paper presents a \emph{p}-multigrid preconditioner strategy applied to the solution of linear system arising from linearly-implicit Rosenbrock-type time discretizations. The algorithm relies on a matrix-free implementation of both the outer FGMRES solver as well as the finest level smoother, while matrix-based GMRES smoothers are employed on coarse levels. Coarse operators of lower polynomial degree are built using a subspace inherited approach.

The performance of the algorithm has been evaluated on test cases of growing complexity. First, we deal with unsteady laminar flows, \textit{i.e.}, the flow around a two-dimensional cylinder and a sphere, showing that the \emph{p}-multigrid preconditioned solver can be used to achieve optimal convergence rates, outperforming standard single-grid preconditioned iterative solvers from a CPU time viewpoint in practical parallel computations. In particular, we consider  and parallel computations (up to 576 cores and 6 elements per partition), with speed-ups ranging from approximatively 1.5 to 3.5. Finally, we perform ILES of the incompressible turbulent flow over a rounded-leading edge plate with different free-stream turbulent intensities. High-order accurate $k=6$ solutions are compared with published numerical results and wind tunnel experiments. The solver strategy is profiled and compared with state-of-the-art single-grid solvers running on large HPC facilities. We show that, if a block-diagonal preconditioner is employed on the finest level, the algorithm reduces the memory footprint of the solver of about 92\% of the standard matrix-based implementation. Interestingly, besides the  memory savings, the \emph{p}-multigrid preconditioned FGMRES solver is also three times faster than the best performing single-grid solver. As proof of concept, we report the ILES of the Boeing Rudimentary Landing Gear at $Re=10^6$. Increasing the complexity of the problems has not required tuning of \emph{p}-multigrid parameters confirming the robustness of the proposed approach.

Future works involve the implementation of an adaptive strategy for the choice of the quadrature degree of exactness, which can be adapted in view of the actual amount of curvature of mesh faces, in order to optimise the computation of the residuals vector and thus the overall performance of the matrix-free algorithm.

\section*{Acknowledgements}\label{sec::acknowledgements}
We acknowledge the CINECA award, under the ISCRA initiative (grant numbers HP10BEE6C4, HP10CPSWZ2, HP10BMA1AP, HP10CKYRYE, HP10CE90VW), for the availability of high performance computing resources and support. Professor Michele Napolitano (Politecnico di Bari) and the co-authors of~\cite{cutrone2008predicting} are also gratefully acknowledged for sharing the experimental data of the T3L1 test case. Dr. Ralf Hartman from DLR, the EU-funded projects ``Advanced Turbulence Simulation for Aerodynamic Application Challenges" (ATAAC) and ``Towards Industrial LES/DNS in Aeronautics -- Paving the Way for Future Accurate CFD" (TILDA) are also acknowledged for sharing the mesh of the Boeing rudimentary landing gear test case.



\appendix
\section{Scaling of the stabilization term} \label{app:stab}
Following the idea proposed by~\cite{botti2017h}, a rescaled Galerkin projection of the stabilization term in order can be introduced to recover the optimal performances of non-inherited \emph{p}-multigrid algorithm.  

As a first point we recall the following bound on the local lifting operator: let $\phi \in L^2(\f)$, for all $\f \in \Ff$ 
\begin{equation}
\label{eq:upliftbound}
\| \mathbf{r}_{\f}^k (\phi) \|_{[L^2 (\Omega)]^d} \leq C_\mathrm{tr} h_{\T,\Tpr}^{-1/2} \| \phi \|_{L^2(\f)}, 
\end{equation}
where $h_{\T,\Tpr} = \min \left( h_\T, h_{\Tpr} \right)$, see \eg \cite[Lemma 2]{Brezzi.Manzini.ea:2000}, \cite[Lemma 7.2]{Toselli03} or \cite[Lemma 4.33 and Lemma 5.18]{DiPiErn11} for a proof. The constant $C_\mathrm{tr}$ depends on $d$, $k$ and the shape regularity of the elements sharing $\f$ and is inherited from the discrete trace inequality: for all $\T \in \Th$, $\f \in \Ff$
\begin{equation}
\label{eq:traceIneq}
\| z_{h} \|_{L^2(\f)} \leq C_{\mathrm{tr}} h_{\T,\Tpr}^{-1/2} \| z_{h} \|_{L^2(\T)}
\end{equation}
As remarked by Di Pietro and Ern \cite[Lemma 1.46]{DiPiErn11} the dependence of $C_{tr}$ on $k$ is a delicate issue that has a precise answer only in specific cases. In this work we follow the estimates given by Hesthaven and Warburton \cite{Warburton03} showing 
that for simplicial meshes $C_{tr}$ scales as $\sqrt{k(k+d)}$ when using complete polynomials of maximum degree $k$. 
This choice turns out to be conservative regarding the dependence on $k$ with respect to estimates derived by Schwab~\cite{Schwab1998} based on tensor product polynomials on mesh elements being affine images of the unit hypercube in $\mathbb{R}^d$, which suggest a $\sqrt{k(k+1)}$ scaling.

Using the Cauchy-Schwarz inequality, for all $\wvec_h, \kvec_h \in [\Poly{d}{k}(\Th)]^{d+1}$ we get
\begin{align}
j_h^{\nu-\mathrm{STB}}({\wvec_h},{\kvec_h})|_{\f \in \Ff} 
     & = \eta_\f \intO \sum_{i,j = 1}^{d+1} \sum_{p,q = 1}^d \; {\frac{\partial \widehat{F}_{p,i}}{ \partial \Bigl( \partial {w}_{j}/ \partial x_q - \eta_\f {r}_{q}^\f({w_{j}) \Bigr) }}}\; 
             \mathbf{r}^{\f}_q(\jump{w_j}) \mathbf{r}^\f_p(\jump{v_i}) \nonumber \\ 
     &\leq {\eta_\f \, C \, {k_\ell(k_\ell+d)} \, h^{-1}_{\T,\T^{'}}} \; {\| \jump{\wvec_h} \|_{L^2(\f)} \; \| \jump{\vVech} \|_{L^2(\f)}} \nonumber \\ 
j^{\nu-\mathrm{STB}}_h({\IntOp_\ell^0 \wvec_h},{\IntOp_\ell^0 \kvec_h})|_{\f \in \Ff} 
     & = \eta_\f \intO \sum_{i,j = 1}^{d+1} \sum_{p,q = 1}^d \; {\frac{\partial \widehat{F}_{p,i}}{ \partial \Bigl( \partial {w}_{j}/ \partial x_q - \eta_\f {r}_{q}^\f({w_{j}) \Bigr) }}} \;
           \mathbf{r}^{\f}_q(\jump{w_j}) \cdot \mathbf{r}^\f_p(\jump{v_i}) \nonumber \\ 
     &\leq {\eta_\f \, C \, {k_0(k_0+d)} \, h^{-1}_{\T,\T^{'}}} \; {\| \jump{\wvec} \|_{L^2(\f)} \; \| \jump{\vvec} \|_{L^2(\f)}} \nonumber  
\end{align}
where $C$ is independent from $h$ and $k$. 
As a result we are able to introduce the scaling factor $\mathcal{S}_0^\ell = {\frac{(k_\ell)(k_\ell+d)}{(k_0)(k_0+d)}}$ 
such that, for all $\wvec_h, \kvec_h \in [\Poly{d}{k}(\Th)]^{d+1}$, it holds
\begin{equation}
j_h^{\nu-\mathrm{STB}}({\wvec_h},{\kvec_h}) \simeq {\mathcal{S}_0^\ell} \; j^{\nu-\mathrm{STB}}_h({\IntOpB_\ell^0 \wvec_h},{\IntOpB_\ell^0 \kvec_h}).
\end{equation}

The viscous Jacobian stabilization operator reads
\begin{equation}
\begin{array}{lll}
(\Jm^{\nu-\mathrm{STB}}_{\ell} (\delta \wvec_h), \kvec_h)_{L^2(\Omega)}  & = {\mathcal{S}_0^\ell} \; j^{\nu-\mathrm{STB}}_h(\IntOpB_\ell^0 (\delta \wvec_{h}), \IntOpB_\ell^0 \kvec_{h})     
                                                                                             & \forall \,  \delta \wvec_h, \kvec_h \in [\Poly{d}{\Kl}(\Th)]^{d+1}
\end{array}
\end{equation}
and, accordingly, the Jacobian stabilization diagonal and off-diagonal block contributions $\SysM{\T,\T}^{\ell,\nu-\mathrm{STB},\IntOp}$ and $\SysM{\T,\T^{'}}^{\ell,\nu-\mathrm{STB},\IntOp}$
can be computed recursively and matrix free by means of a rescaled Galerkin projection.
The rescaled-inherited blocks of the Jacobian matrix are computed as follows
\begin{align}
\SysM{\T,\T}^{\ell+1,\IntOp}      &=  \M_{\ell+1,\ell}^\T \; \left(\SysM{\T, \T}^{\ell,!\nu,\nu\backslash\mathrm{STB},\IntOp}\right) \;     \left(\M_{ \ell+1,\ell}^{\Tpr}\right)^t
                                     + \mathcal{S}_\ell^{\ell+1} \, \M_{\ell+1,\ell}^\T \; \left(\SysM{\T, \T}^{\ell,\nu-\mathrm{STB},\IntOp}\right) \;      \left(\M_{ \ell+1,\ell}^{\Tpr}\right)^t, \\
\SysM{\T,\T^{'}}^{\ell+1,\IntOp}  &=  \M_{\ell+1,\ell}^\T \; \left(\SysM{\T, \T^{'}}^{\ell,!\nu,\nu\backslash\mathrm{STB},\IntOp}\right) \;  \left(\M_{ \ell+1,\ell}^{\Tpr}\right)^t 
                                     +  \mathcal{S}_\ell^{\ell+1} \, \M_{\ell+1,\ell}^\T \; \left(\SysM{\T, \T^{'}}^{\ell,\nu-\mathrm{STB},\IntOp}\right) \;  \left(\M_{ \ell+1,\ell}^{\Tpr}\right)^t, \label{eq:GalProjResc}
\end{align}
where $\mathcal{S}_\ell^{\ell+1} = \frac{(k_{\ell+1})(k_{\ell+1}+d)}{(k_\ell)(k_\ell+d)}$.
We remark that $\SysM{\T,\T}^{\ell,!\nu,\nu\backslash\mathrm{STB}}$ and $\SysM{\T,\T^{'}}^{\ell,!\nu,\nu\backslash\mathrm{STB}}$
are the Jacobian blocks corresponding to inviscid contributions plus the viscous contributions without the stabilization terms.

\section{Assessment of the stabilization scaling on a Poisson Problem}\label{sec:PoissResults}
Since the stabilization scaling influences only the elliptic part of coarse grid operators, it is convenient to assess its effectiveness by the numerical solution of a Poisson problem, representative of diffusion dominated regimes. In particular we consider the performance of a \emph{p}-multigrid preconditioned FMGRES solver applied to a high-order $k=6$ BR2 dG discretization over three $h$-refined mesh sequences of the bi-unit square $\Omega = [-1,1]^2$:
\begin{inparaenum}[i)]
\item a regular Delaunay triangular mesh sequence (reg-tri),
\item a distorted quadrilateral mesh sequence (dist-quad) obtained by randomly perturbing the nodes of a Cartesian grid,
\item a distorted and graded triangular mesh sequence (grad-tri) where the elements shrink close to the domain boundaries 
      mimicking the end-points clustering of one-dimensional Gaussian quadrature rules in each Cartesian direction.
\end{inparaenum}
Dirichlet boundary conditions and the forcing term are imposed according to the smooth exact solution $u = e^{-2.5\left((x-1)^2+(y-1)^2\right)}$. The potential field rapidly varies in the proximity of the upper-right corner of the square in order to replicate the presence of a boundary layer. 

The \emph{p}-multigrid preconditioner options are as follows: we consider a three-levels ($L=2$) and a six-levels ($L=5$) $\mathcal{V}$-cycle iteration with ILU(0) right-preconditioned GMRES smoothers on each level. On all levels but the coarsest (that is for $\ell<L$) we perform a single smoothing iteration. On the coarse level we set the relative residual tolerance to $10^{-3}$ and impose a maximum number of iterations of 40 or 400. Polynomial degree coarsening on six-levels is achieved by recursively reducing the polynomial degree by one, that is $\Kl {=} 6 {-} \ell$. On the three level strategy the coarsening strategy is more aggressive: we drop to $k{=}3$ on the first level and we employ a second-order $k{=}1$ dG discretization on the coarsest level. Interestingly, this latter setup seeks to replicate the four-fold degrees of freedom decrease of \emph{h}-multigrid strategies in two space dimensions.

In Table \ref{tab:Poisson3L} and Table \ref{tab:Poisson6L} we 
consider the three- and the six-levels $\mathcal{V}$-cycle iterations, respectively, and we assess the benefits of stabilization scaling (scaling on) with respect to standard inherited-\emph{p}-multigrid coarse grid operators (scaling off). Execution time gains are remarkable on regular triangular and distorted quadrilateral mesh sequences (solution time speedup of 2.4 and 2.2 on average, respectively) but still present on the graded triangular mesh sequence (50\% faster solution process on average). Performance of iterative solver can be evaluated in terms of convergence factor, that is the average residual decrease per iteration, which can be computed as follows $$ \rho = e^{\left(\frac{1}{N_\mathrm{it}} ln \frac{\| d_{N_\mathrm{it}} \|}{\| d_{0}\|}\right)},$$ where $N_\mathrm{it}$ is the number of iterations required to reach the prescribed residual drop, and $d_{i}$ is the defect (or residual) of the linear system solution at the i-th iteration. It is interesting to remark that stabilization scaling always improves the convergence factor of the coarse grid solver having a positive impact on the performance of the algorithm on diffusion dominated regimes, in particular one of two following situations might occur.
\begin{enumerate}
\item The prescribed residual drop of $10^{-3}$ is attained in a smaller number of coarse solver iterations. This is typically observed when the maximum number of iterations is set to 400.
\item The prescribed maximum number of iteration of the coarse solver is attained leading to a tinier defect for the rescaled stabilization algorithm. Accordingly, convergence of the outer solver is improved and a smaller number of FGMRES iterations is required to solve the linear system. This is typically observed when the maximum number of iterations of the coarse solver is set to 40. 
\end{enumerate}
It is worth noting that, when this technique is employed for convection-dominated regimes, the advantages arising from an improved coarse space viscous side are less dominant on the overall efficiency of the multigrid algorithm, as proved in Section~\ref{sec:INSResults}.

We remark that uniform convergence with respect to the mesh density is obtained on regular triangular and distorted quadrilateral mesh sequences when employing a sufficiently high number of GMRES iterations on the coarse level. On the distorted and graded triangular mesh sequence the number of FGMRES iterations increases with the mesh density due to the presence of increasingly stretched elements close to the domain boundaries. Note that the number of iteration increase is less pronounced when employing six-levels instead of three-levels for the $\mathcal{V}$-cycle iteration.

To conclude, we mention that the number of iterations of rescaled-inherited and non-inherited multigrid has been checked to be equal on all but the finest grids of the distorted and graded triangular mesh sequence, where the former is slightly sub-optimal as compared to the latter (by at most 20\%). This confirms that stabilization terms scaling is almost able to recover the convergence rates of non-inherited multigrid while also cutting down assembly costs for diffusion dominated regimes.

\begin{table}
\begin{tabular}{l | c c c c | c c c c | c c}
\cline{2-11}
\multicolumn{1}{c}{\multirow{4}{*}{}} & \multicolumn{3}{c}{Solver} & \multicolumn{1}{c}{$\ell$} & $\Kl$ & rTol & ITs & \multicolumn{3}{c}{Smoother} \\
  \cline{2-11}
 \multicolumn{1}{c}{} &  \multicolumn{3}{c}{\multirow{3}{*}{FGMRES[MG${}_{\mathcal{V}}$], $L=2$}} &\multicolumn{1}{c}{0} &6 &$-$       &1        & \multicolumn{3}{c}{\multirow{3}{*}{GMRES[ILU(0)]}}\\
 \multicolumn{1}{c}{}& \multicolumn{3}{c}{}                                                       &\multicolumn{1}{c}{1} &3 &$-$       &1        & \multicolumn{3}{c}{} \\
\multicolumn{1}{c}{} & \multicolumn{3}{c}{}                                                       &\multicolumn{1}{c}{2} &1 &$10^{-3}$ &400      & \multicolumn{3}{c}{} \\
  \hline
grid & \multicolumn{4}{c|}{scaling off}   & \multicolumn{4}{c|}{scaling on} & \multicolumn{2}{c}{speedup} \\
\hline
reg-tri  & $\rho$ & $\rho_c$ & ITs & ITs${}_c$ & $\rho$ & $\rho_c$ & ITs & ITs${}_c$ & Tot & Sol \\
39${}^2$$\cdot$2 & 0.0822 & 0.957 & 10 & 157 & 0.112 & 0.833 & 11 & 38 & 1.3 & 1.7 \\
79${}^2$$\cdot$2 & 0.0711 & 0.969 & 9 & 223 & 0.108 & 0.929 & 11 & 95 & 1.7 & 2.3 \\
158${}^2$$\cdot$2 & 0.0847 & 0.99 & 10 & 399 & 0.117 & 0.946 & 11 & 125 & 2 & 2.5 \\
311${}^2$$\cdot$2 & 0.133 & 0.997 & 12 & 399 & 0.107 & 0.982 & 11 & 385 & 1.2 & 1.2 \\
\hline
dist-quad  & $\rho$ & $\rho_c$ & ITs & ITs${}_c$ & $\rho$ & $\rho_c$ & ITs & ITs${}_c$ & Tot & Sol  \\
32${}^2$ & 0.0718 & 0.913 & 9 & 76 & 0.0409 & 0.764 & 8 & 26 & 1.2 & 1.5 \\
64${}^2$ & 0.0694 & 0.956 & 9 & 155 & 0.0405 & 0.865 & 8 & 48 & 1.4 & 2 \\
128${}^2$ & 0.0641 & 0.966 & 9 & 200 & 0.0369 & 0.881 & 7 & 55 & 1.9 & 3.3 \\
256${}^2$ & 0.0606 & 0.989 & 9 & 399 & 0.0279 & 0.957 & 7 & 159 & 2.1 & 3 \\
\hline
grad-tri & $\rho$ & $\rho_c$ & ITs & ITs${}_c$ & $\rho$ & $\rho_c$ & ITs & ITs${}_c$ & Tot & Sol  \\
32${}^2$$\cdot$2 & 0.141 & 0.909 & 12 & 73 & 0.165 & 0.747 & 13 & 24 & 1.1 & 1.1 \\
64${}^2$$\cdot$2 & 0.204 & 0.929 & 15 & 94 & 0.214 & 0.774 & 15 & 27 & 1.2 & 1.4 \\
128${}^2$$\cdot$2 & 0.285 & 0.96 & 19 & 170 & 0.315 & 0.925 & 20 & 89 & 1.5 & 1.8 \\
256${}^2$$\cdot$2 & 0.359 & 0.965 & 23 & 194 & 0.436 & 0.944 & 28 & 120 & 1.3 & 1.4 \\
\hline
\multicolumn{1}{c}{\multirow{4}{*}{}} & \multicolumn{3}{c}{Solver} & \multicolumn{1}{c}{$\ell$} & $\Kl$ & rTol & ITs & \multicolumn{3}{c}{Smoother} \\
  \cline{2-11}
 \multicolumn{1}{c}{} &  \multicolumn{3}{c}{\multirow{3}{*}{FGMRES[MG${}_{\mathcal{V}}$], $L=2$}} &\multicolumn{1}{c}{0} &6 &$-$       &1        & \multicolumn{3}{c}{\multirow{3}{*}{GMRES[ILU(0)]}}\\
 \multicolumn{1}{c}{}& \multicolumn{3}{c}{}                                                       &\multicolumn{1}{c}{1} &3 &$-$       &1        & \multicolumn{3}{c}{} \\
\multicolumn{1}{c}{} & \multicolumn{3}{c}{}                                                       &\multicolumn{1}{c}{2} &1 &$10^{-3}$ &40       & \multicolumn{3}{c}{} \\
  \hline
grid & \multicolumn{4}{c|}{scaling off}   & \multicolumn{4}{c|}{scaling on} & \multicolumn{2}{c}{speedup} \\
\hline
reg-tri  & $\rho$ & $\rho_c$ & ITs & ITs${}_c$ & $\rho$ & $\rho_c$ & ITs & ITs${}_c$ & Tot & Sol \\
39${}^2$$\cdot$2 & 0.194 & 0.962 & 15 & 39 & 0.112 & 0.909 & 11 & 39 & 1.1 & 1.3 \\
79${}^2$$\cdot$2 & 0.356 & 0.983 & 23 & 39 & 0.142 & 0.954 & 12 & 39 & 1.4 & 1.8 \\
158${}^2$$\cdot$2 & 0.665 & 0.993 & 56 & 39 & 0.226 & 0.985 & 16 & 39 & 2.2 & 3.2 \\
311${}^2$$\cdot$2 & 0.819 & 0.994 & 113 & 39 & 0.408 & 0.986 & 26 & 39 & 3.1 & 4.1 \\
\hline
dist-quad & $\rho$ & $\rho_c$ & ITs & ITs${}_c$ & $\rho$ & $\rho_c$ & ITs & ITs${}_c$ & Tot & Sol  \\
32${}^2$ & 0.0939 & 0.962 & 10 & 39 & 0.0409 & 0.72 & 8 & 21 & 1.1 & 1.3 \\
64${}^2$ & 0.162 & 0.982 & 13 & 39 & 0.0406 & 0.877 & 8 & 39 & 1.2 & 1.5 \\
128${}^2$ & 0.351 & 0.976 & 23 & 39 & 0.0669 & 0.958 & 9 & 39 & 1.5 & 2.3 \\
256${}^2$ & 0.621 & 0.993 & 48 & 39 & 0.148 & 0.958 & 13 & 39 & 2.2 & 3.3 \\
\hline
grad-tri & $\rho$ & $\rho_c$ & ITs & ITs${}_c$ & $\rho$ & $\rho_c$ & ITs & ITs${}_c$ & Tot & Sol  \\
32${}^2$$\cdot$2 & 0.146 & 0.898 & 12 & 39 & 0.164 & 0.765 & 13 & 26 & 0.99 & 0.99 \\
64${}^2$$\cdot$2 & 0.215 & 0.971 & 15 & 39 & 0.214 & 0.867 & 15 & 39 & 1 & 1 \\
128${}^2$$\cdot$2 & 0.381 & 0.988 & 24 & 39 & 0.315 & 0.911 & 20 & 39 & 1.1 & 1.2 \\
256${}^2$$\cdot$2 & 0.579 & 0.989 & 42 & 39 & 0.437 & 0.933 & 28 & 39 & 1.3 & 1.5 \\
\hline
\end{tabular}
\caption{$k=6$ BR2 discretization of the Laplace equation, three-levels \emph{p}-multigrid preconditioner performance
         on three $h$-refined mesh sequences, with and without stabilization scaling.
         Comparison of convergence rates of the outer solver and the coarse smoother ($\rho$ and $\rho_c$, respectively),
         comparison of the number of iterations of the outer solver and the coarse smoother (ITs and ITs${}_c$, respectively),
         and evaluation of the speedup $\left(\frac{\text{wall clock time scaling off}}{\text{wall clock time scaling on}}\right)$
         considering solution CPU time and solution plus assembly CPU time (Sol and Tot, respectively). \label{tab:Poisson3L}}
\end{table}

\begin{table}
\begin{tabular}{l | c c c c | c c c c | c c}
\cline{2-11}
\multicolumn{1}{c}{\multirow{4}{*}{}} & \multicolumn{3}{c}{Solver} & \multicolumn{1}{c}{$\ell$} & $\Kl$ & rTol & ITs & \multicolumn{3}{c}{Smoother} \\
  \cline{2-11}
\multicolumn{1}{c}{}  &  \multicolumn{3}{c}{\multirow{2}{*}{FGMRES[MG${}_{\mathcal{V}}$], $L=5$}} &\multicolumn{1}{c}{0,...,4} &$6-\ell$ &$-$  &1        & \multicolumn{3}{c}{\multirow{2}{*}{GMRES[ILU(0)]}}\\
\multicolumn{1}{c}{} & \multicolumn{3}{c}{}                                                       &\multicolumn{1}{c}{5}         &1 &$10^{-3}$ &400        & \multicolumn{3}{c}{} \\
\hline
 grid  & \multicolumn{4}{c|}{scaling off}   & \multicolumn{4}{c|}{scaling on} & \multicolumn{2}{c}{speedup} \\
  \hline
reg-tri & $\rho$ & $\rho_c$ & ITs & ITs${}_c$ & $\rho$ & $\rho_c$ & ITs & ITs${}_c$ & Tot & Sol \\
39${}^2$$\cdot$2 & 0.0288 & 0.965 & 7 & 196 & 0.0176 & 0.885 & 6 & 57 & 1.5 & 2 \\
79${}^2$$\cdot$2 & 0.0281 & 0.982 & 7 & 389 & 0.0163 & 0.913 & 6 & 76 & 1.8 & 2.5 \\
158${}^2$$\cdot$2 & 0.0502 & 0.996 & 8 & 399 & 0.0186 & 0.962 & 6 & 177 & 1.8 & 2.2 \\
311${}^2$$\cdot$2 & 0.102 & 0.997 & 11 & 399 & 0.0161 & 0.99 & 6 & 399 & 1.7 & 1.9 \\
\hline
dist-quad  & $\rho$ & $\rho_c$ & ITs & ITs${}_c$ & $\rho$ & $\rho_c$ & ITs & ITs${}_c$ & Tot & Sol  \\
32${}^2$ & 0.0274 & 0.908 & 7 & 72 & 0.00659 & 0.786 & 5 & 29 & 1.2 & 1.6 \\
64${}^2$ & 0.0267 & 0.963 & 7 & 186 & 0.0063 & 0.876 & 5 & 53 & 1.4 & 2 \\
128${}^2$ & 0.0215 & 0.98 & 6 & 350 & 0.00623 & 0.927 & 5 & 92 & 1.7 & 2.4 \\
256${}^2$ & 0.0325 & 0.995 & 7 & 399 & 0.00403 & 0.967 & 5 & 206 & 1.8 & 2.4 \\
\hline
grad-tri & $\rho$ & $\rho_c$ & ITs & ITs${}_c$ & $\rho$ & $\rho_c$ & ITs & ITs${}_c$ & Tot & Sol  \\
32${}^2$$\cdot$2 & 0.0684 & 0.887 & 9 & 58 & 0.0443 & 0.79 & 8 & 30 & 1.1 & 1.3 \\
64${}^2$$\cdot$2 & 0.0907 & 0.957 & 10 & 157 & 0.0728 & 0.84 & 9 & 40 & 1.4 & 1.8 \\
128${}^2$$\cdot$2 & 0.146 & 0.958 & 12 & 163 & 0.117 & 0.93 & 11 & 96 & 1.4 & 1.6 \\
256${}^2$$\cdot$2 & 0.208 & 0.982 & 15 & 371 & 0.168 & 0.955 & 13 & 150 & 1.8 & 2.1 \\
\hline
\multicolumn{1}{c}{\multirow{4}{*}{}} & \multicolumn{3}{c}{Solver} & \multicolumn{1}{c}{$\ell$} & $\Kl$ & rTol & ITs & \multicolumn{3}{c}{Smoother} \\
  \cline{2-11}
\multicolumn{1}{c}{}  &  \multicolumn{3}{c}{\multirow{2}{*}{FGMRES[MG${}_{\mathcal{V}}$], $L=5$}} &\multicolumn{1}{c}{0,...,4} &$6-\ell$ &$-$  &1        & \multicolumn{3}{c}{\multirow{2}{*}{GMRES[ILU(0)]}}\\
\multicolumn{1}{c}{} & \multicolumn{3}{c}{}                                                       &\multicolumn{1}{c}{5}         &1 &$10^{-3}$ &40         & \multicolumn{3}{c}{} \\
\hline
grid & \multicolumn{4}{c|}{scaling off}   & \multicolumn{4}{c|}{scaling on} & \multicolumn{2}{c}{speedup} \\
  \hline
reg-tri & $\rho$ & $\rho_c$ & ITs & ITs${}_c$ & $\rho$ & $\rho_c$ & ITs & ITs${}_c$ & Tot & Sol \\
39${}^2$$\cdot$2 & 0.166 & 0.972 & 14 & 39 & 0.0279 & 0.916 & 7 & 39 & 1.4 & 1.8 \\
79${}^2$$\cdot$2 & 0.339 & 0.985 & 22 & 39 & 0.0664 & 0.94 & 9 & 39 & 1.6 & 2.2 \\
158${}^2$$\cdot$2 & 0.661 & 0.991 & 55 & 39 & 0.206 & 0.983 & 15 & 39 & 2.5 & 3.3 \\
311${}^2$$\cdot$2 & 0.829 & 0.988 & 122 & 39 & 0.391 & 0.979 & 25 & 39 & 3.7 & 4.6 \\
\hline
dist-quad & $\rho$ & $\rho_c$ & ITs & ITs${}_c$ & $\rho$ & $\rho_c$ & ITs & ITs${}_c$ & Tot & Sol  \\
32${}^2$ & 0.0581 & 0.956 & 9 & 39 & 0.00658 & 0.794 & 5 & 31 & 1.2 & 1.6 \\
64${}^2$ & 0.139 & 0.979 & 12 & 39 & 0.00959 & 0.928 & 5 & 39 & 1.4 & 2 \\
128${}^2$ & 0.337 & 0.982 & 22 & 39 & 0.0442 & 0.964 & 8 & 39 & 1.7 & 2.4 \\
256${}^2$ & 0.613 & 0.985 & 47 & 39 & 0.137 & 0.959 & 12 & 39 & 2.5 & 3.5 \\
\hline
grad-tri  & $\rho$ & $\rho_c$ & ITs & ITs${}_c$ & $\rho$ & $\rho_c$ & ITs & ITs${}_c$ & Tot & Sol  \\
32${}^2$$\cdot$2 & 0.0743 & 0.924 & 9 & 39 & 0.0443 & 0.82 & 8 & 35 & 1.1 & 1.1 \\
64${}^2$$\cdot$2 & 0.143 & 0.975 & 12 & 39 & 0.0726 & 0.89 & 9 & 39 & 1.1 & 1.3 \\
128${}^2$$\cdot$2 & 0.343 & 0.989 & 22 & 39 & 0.118 & 0.973 & 11 & 39 & 1.5 & 1.8 \\
256${}^2$$\cdot$2 & 0.564 & 0.991 & 40 & 39 & 0.215 & 0.99 & 15 & 39 & 2 & 2.5 \\
\hline
\end{tabular}
\caption{$k=6$ BR2 discretization of the Laplace equation, six-levels \emph{p}-multigrid preconditioner performance on three $h$-refined mesh sequences, with and without stabilization scaling. Comparison of convergence rates of the outer solver and the coarse smoother ($\rho$ and $\rho_c$, respectively), comparison of the number of iterations of the outer solver and the coarse smoother (ITs and ITs${}_c$, respectively), and evaluation of the speedup $\left(\frac{\text{wall clock time scaling off}}{\text{wall clock time scaling on}}\right)$ considering solution CPU time and solution plus assembly CPU time (Sol and Tot, respectively).\label{tab:Poisson6L}}
\end{table}


\clearpage
\section*{References}


%
%
%
%
%
%

\end{document}